# AI-driven structural informatics reveals diverse glycosylation machinery across double-stranded DNA viruses

Lan Na[1,2]

1: Complex carbohydrate Research Center, University of Georgia, Athens, GA 30602, USA

2: Department of Biochemistry and Molecular Biology, University of Georgia, Athens, GA 30602, USA.

Corresponding author

Correspondence to Lan Na (Lan.Na@uga.edu)

## Abstract

Glycosylation is widespread in biology and shapes molecular recognition, yet the diversity and biosynthetic organization of viral glycosyltransferases (GTs) remain poorly understood. Here we integrate structural-alphabet-guided searches and structure similarity networks with genomic-neighbourhood and biosynthetic gene cluster embedding analyses to examine viral GT-A proteins. This approach reveals extensive structural diversity overlooked by sequence-based annotation, including previously uncharacterized viral GT lineages. Most strikingly, we identified a jumbo-phage locus encoding five tandem glycosyltransferases alongside hydroxymethyl-pyrimidine biosynthetic and sulfation-associated enzymes, suggesting a previously unrecognized form of viral DNA modification involving a sulfated pentasaccharide. Beyond these examples, predicted saccharide-associated loci span divergent double-stranded DNA viruses, from giant viruses to tailed phages, and occupy multiple regions of the reference biosynthetic landscape. Together, our findings uncover an underappreciated repertoire of virus-encoded glycosylation machinery and suggest that viral glycan biosynthesis encompasses diverse molecular contexts and potential biological functions.

## Introduction

Viruses are often viewed as genetically compact entities that depend extensively on host biosynthetic machinery. This view has been widely acknowledged in glycobiology, where viral glycosylation has largely been understood as a host-dependent process: in most well-studied eukaryotic viruses, glycans on viral proteins are installed or remodeled by host glycosyltransferases and glycosidases, and consequently resemble those of the infected cell (Speciale et al., 2022). Yet a growing body of work has shown that some viruses, particularly large and giant DNA viruses, encode enzymes involved in sugar-nucleotide biosynthesis, glycan assembly and glycan modification, suggesting that viruses are not exclusively dependent on host enzymes, rather it can possess a far autonomous than previously appreciated (Speciale et al., 2022; Notaro et al., 2022).

Recent advances in AI-based protein structure prediction and structure-guided comparison offer a way to overcome a limitation that are found in sequence-based annotation (van Kempen et al., 2024; Mifsud et al., 2024). Recent study highlights the emphasis of AI based structural prediction and computational approaches (Holmes et al., 2024). For example, Flaviridae evolution was elaborated to glycoprotein evolution through structure prediction, structure-guided phylogenetic analysis (Mifsud et al., 2024). Because structure is generally more conserved than primary sequence, fold-based analysis can reveal hidden evolutionary and functional relationships that remain invisible to sequence searches alone. This is especially relevant in viruses, where much of the biochemical landscape lies in the “dark matter” of poorly annotated proteins. We therefore reasoned that a structure-first framework could expose previously unrecognized viral glycosyltransferases, organize them into evolutionarily meaningful groups and connect candidate enzymes to broader glycan biosynthetic gene neighborhoods.

Here we present an integrative platform to decode the unknown viral glycome by combining structure-guided discovery, comparative analysis. We systematically mine viral proteomes for glycosyltransferase-like proteins, classify candidate enzymes using structural and sequence-level relationships, and annotate putative viral glycan biosynthetic gene clusters in their genomic context. This combined strategy moves beyond isolated annotation and instead treats viral glycosylation as an evolvable, pathway-level feature of viral genomes.

Using this framework, we identify putative glycosyltransferases across diverse viral taxa and uncover striking structural diversity, including viral clusters that do not map cleanly onto currently recognized CAZy families. These findings suggest that virus-encoded glycosylation is not limited to a few exceptional model systems, but instead represents a broader and underexplored dimension of viral biology. More broadly, our work establishes a generalizable route for discovering hidden carbohydrate-active enzymes in viral genomes and positions the viral glycome as a source of new evolutionary insight, biochemical function and potential biotechnological utility.

## Main

### AI-Driven, Structure-first workflows reveals a hidden viral glycosyltransferase space

To systematically explore the extent of glycosyltransferase-like proteins in viruses, we established an AI-driven, structure-guided two-stage discovery framework that integrates viral proteome mining, structure-based similarity search, structural clustering (**Figure 1**). Because viral proteins can diverge substantially at the sequence level for their rapid evolution, we used protein structural information to identify GT-like folds that may be missed by sequence-based rapidly evolving viral proteins. We first defined a minimal structural core from reference glycosyltransferases and encoded this query using the AI-learned structural alphabet 3Di (van Kempen et al., 2024), enabling fold-aware searches against ~300,000 synthetic structures derived from viral proteomes. The searches recovered 1,727 GT-A-query-derived candidates, revealing a broad viral structural space associated with established glycosyltransferase folds.

We next applied permissive structural clustering as a false-positive hits removal step to organize the GT-A-query-derived candidates. Clustering was performed without a minimum sequence-identity requirement and with a low symmetric coverage threshold, thereby allowing highly sequence-divergent proteins and proteins with substantial differences in length or domain architecture to remain connected through shared structural features. We hypothesized that genuine GT-A-like proteins would repeatedly recover a conserved fold-level architecture and therefore preferentially form multi-member clusters, whereas false positive hits including generic nucleotide-binding elements, partial structural cores or unrelated Rossmann-like proteins would remain dispersed among small clusters or singletons. All 1,727 candidates were assigned to 761 clusters, including 115 multi-member clusters containing 1,081 proteins and 646 singletons (**Supplementary Figure. 1**). Cluster sizes were highly uneven. The largest cluster contained 594 proteins, whereas the second-largest contained only 43. This marked difference prompted us to examine whether the dominant cluster represented a recurrent GT-A-like structural group.

To assess whether the dominant cluster captured known viral GT architectures, we performed a positive-control spike-in benchmark using six non-redundant, previously characterized viral GTs (A197, CvHAS, gp110, A111/114R, A064R, and L230; **Supplementary Figure. 2 a,b**). These controls were added to the original 1,727-protein dataset and subjected to structural clustering using the same permissive parameters. All six controls were assigned to the dominant cluster represented by A0A3Q8Q3U2 (**Supplementary Figure. 2a**). We next examined whether the controls shared direct structural similarity with the original members of this cluster, rather than being grouped together solely through permissive clustering. To test this, we aligned each control against the cluster representative, A0A3Q8Q3U2, and all 594 original cluster members (**Supplementary Figure. 2c**). Each control showed a high-coverage structural match to at least one original member (query coverage, 85.3–100%; target coverage, 90.4–100%; min(qTM, tTM), 0.625–0.956). In particular, A197 and gp110 showed substantial structural similarity (min(qTM, tTM) = 0.674 and 0.797, respectively) despite low sequence identities (19.4% and 21.1%). Together, these results show that the dominant cluster contains established viral GTs with detectable structural similarity to its original members, even when their sequence identities are low.

Having established that the dominant cluster recovered known viral GT architectures, we next tested whether recurrent GT-like structures were also present outside the dominant cluster. We independently performed the Foldseek structural search of all 114 non-giant cluster representatives and classified four as GT-like, 23 as non-GT and 87 as ambiguous or unresolved (**Supplementary Figure. 3 f**). Of these 114 clusters, 51 had direct

representative–member alignments in the initial Foldseek output, whereas 63 lacked such records. We therefore independently compared all protein pairs within the latter 63 clusters and found that 54 contained no pairs meeting both the predefined TM-score and alignment-coverage thresholds (**Supplementary Figure. 3d,e**). These findings supported prioritizing the dominant cluster for downstream analysis, presenting GT in giant cluster.

Following our analysis of the multi-member clusters, we examined the remaining 646 proteins assigned to singleton clusters across a total 761 clusters during the initial structural clustering. Because these proteins did not cluster with other viral candidates, we investigated whether singleton assignments were associated with short or low-confidence structural models. Of the 646 singletons, 328 (50.8%) were shorter than 122 amino acids and had a median pLDDT below 70, whereas 93 (14.4%) were at least 122 amino acids long with a median pLDDT of 70 or greater (**Supplementary Figure. 4d**). These results suggest that a substantial fraction of singletons consisted of short or poorly resolved structural models, although singleton status alone does not exclude potentially divergent GT-like proteins.

Together, these three analyses supported the enrichment of recurrent GT-A-like structural signal in the dominant cluster, whereas smaller clusters and singletons frequently showed uncertain functional assignments or limited structural support. We therefore selected the dominant cluster as the primary input for high-resolution all-versus-all structural alignment and downstream structure-similarity network analysis.

**Structure similarity networks reveal sequence, structural, and evolutionary divergence of viral glycosyltransferases across the GT-A landscape**

Many viral GT candidates remained unclassified by conventional sequence-based annotation, leaving their relationships to established glycosyltransferase families unresolved. We therefore hypothesized that viral GT lineages could be resolved within the broader GT-A structural landscape. To test this hypothesis, we performed all-versus-all structural comparisons of 4,854 proteins, combining viral GT candidates with reference GT-A proteins from across cellular life. The resulting network retained 224,825 pairwise structural relationships, which were resolved into 80 multi-protein structural communities by weighted Leiden partitioning (**Figure. 2A**). Strikingly, although CAZy annotations were not used for community detection, the structural network largely recapitulated sequence-defined family organization: more than 84% of evaluable communities and CAZy families exhibited ≥80% purity and concentration, respectively (**Supplementary Figure. 5**). Together, these results provide a structural framework for inferring the potential functions of uncharacterized viral GTs through their relationships to established GT families.

We found that 383 of the 687 viral proteins (55.7%) lacked a CAZy-family assignment (**Figure. 2B**). Yet, their structural relationships revealed a more complex picture that are not captured by sequence-based analysis alone. Notably, nearly one-third of these unassigned viral GTs (110 of 383, 28.7%) showed direct, family-specific structural connections to CAZy-annotated GTs. Yet, their structural relationships displaying various structural landscape: Some unassigned viral proteins occupied weakly connected family peripheries (with weak edge), others connected distinct structural communities, while a subset formed virus-enriched groups without CAZy-annotated members. Thus, CAZy-unassigned viral GTs do not represent a single unresolved group. Instead, structural mapping distinguishes viral GTs with recognizable relationships to known families from those occupying previously uncharacterized regions of the GT-A structural landscape.

Two viral glycosyltransferases shows distinct relationships between sequence divergence, structural conservation and enzymatic function. F2Y1U4 from Organic Lake phycodnavirus 2 lacked a CAZy-family assignment but placed within the GT31 structural landscape, which encompasses diverse glycosyltransferases involved in glycoprotein and glycolipid biosynthesis. To determine whether F2Y1U4 preserved a GT31-like architecture, we defined a conserved structural core from 17 GT31 reference proteins and mapped it onto F2Y1U4. Despite a mean sequence identity of only 16.42% across four GT31 reference subclusters, F2Y1U4 retained structural correspondence at 93 of 98 strict pan-GT31 core positions (**Supplementary Figure. 6A**). This supports a previously unrecognized structural relationship to GT31, although its precise enzymatic function remains unresolved.

R699 from Acanthamoeba polyphaga mimivirus provided a contrasting example. R699 and the Caenorhabditis elegans PLOD-like glycosyltransferase Q20679 (LET-268) both catalyse UDP-glucose-dependent glucosylation of galactosylhydroxylysine in collagen (Wang et al., 2002; Wu et al., 2022), yet occupied distinct positions within the structural network. Notably, R699 connected two structural communities, C005 and C009 **(Supplementary Figure. 6B**). Mapping the independently defined structural cores of these communities onto R699 revealed that their shared structural support was concentrated predominantly within the non-UDP-binding domain, rather than the UDP-binding catalytic domain. This indicates that R699's intercommunity connectivity is largely associated with conservation of its accessory-domain architecture, despite its shared enzymatic activity with cellular collagen glucosyltransferases.

Further, we identified three distinctive CAZy-unassigned, viral GT structural groups: Community 027 (C027), cluster 7 and 8. C027 was particularly unusual, forming an isolated seven-member viral structural community. Yet, C027 indeed shared structural architecture and conserved putative active-site features (**Figure 2D**). We further identified 27 of C027 viral homologues in UniRef90, revealing that C027 homologues were found across divergent eukrayotes infecting ds DNA viruses including giant viruses, from insect-infecting baculoviruses to bird-infecting avipoxvirues (**Supplementary Figure 7**).

CAZy-unassigned viral proteins were not randomly dispersed across GT structural space, rather local network topology resolved them into discrete, virus-enriched lineages, including cluster 7 and 8 (**Figure. 2c**). Cluster 7 formed a cohesive viral subnetwork within community 006, a mixed community containing GT8, GT77 and GT116. To examine evolutionary relationships obscured by sequence divergence, we used FoldTree, a structure-informed phylogenetic method (Moi et al., 2025). Cluster 7 grouped with environmental MGnify homologues on a distinct branch that joined other GT-like lineages deeper in the tree (Supplementary Figure. X-2-4). This suggests that Cluster 7 has diverged within a broader GT-like structural lineage rather than representing an isolated new fold. Conserved residues mapped near the putative nucleotide-sugar-binding and catalytic regions (**Figure. 2d**), while Cluster 7-containing loci included genes associated with sugar biosynthesis, carbohydrate modification and virion structure. These observations also suggest a potential role in virion glycosylation.

Cluster 8 formed another compact six-protein module within C001 (**Figure. 2c,d**), but its genomic neighbourhoods pointed to different biosynthetic contexts. Some members occurred alongside hydroxymethyl-pyrimidine and DNA-modification enzymes, whereas others were associated with virion structural and glycan-related genes (**Figure. 4**). This suggests that structurally related viral GTs may participate in distinct glycosylation pathways. We therefore focused on the DNA-associated loci, where neighbouring enzymes provided a chemical basis for investigating glycosylation of modified viral DNA.

**A divergent viral glycosyltransferase lineage occupies distinct DNA-modification and virion-associated genomic contexts**

Within the viral GT structure-similarity network, Cluster 8 emerged as a coherent group of CAZy-unassigned viral proteins (**Supplementary Figure. 9**). Despite substantial sequence divergence, Cluster 8 members retained high structural similarity (**Supplementary Figure. 11**) and shared a conserved GT-A-like structural core (**Figure. 3A,B**). Sequence- and structure-based conservation analyses further revealed that the central scaffold was preserved, whereas peripheral regions were more variable, particularly surface-exposed loops **(Figure. 3D,E**). Together, these findings support Cluster 8 as a structurally coherent, previously uncharacterized viral GT-like lineage.

Next, we analyzed whether cluster 8 members shared a conserved genomic context, we extracted local gene neighborhoods around each cluster 8 member and compared their gene neighborhoods (**Figure 4**). Neighboring genes were classified by predicted functions in hydroxymethyl-base precursor formation, pyrimidine/nucleotide metabolism, DNA metabolism, glycan biosynthesis and virion structure, along with uncharacterized ORFs. This comparison revealed at least two distinct genomic contexts rather than a single conserved gene organization.

In one subset, Cluster 8-containing gene neighborhoods showed disntict combinations of DNA-modification, glycan-biosynthetic and virion-associated genes. This contained predicted dCMP/dUMP hydroxymethylase-like enzymes and other hydroxymethyl-pyrimidine-associated functions, together with DNA-metabolism genes and additional GT-like proteins. These associations suggest a potential role in viral DNA glycosylation, with hydroxymethylated pyrimidines serving as candidate glycosyl acceptors.

In contrast, other Cluster 8-containing loci lacked comparable hydroxymethyl-pyrimidine precursor signatures but were associated with glycan-related enzymes and virion structural genes, including capsid, portal and tail proteins. This distinct genomic organization suggests a potential association with virion glycosylation rather than DNA modification, although the identity of the acceptor substrates remains unexplored. Thus, despite their shared structural architecture, Cluster 8 proteins occur in different glycosylation-associated genomic environments, raising the possibility that this divergent lineage participates in distinct viral biosynthetic processes. We therefore focused subsequent mechanistic analysis on the DNA-associated subset, where the accompanying precursor enzymes provide a chemical basis for investigating putative DNA-glycosylation pathways.

**A jumbo-phage multi-GT locus suggests a sulfated DNA-linked pentasaccharide biosynthetic pathway**

Comparative analysis of DNA-modification-associated Cluster 8 loci identified an unusual glycan-biosynthetic module in jumbo phage B2ZYD6 (**Figure. 5**). The Type III locus encodes five tandem GT-like proteins (BAG41682–BAG41686), a putative hydroxymethyl-pyrimidine precursor enzyme, and sulfation-associated proteins (BAG41681 and BAG41687). In contrast to canonical T4-like DNA-glucosylation systems, this combination suggests the capacity to assemble an extended DNA-linked glycan and further modify it through sulfation. One possible product is a sulfated pentasaccharide attached to a hydroxymethylated pyrimidine. Sulfated N-glycans have previously been identified on viral envelope glycoproteins (Shilatifard et al., 1993; Karaivanova and Spiro, 1998), but sulfation of DNA-linked glycans in bacteriophages has not, to our knowledge, been demonstrated. Such chemical elaboration could alter how bacterial defence systems recognize modified phage DNA, potentially contributing to the ongoing host–phage evolutionary conflict.

The Type IV locus, by contrast, contained a Cluster 8 GT alongside an α-glucosyltransferase-like enzyme related to characterized T4 enzymes that install the first glucose onto 5hmC-containing DNA ((Sommer et al., 2004; Larivière et al., 2005). This arrangement suggests a possible sequential glycosylation pathway, in which the Cluster 8 enzyme acts downstream of the initial glucosylation step (**Figure. 5**).

To investigate whether these Cluster 8-associated loci could provide suitable DNA-glycosylation acceptors, we examined their hydroxymethyl-pyrimidine precursor enzymes through structure-guided active-site analysis (**Supplementary Figures. 13 and 14**). Type I and IV loci encoded dCMP hydroxymethylase-like enzymes retaining the catalytic residues and cytosine-recognition Asp characteristic of dCMP-directed hydroxymethylation, consistent with the formation of 5hmC-type precursors. Type II loci instead contained TET/JBP-like dioxygenases, suggesting a possible post-replicative route to hydroxymethylated pyrimidines. In the Type III locus, BAG41679 retained a folate-dependent pyrimidine C5-hydroxymethylase scaffold but carried Asn242 in place of the Asp residue associated with cytosine recognition, suggesting that BAG41679 uses dUMP rather than d CMP.

We next investigated which enzyme in the expanded B2ZYD6 locus could initiate sugar installation onto hydroxymethylated DNA. Structure-guided comparison with characterized DNA glycosyltransferases identified two key structural elements in BAG41686: the αF helix and Loop X (**Figure. 6B**). The αF helix contains the predicted catalytic base and an adjacent aromatic residue positioned to form a π–π stacking interaction with the flipped 5hmU pyrimidine ring (Figure. 6C). This spatial pattern, also found in characterized first-sugar-installing viral GTs including contig 32 (Pyle et al., 2024), could position the hydroxymethyl oxygen towards the anomeric carbon of the nucleotide-sugar donor, while the nearby catalytic base facilitates proton abstraction and glycosyl transfer. Loop X retains a configure ration compatible with recognition of the DNA deoxyribose, further contributing the acceptor recognition.

To investigate how Cluster 8-associated GTs might catalyse the first and subsequent sugar-transfer steps in DNA glycosylation, we compared the active-site architectures of BAG41686 and AKJ73160 from the Type III and IV loci, respectively, with those of characterized DNA glycosyltransferases (**Figure. 6**). In the expanded B2ZYD6 Type III locus, BAG41686 retained two key structural elements involved in direct hydroxymethyl-base recognition: the αF helix and Loop X (**Figure. 6B**). The αF helix contains the predicted catalytic base and an adjacent aromatic residue positioned for π–π stacking with the flipped 5hmU pyrimidine ring (**Figure. 6C**). This arrangement, also found in characterized first-sugar-installing viral GTs such as contig 32 (Pyle et al., 2024), could position the hydroxymethyl oxygen towards the anomeric carbon of the nucleotide-sugar donor, while the nearby catalytic base facilitates proton abstraction and glycosyl transfer. Loop X further contributes to acceptor positioning through interactions with the DNA deoxyribose. Together, these features support BAG41686 as a candidate first-sugar installer.

In contrast, the Type IV locus encodes AKJ73146, a T4-like α-glucosyltransferase predicted to install the first glucose onto 5hmC-containing DNA (**Supplementary Figure. 16**), alongside the Cluster 8 enzyme AKJ73160. Unlike BAG41686 and other first-sugar-installing GTs, AKJ73160 lacks the αF-associated aromatic residue involved in direct hydroxymethyl-base recognition (**Figure. 6B,D**). Instead, the aromatic contact with the underlying 5hmC base is shifted to F151 in Loop X, while Asn174 is positioned near the glucose moiety already attached to 5hmC. This rearrangement of the acceptor-binding pocket could accommodate a pre-glucosylated DNA substrate and orient a hydroxyl group of the attached glucose for a subsequent glycosyl transfer. Thus, the contrasting active-site architectures suggest that Cluster 8 GTs may recognize different acceptor states, with BAG41686 initiating glycosylation directly on a hydroxymethylated base and AKJ73160 potentially extending an existing DNA-linked glycan.

The identification of BAG41686 as a candidate first-sugar installer provides a starting point for interpreting the expanded B2ZYD6 biosynthetic locus. The four additional GT-like proteins (BAG41682–BAG41685) could contribute to further glycan elaboration, although their individual roles remain unresolved. Notably, this locus also encodes a putative carbohydrate sulfotransferase (BAG41681) and an adenylyl-sulfate kinase-like enzyme (BAG41687) associated with PAPS biosynthesis, providing a possible route for sulfation of the DNA-linked glycan. Together, these features suggest a pathway capable of assembling an extended oligosaccharide and introducing additional chemical modifications, with a sulfated pentasaccharide representing one possible product. This combination of glycan assembly and secondary modification suggests that phage DNA may support a broader repertoire of carbohydrate chemistry than is currently characterized.

**Embedding analysis suggests broad spread of viral glycosylation module across dsDNA**

Having established the potential role of cluster 8 associated with DNA modification, we next asked whether comparable viral GT-associated gene neighborhoods occur across diverse viral genomes. To address this question, we extracted 11 gene window and 15 gene window centered on viral GTs, representing a total 466 genome units, and analyzed them using BGC-Prophet, a deep-learning model for predicting classes of biosynthetic gene clusters. Of the 1,477 evaluated viral GT-centered windows, 511 were classified as Saccharide-associated, comprising 246 15-gene and 265 11-gene windows. To place these findings in a broader biosynthetic context, we carried out a joint embedding analysis of a viral GT-associated loci and reference BGCs from MIBiG database. Leiden clustering identified total 10 global cluster (GC) that well corresponded to annotated biosynthetic product classes, suggesting that the embedding captured chemical difference of BGC classes among reference BGCs (**Supplementary Figure 18A**). We then calculated cosine similarities between viral and reference BGCs embeddings and selected viral loci for which at least five of ten nearest reference BGCs belonged to the Saccharide class (P@10 ≥ 0.5). 325 positive 15 -gene window results were identified as Saccharide-positive. Together, these observations raise the possibility that viral GTs participate in coordinated biosynthetic modules rather than acting solely as isolated enzymes, potentially contributing to diverse viral glycan structures.

To determine whether putative saccharide-biosynthetic loci were restricted to particular viral lineages or distributed more broadly, we examined their taxonomic distribution (**Fig. 7B**). Among the 242 candidate viral genome units, 154 belonged to Caudoviricetes and 67 to Megaviricetes, spanning tailed phages and large eukaryotic DNA viruses. Within Megaviricetes, candidate loci were identified in several families, including Mimiviridae (n = 20), Phycodnaviridae (n = 19) and Pithoviridae (n = 8). Notably, three additional candidate genome units were assigned to Adnaviria, a distinct realm of archaeal dsDNA viruses. Together, these findings suggest that saccharide-associated biosynthetic potential extends across distantly related DNA viral lineages. This broad taxonomic distribution prompted us to examine whether these viral loci share biosynthetic signatures with characterized saccharide BGCs.

To investigate whether putative viral saccharide-biosynthetic loci share features with characterized glycan biosynthetic pathways, we examined their distribution within the BGC embedding space (**Fig. 7C**). Leiden clustering partitioned 287 viral GT-associated loci into five communities (VC01–VC05), which occupied different regions of the joint embedding landscape. Among 195 Saccharide-annotated reference BGCs, 124 (63.6%) were assigned to two global communities, GC08 (n = 100) and GC05 (n = 24; Supplementary Fig. 19). Notably, 72% of the Saccharide-associated references in GC08 were linked to capsule, extracellular polysaccharide, LPS/LOS/O-antigen or other surface-glycan biosynthesis, whereas GC05 contained a greater proportion of aminoglycoside and glycosylated specialized metabolite references. The viral communities also showed different distributions across these global communities, with VC01, VC04 and VC05 represented in GC08, and VC02 and VC03 in GC05. Together, these observations suggest that viral GT-associated loci encompass diverse biosynthetic signatures, with a subset potentially contributing to extracellular or surface-associated glycan biosynthesis.

To investigate the genomic contexts of structurally defined Cluster 7 proteins, we compared representative loci from two cyanophages and two giant viruses (**Fig. 7D**). In S-SKS1 and S-PRM1, six and three neighbouring proteins, respectively, showed sequence relationships to proteins encoded by the Vibrio parahaemolyticus O/K-antigen-associated locus BGC0000780, including nucleotide-sugar precursor enzyme homologues. In giant viruses, CroV and Tupanvirus encoded Cluster 7 proteins within GT-rich genomic regions, where two additional pairs of GT-like proteins showed direct interviral sequence similarity. Both giant-virus regions also showed embedding similarity to a Staphylococcus aureus capsular-polysaccharide reference Together, these findings reveal that Cluster 7 represents a structurally conserved viral GT lineage embedded within diverse combinations of glycosyltransferases, sugar-precursor biosynthetic enzymes and glycan-modifying proteins, highlighting the remarkable biosynthetic complexity of viral glycosylation machinery.

## Discussion

A few Viruses have been known to synthesize glycans without relying entirely on host processing (Speciale et al., 2020, 2022; Freeman et al., 2023; Pyle et al., 2024). What has remained difficult to assess is the diversity of the enzymes involved, particularly when viral sequences have diverged beyond conventional family recognition. By comparing viral GT-like structures with reference glycosyltransferase from across cellular life, we found that the viral GTs positioned unique : some viral proteins retain recognizable family-level architectures, whereas others form structurally coherent groups with little direct support from annotated GTs. The finding that 110 CAZy-unassigned viral proteins have direct family-specific structural associations provides a starting point for functional investigation.

F2Y1U4 preserves much of the GT31 structural core at low sequence identity, whereas the position of the characterized viral GT R699 between structural communities is explained largely by conservation of an accessory domain. Cluster 7 occupies another position in this landscape: structural phylogeny links its viral proteins and environmental homologues to a wider GT-like group without placing them confidently in an established CAZy family. C027 requires particular care. Some of its homologues overlap the previously described Apsup/Ld109/Ac112–113 group, and Ld109 suppresses apoptosis during baculovirus infection (Yamada et al., 2011, 2013).

Genomic context adds a different kind of evidence. Cluster 8 is found in more than one genomic setting. Some loci contain hydroxymethyl-pyrimidine precursor and DNA-metabolism genes, while others contain virion-associated and glycan-related genes. Even the DNA-associated loci differ in their predicted precursor routes and arrangement of GT-like proteins. This suggests that related viral enzyme architectures may be deployed in distinct biosynthetic settings, but gene proximity cannot establish a shared product, acceptor or reaction sequence. The B2ZYD6 Type III locus offers the most distinctive chemical hypothesis. Five tandem GT-like genes occur alongside a putative hydroxymethyl-pyrimidine precursor enzyme and proteins associated with sulfate activation and transfer. BAG41679 has an active-site substitution compatible with a uracil-directed precursor route, and the modeled pocket of BAG41686 resembles those of enzymes that glycosylate hydroxymethylated DNA bases directly. In a different Cluster 8 context, AKJ73160 lacks the same base-stacking arrangement and may instead recognize an already glucosylated acceptor. These contrasting pockets suggest possible initiation and extension roles for related GT-like proteins. Recent work has established mono-, di- and trisaccharides on phage DNA, as well as glycosylated 5-hydroxycytosine produced through a distinct precursor route (Pyle et al., 2024; Lee et al., 2026). The distinguishing feature of B2ZYD6 is the proposed coupling of multiple glycosylation steps with sulfate-related chemistry.

Whether these predicted modifications protect viral genomes or alter particle recognition also remains open. Phage DNA sugars can reduce sensitivity to some host defenses while creating substrates for others (Mahler et al., 2025), and experimentally demonstrated virion glycosylation can alter immune recognition (Freeman et al., 2023). We identified putative pentasaccahride sulfated mode on jumbo phage DNA. High-throughput experimentation could facilitate the discovery and functional characterization of previously unrecognized viral glycosylation pathways and their associated glycan products.

Beyond the structurally defined viral GT lineages, our genome-wide analysis suggests that putative glycan biosynthetic capacity is distributed across diverse DNA viruses. Candidate saccharide-associated loci were identified in both Caudoviricetes and Megaviricetes, with additional candidates in archaeal viruses. Embedding-based comparisons further revealed that viral GT-associated gene neighbourhoods occupy regions associated with different classes of characterized saccharide biosynthetic gene clusters, including pathways involved in capsule, extracellular polysaccharide and other surface-glycan biosynthesis. These associations raise the possibility that viral glycosylation extends beyond DNA modification to encompass a broader range of glycan biosynthetic processes. These findings extend the structural analysis to a broader genomic landscape and identify candidate biosynthetic loci for further biochemical investigation.

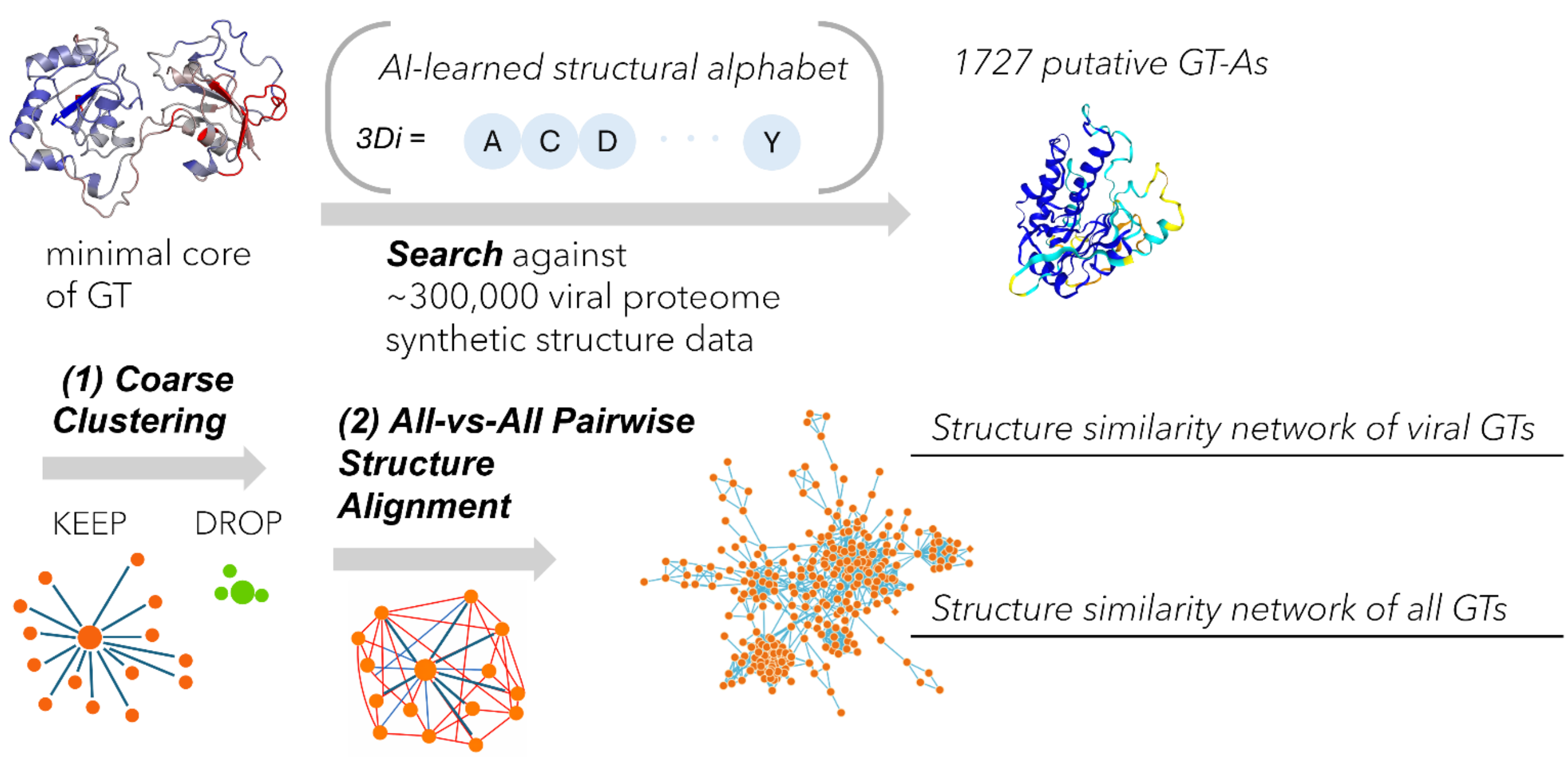

**Figure 1**. **Structure-guided discovery and clustering of viral glycosyltransferase candidates.**

A minimal structural core derived from reference GT-A glycosyltransferases was searched against approximately 300,000 predicted viral protein structures using Foldseek (van Kempen et al., 2024). The search recovered 1,727 GT-A-query-derived candidates, which were grouped into 761 clusters comprising 115 multi-member clusters and 646 singletons. The largest cluster contained 594 proteins. The dominant cluster was subjected to all-versus-all structural alignment to generate a viral glycosyltransferase candidate structure-similarity network.

**A.**

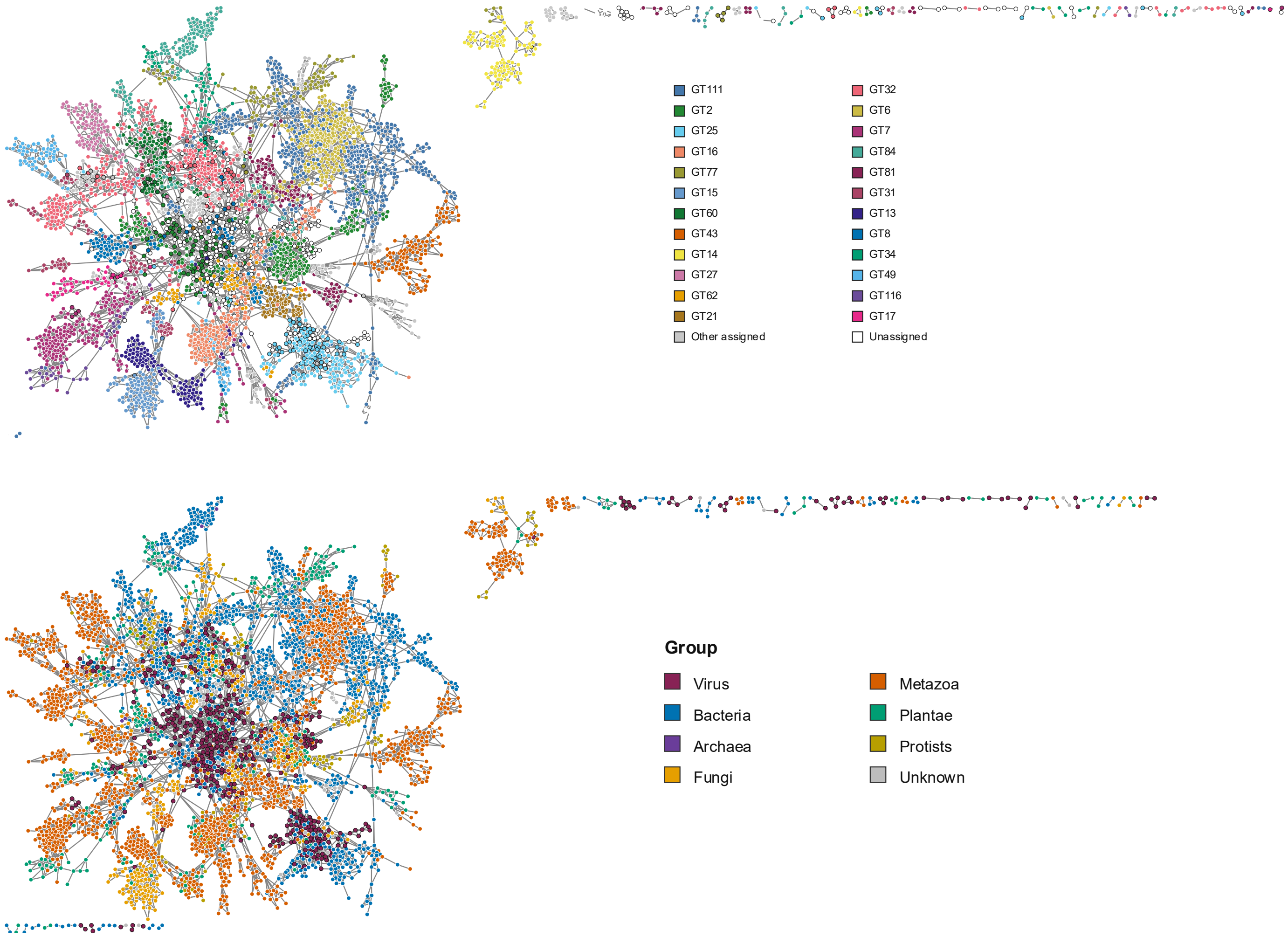
GT111
GT2
GT25
GT16
GT77
GT15
GT60
GT43
GT14
GT27
GT62
GT21
Other assigned
GT32
GT6
GT7
GT84
GT81
GT31
GT13
GT8
GT34
GT49
GT116
GT17
Unassigned
Group
Virus
Bacteria
Archaea
Fungi
Metazoa
Plantae
Protists
Unknown

**b** **Structural placement of viral proteins**

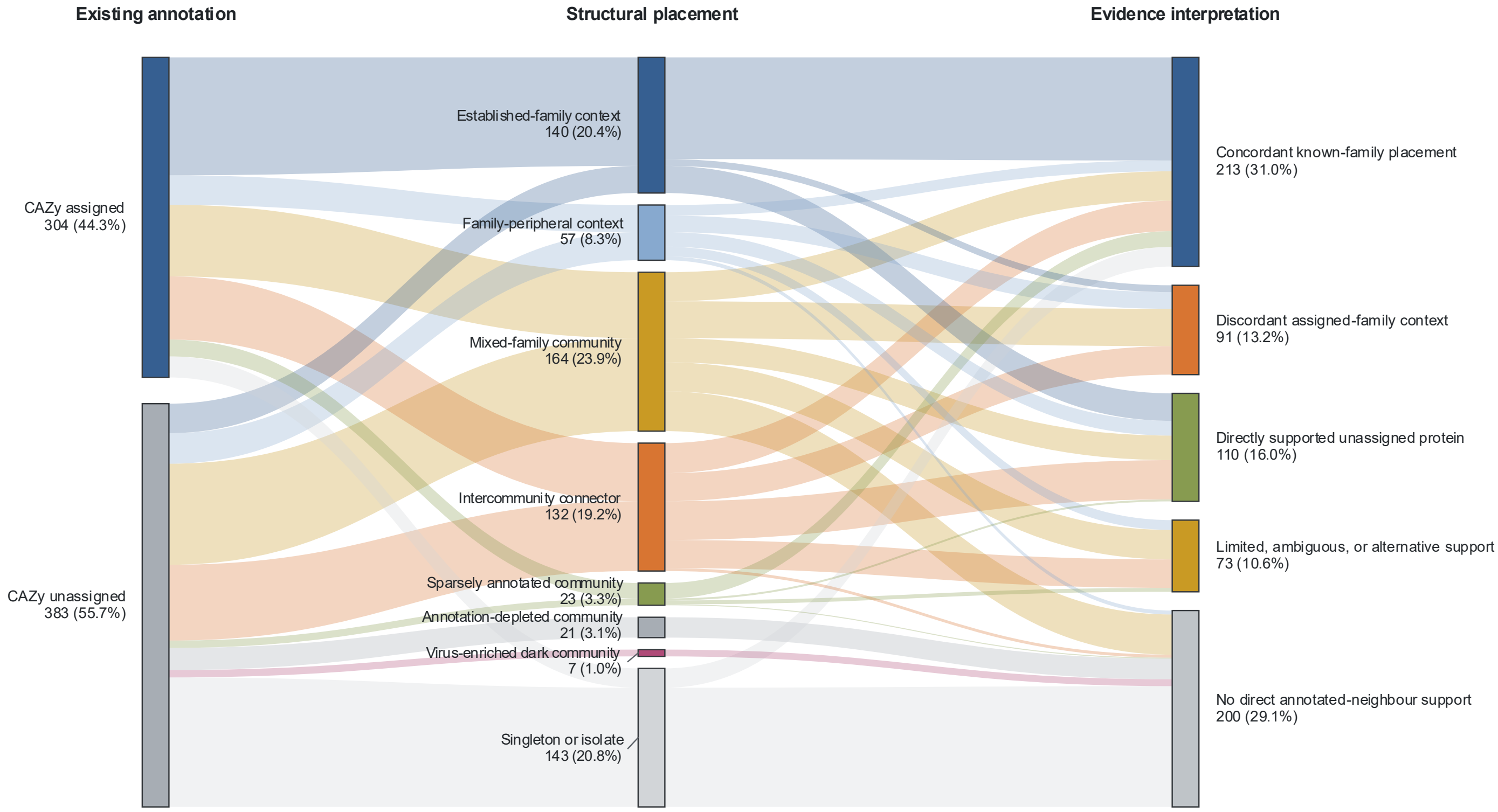

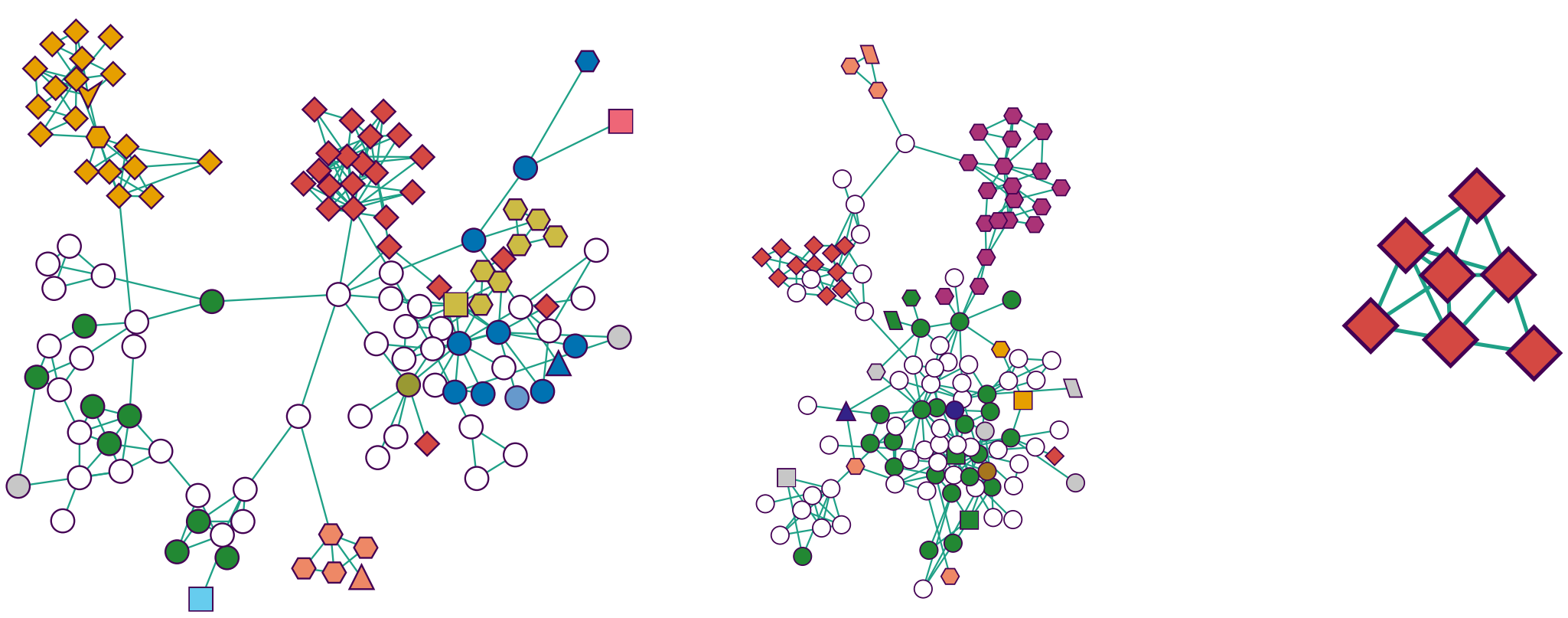

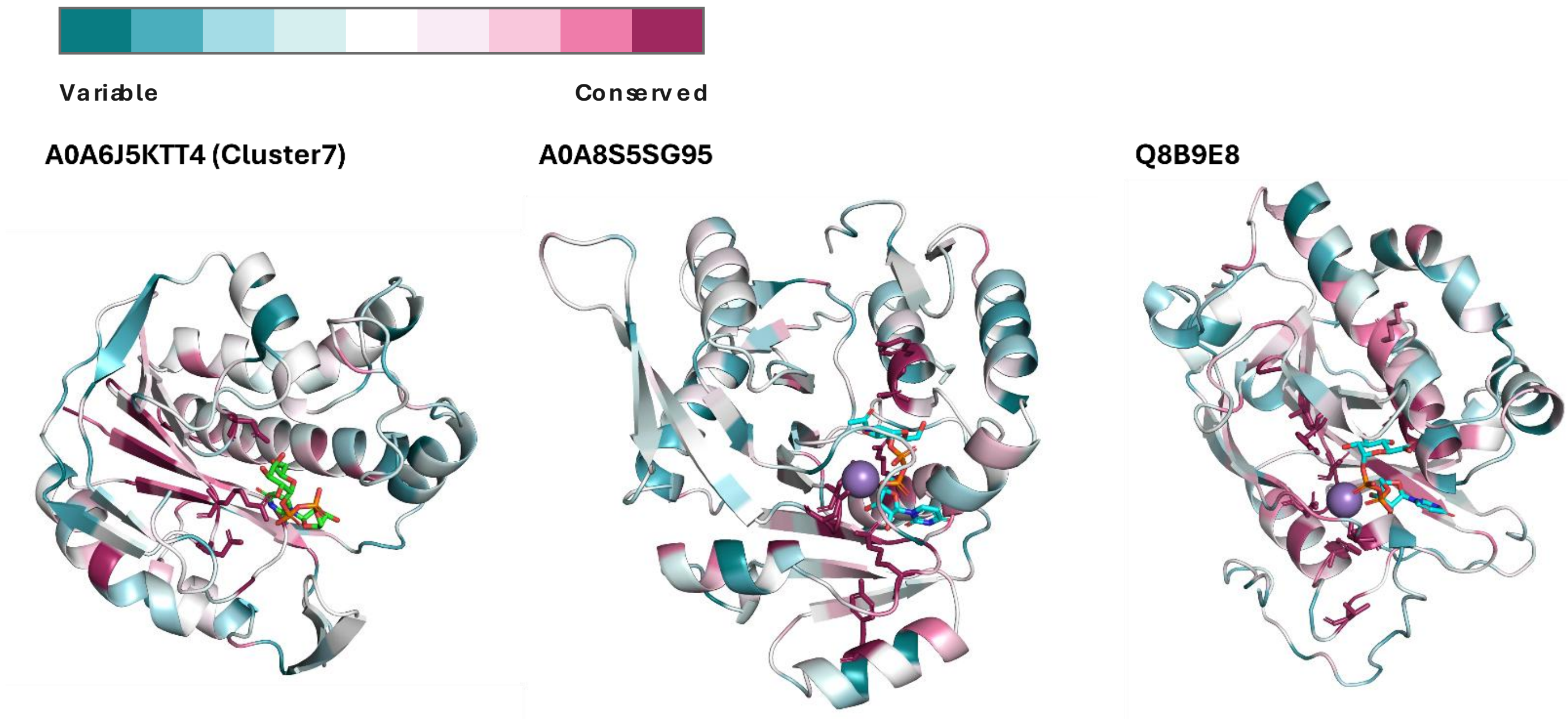

**Figure. 2** A. Viral glycosyltransferase candidates span the GT-A structural landscape. Matched views of the structure-similarity network coloured by CAZy base-family assignment (left) or broad taxonomic origin (right). Nodes represent protein structures and edges indicate structural similarity. Less abundant CAZy-assigned families are shown in grey and unassigned proteins in white. Viral proteins are enlarged and outlined in black. The same node coordinates are used in both views; isolated proteins are not shown.

**B. Structural placement and CAZy-family associations of viral GTs.** Alluvial diagram classifying 687 viral proteins by existing CAZy annotation, structural context and family-association evidence. Ribbon widths indicate protein counts.; Structural contexts were classified according to CAZy-family composition and network connectivity. Established-family communities contained at least five CAZy-annotated proteins, with ≥80% belonging to one family, whereas family-peripheral proteins showed weak connectivity within these communities (lowest quartile of within-community edge strength). Mixed-family communities contained substantial representation from at least two CAZy families, while intercommunity connectors shared appreciable structural edge weight across multiple communities. Sparsely annotated communities contained only 1–4 CAZy-annotated proteins, whereas annotation-depleted communities contained none. Virus-enriched dark communities were annotation-depleted communities significantly enriched in viral proteins. Singletons or isolates formed single-member communities or lacked retained network connections. Family-association evidence was assessed differently according to annotation status. For CAZy-assigned proteins, concordance or discordance indicates whether the existing family annotation matches the dominant CAZy family of their structural community. For CAZy-unassigned proteins, direct family support indicates that multiple annotated structural neighbours provide sufficiently strong and consistent evidence for association with a particular CAZy family. Limited, ambiguous or alternative support indicates insufficient, competing or community-discordant family evidence, respectively. No direct support indicates the absence of directly connected CAZy-annotated neighbours within the same community. These classifications describe structural family associations, not formal CAZy reassignments or experimentally established functions.

**C.** Local structural neighbourhoods of viral Clusters 7, 8, and C027(top). Nodes are coloured by CAZy-family assignment: GT2, dark green; GT6, mustard; GT7, magenta; GT8, blue; GT13, purple; GT15, steel blue; GT16, coral; GT21, brown; GT25, cyan; GT32, rose; GT62, orange; and GT77, olive. GT12, GT40, GT71, GT82 and GT109 are shown in grey, and CAZy-unassigned proteins in white. Node shapes denote taxonomic origin: bacteria, rectangles; fungi, diamonds; metazoa, hexagons; plants, triangles; viruses, ellipses; protists;

parallelogram; and proteins of unknown origin, V-shaped nodes. Cluster 7 and Cluster 8 members are highlighted as red diamonds. Lines indicate retained pairwise structural-similarity relationships. **D. Sequence conservation across representative dark viral glycosyltransferase lineages (Bottom).** Residue-level conservation derived from the corresponding multiple-sequence alignments was projected onto representative structures of Cluster 7, Cluster 8 and C027 (left to right). Residues are coloured from cyan, indicating variable positions, through white to magenta, indicating highly conserved positions. Bound or modelled nucleotide-sugar ligands are shown as sticks, and catalytic metal ions as purple spheres.

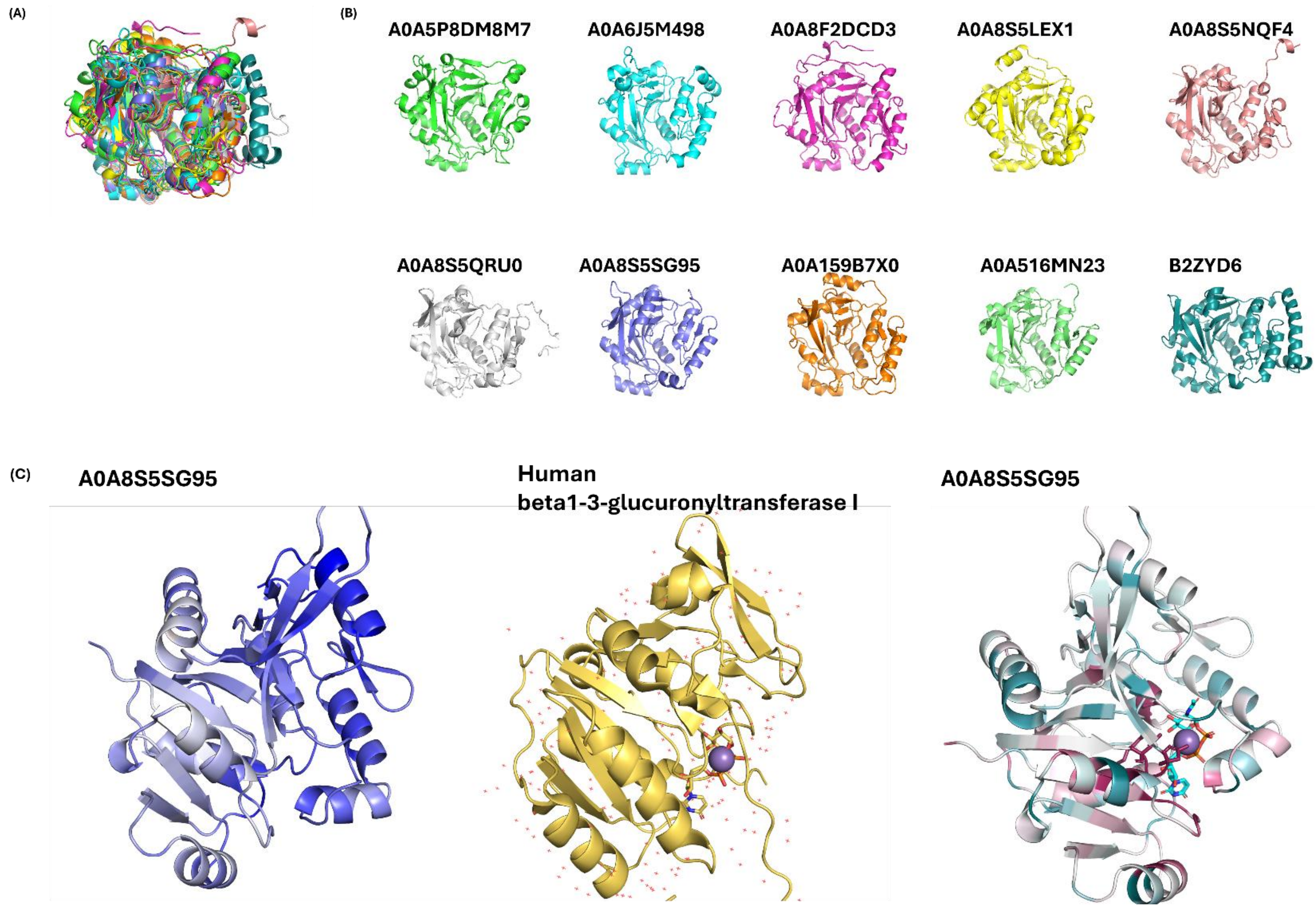

**Figure 3. Structural coherence and conserved core architecture define cluster 8 as a distinct GT-like family.**

**A.** Structural superposition of ten selected Cluster 8-associated viral proteins, revealing a shared GT-A-like core with variation in peripheral structural elements. Individual proteins are coloured as in B.

**B.** Predicted structures of the ten proteins included in A, displayed individually and labelled with their corresponding protein identifiers.

**C.** Structural comparison of the representative Cluster 8 protein A0A8S5SG95 (left) with human β1,3-glucuronyltransferase I (GlcAT-I; centre), a characterized GT43 enzyme adopting the GT-A fold. Residue-level sequence conservation derived from the corresponding multiple-sequence alignment was projected onto A0A8S5SG95 (right). Residues are coloured from cyan (variable positions), through white (intermediate conservation), to magenta (highly conserved positions).

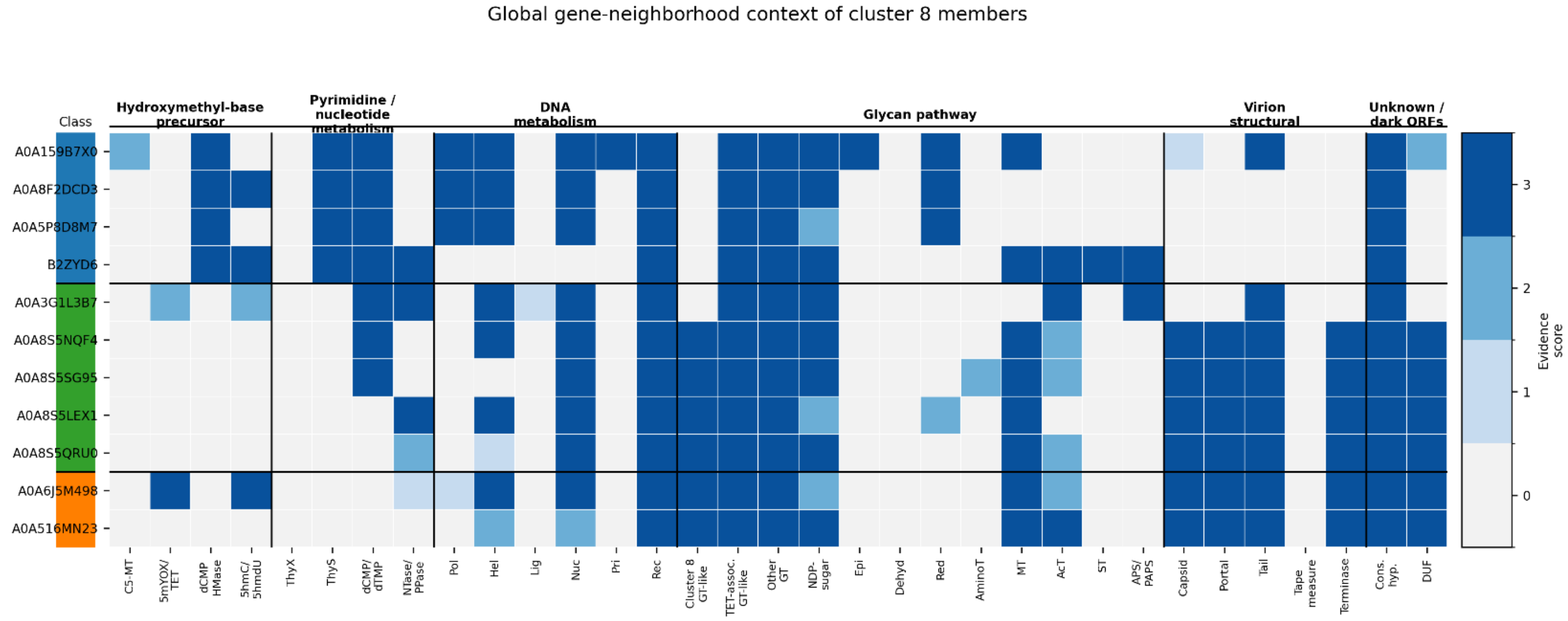

**Figure 4. Distinct genomic context of Cluster 8 member**

Heatmap of 11 viral gene neighborhoods containing Cluster 8 GT-like proteins (rows), ordered by inferred genomic context. Columns represent functional gene classes associated with hydroxymethyl-pyrimidine formation, nucleotide and DNA metabolism, glycan biosynthesis, virion structure, and uncharacterized proteins. Cell colour indicates the highest Foldseek-based annotation confidence among neighboring open reading frames assigned to each gene class: 0, no reliable evidence; 1, weak; 2, medium; and 3, high confidence.

Shortened labels in axis are: C5-cytosine methyltransferase / C5 methyltransferase (C5-MT), 5-methylpyrimidine dioxygenase / TET-like dioxygenase (5mYOX / TET), dCMP hydroxymethylase (dCMP HMase), Flavin-dependent thymidylate synthase (ThyX), Classical thymidylate synthase (ThyS), dCMP/dTMP-related pyrimidine metabolism enzyme (dCMP / dTMP), Nucleotidyltransferase / pyrophosphorylase (NTase / PPase). DNA polymerase (Pol), DNA helicase (Hel), DNA ligase (Lig), Nuclease / endonuclease / exonuclease (Nuc), DNA primase (Pri), Recombinase (Rec), Cluster 8 glycosyltransferase-like protein (Cluster 8 GT-like), TET-associated glycosyltransferase-like protein (TET-assoc. GT-like), Additional glycosyltransferase (Other GT), NDP-sugar biosynthesis enzyme / nucleotide-sugar enzyme (NDP-sugar), Epimerase (Epi), Dehydratase (Dehyd), Reductase (Red), Aminotransferase (AminoT), Methyltransferase (MT), Acetyltransferase (AcT), Sulfotransferase (ST), adenylyl-sulfate kinase or PAPS biosynthesis enzyme (PAPS), Capsid protein (Capsid), Portal protein (Portal), Tail (Tail protein), Tail tape-meaure protein (Tape measure), Terminase subunit (Terminase), Hypothetical protein (hyp), Domain of unknown function-containing protein (DUF).

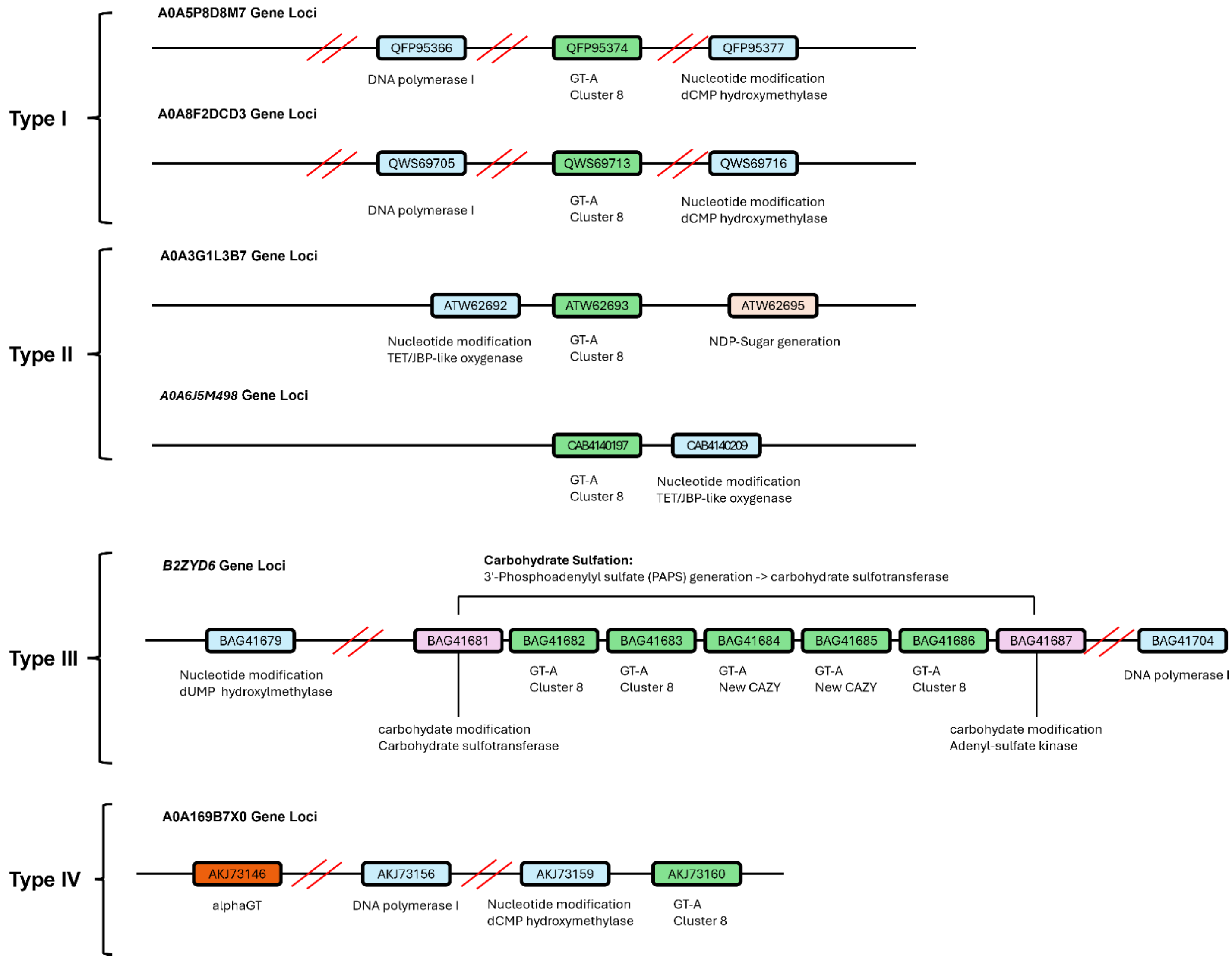

**Figure 5. DNA-associated cluster 8 loci partition into multiple hydroxymethyl-pyrimidine-linked architectures.**
Comparative gene-neighborhood diagrams of the DNA-modification-associated subset of cluster 8. Loci are grouped by pathway architecture according to the organization of hydroxymethyl-pyrimidine precursor genes, DNA-metabolism genes, cluster 8 GTs, additional GTs, and carbohydrate-modifying enzymes. Compact Type I/II loci contain cluster 8 GTs near pyrimidine- or DNA-metabolism genes, suggesting minimal DNA-modification-associated modules. Type III loci contain an expanded multi-GT architecture with a predicted hydroxymethyl-pyrimidine precursor enzyme and multiple tandem GT-like genes, suggesting a noncanonical DNA glycan assembly pathway. Type IV loci contain an α-GT-like enzyme together with a dCMP hydroxymethylase-like enzyme and DNA polymerase context, resembling a T4-like DNA glucosylation architecture. Red slashes indicate intervening genes omitted for display.

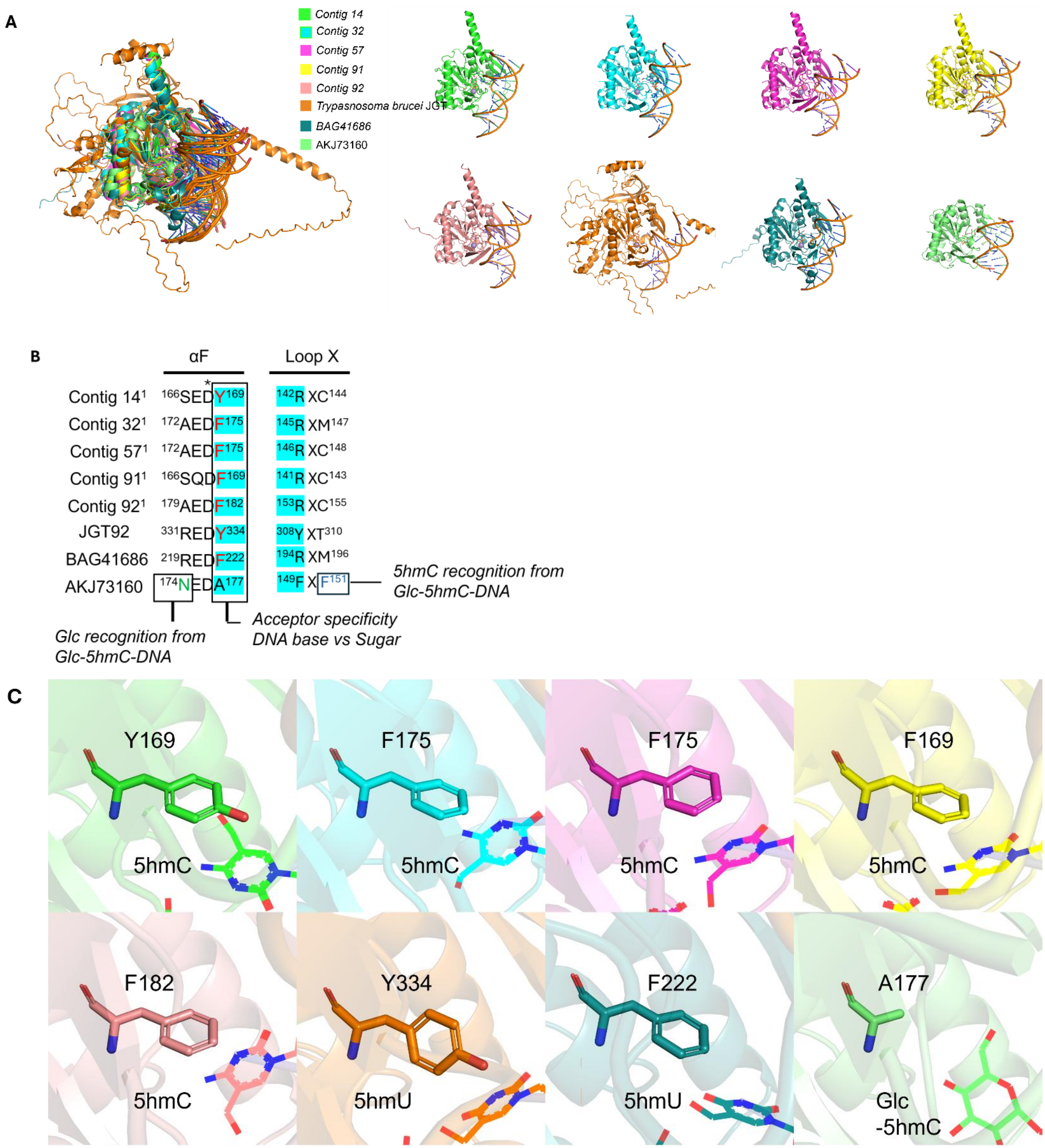
A
Contig 14
Contig 32
Contig 57
Contig 91
Contig 92
Trypasnosoma brucei JGT
BAG41686
AKJ73160
B
αF
Loop X
Contig 14[1] 166SEDY169 142RXC144
Contig 32[1] 172AEDF175 145RXM147
Contig 57[1] 172AEDF175 146RXC148
Contig 91[1] 166SQDF169 141RXC143
Contig 92[1] 179AEDF182 153RXC155
JGT92 331REDY334 308YXT310
BAG41686 219REDF222 194RXM196
AKJ73160 174NEDA177 149FXF151
5hmC recognition from Glc-5hmC-DNA
Acceptor specificity DNA base vs Sugar
Glc recognition from Glc-5hmC-DNA
C
Y169
5hmC
F175
5hmC
F175
5hmC
F169
5hmC
F182
5hmC
Y334
5hmU
F222
5hmU
A177
Glc -5hmC

D

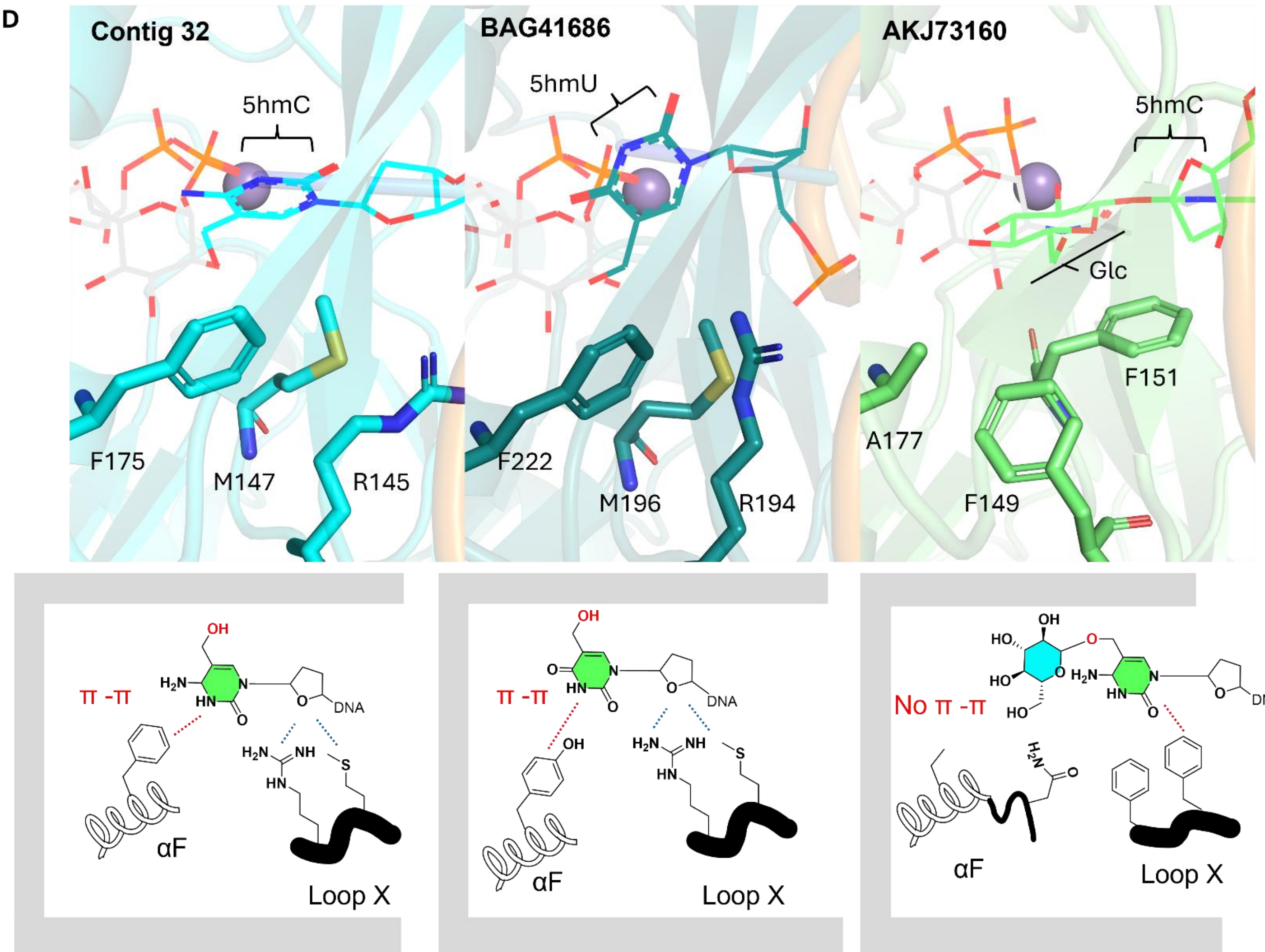
Contig 32
BAG41686
AKJ73160
5hmC
5hmU
5hmC
Glc
F175
M147
R145
F222
M196
R194
A177
F151
F149
π -π
π -π
No π -π
OH
DNA
αF
Loop X
αF
Loop X
αF
Loop X

E

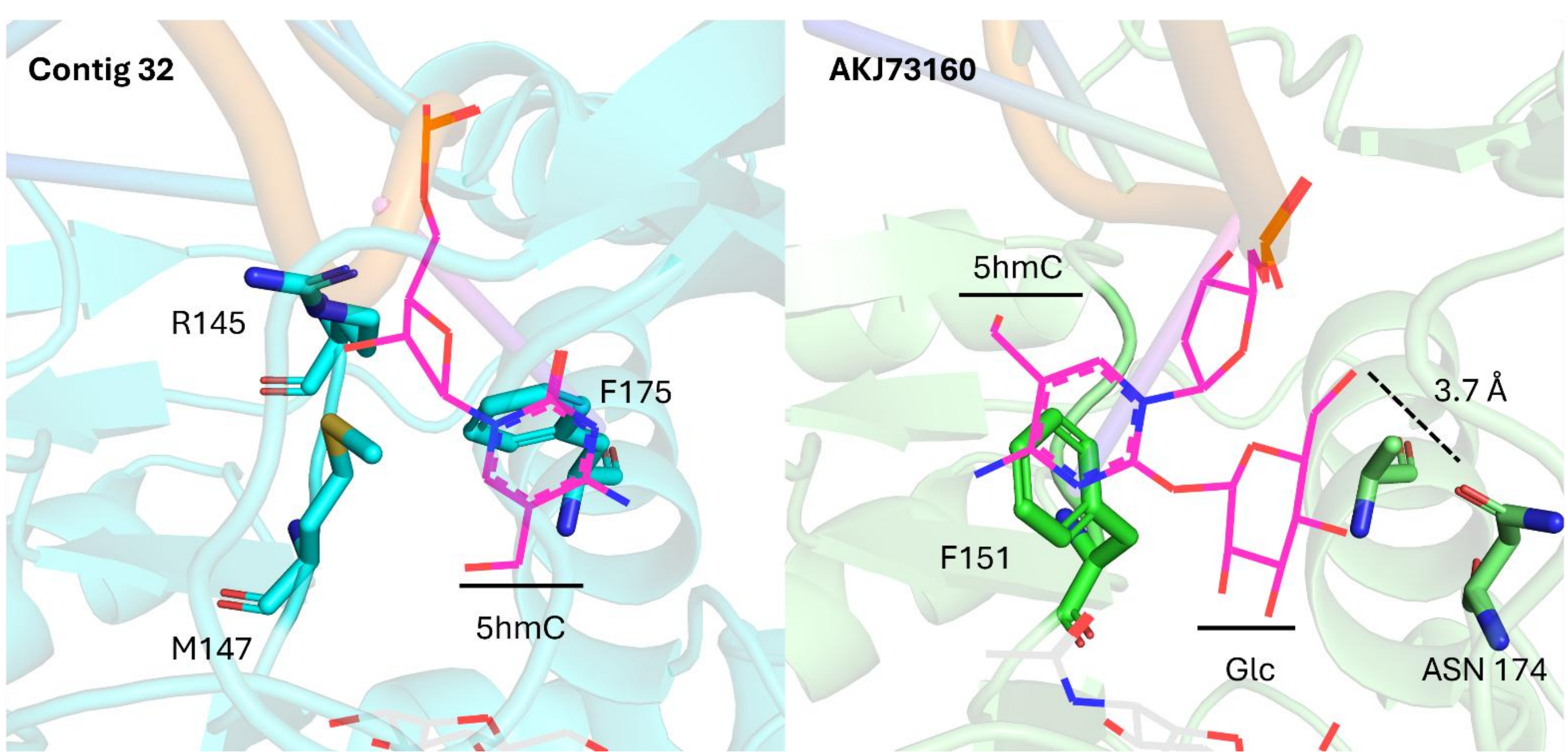
Contig 32
AKJ73160
R145
F175
M147
5hmC
5hmC
3.7 Å
F151
Glc
ASN 174

**Figure 6. Structural rewiring of cluster 8 DNA glycosyltransferases supports first-sugar and glycan-extension modes**

A. **Structural comparison of representative cluster 8 GTs with known hydroxymethyl-base-directed DNA glycosyltransferases.** AlphaFold3 ternary models of representative DNA glycosyltransferases were generated with nucleotide-sugar donor and modified DNA acceptor substrates and structurally aligned. The comparison includes Pyle 2024 DNA GT references, including contigs 14, 32, 57, 91, and 92, the protozoan J-base glucosyltransferase from *Trypanosoma brucei* JGT, and the cluster 8 representatives BAG41686 and AKJ73160. Models were prepared with 5hmC-containing DNA acceptors for the Pyle reference GTs and AKJ73160, whereas *T. brucei* JGT and BAG41686 were modeled with 5hmU-containing DNA acceptors. The global superposition shows that BAG41686 and AKJ73160 adopt a DNA-GT-like structural scaffold compatible with nucleotide-sugar donor binding and modified-DNA acceptor engagement. This shared architecture supports comparison of their active-site features with established first-sugar-installing DNA GTs.
B. **Active-site comparison identifies a conserved hydroxymethyl-base recognition motif in direct DNA GTs but not in AKJ73160.**Structure-guided multiple alignment and active-site mapping of representative hydroxymethyl-base-directed DNA glycosyltransferases, including Pyle 2024 reference GTs, *Trypanosoma brucei* JGT, BAG41686, and AKJ73160. In the known direct DNA GTs and in BAG41686, the central helix (αF), which contains the catalytic base (asterisk), also carries an adjacent aromatic residue. This aromatic position is spatially conserved and is predicted to contribute to acceptor recognition through π-stacking with the hydroxymethylated pyrimidine base. A second conserved feature is Loop X, which in the direct DNA GTs contains residues positioned to engage the deoxyribose moiety of the flipped base. In contrast, AKJ73160 lacks the αF-associated aromatic residue found in direct hydroxymethyl-base GTs and instead contains a acceptor pocket, including Asn174, consistent with recognition of a sugar-modified acceptor rather than a naked hydroxymethylated base.
C. **Aromatic stacking interactions support direct recognition of hydroxymethylated pyrimidine bases by BAG41686 but not by AKJ73160.** Close-up views of the acceptor-binding pocket show that the known direct DNA GTs and BAG41686 position a conserved aromatic side chain adjacent to the flipped 5hmC or 5hmU base. This arrangement is consistent with π-stacking-mediated stabilization of the modified pyrimidine acceptor above the nucleotide-sugar donor, providing a geometry compatible with direct β-glycosyl transfer to the hydroxymethyl handle. BAG41686 preserves this architectural feature for 5hmU recognition. In contrast, AKJ73160 lacks the analogous aromatic base-stacking residue, supporting the conclusion that it is not optimized to recognize an unmodified hydroxymethylated pyrimidine base as the primary acceptor.
D. **Acceptor-recognition logic is shifted in AKJ73160 from hydroxymethyl-base recognition to Glc-5hmC recognition.** Representative active-site views comparing a known direct 5hmC-DNA GT (contig 32), BAG41686, and AKJ73160. In contig 32 and BAG41686, the central αF aromatic residue and Loop X residues are positioned to recognize a flipped hydroxymethylated base and its deoxyribose moiety, respectively, thereby supporting direct glycosylation of 5hmC or 5hmU. In AKJ73160, this architecture is: the αF-associated aromatic base-recognition feature is lost, whereas the acceptor pocket contains residues including Asn174 and aromatic residues in the Loop X region that are better positioned to engage a glucosylated 5hmC acceptor. This arrangement supports a model in which AKJ73160 acts after initial glucosylation, recognizing Glc-5hmC rather than naked 5hmC and thereby functioning as a candidate second-sugar installer or DNA-glycan extension enzyme.
E. **AKJ73160 exhibits a shifted acceptor-recognition mode relative to direct 5hmC DNA GTs.** Structural superposition of contig 32 and AKJ73160 shows that contig 32 recognizes the flipped 5hmC base through π-stacking with F175, a hallmark of direct hydroxymethyl-base glycosyltransferases. In AKJ73160, this interaction is lost. Instead, Asn174 is positioned to recognize the glucose moiety of Glc-5hmC, while Loop X residue F151 provides an aromatic contact to the 5hmC base. This redistribution of acceptor contacts indicates that AKJ73160 is adapted to recognize Glc-5hmC rather than naked 5hmC, supporting a downstream glycan-extension role.

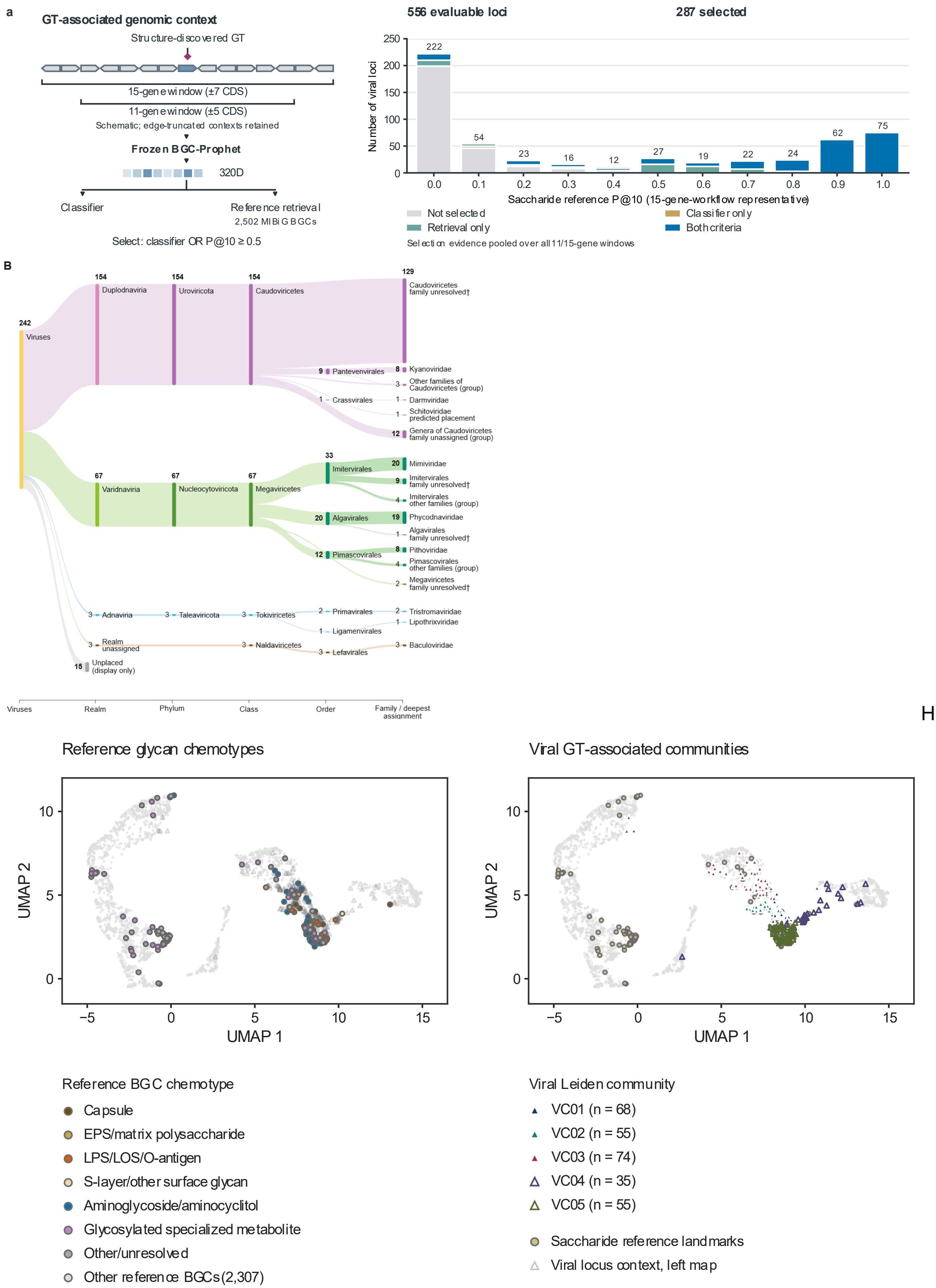
a
GT-associated genomic context
Structure-discovered GT
15-gene window (±7 CDS)
11-gene window (±5 CDS)
Schematic; edge-truncated contexts retained
Frozen BGC-Prophet
320D
Classifier
Reference retrieval
2,502 MIBiG BGCs
Select: classifier OR P@10 ≥ 0.5
556 evaluable loci
287 selected
Number of viral loci
222
54
23
16
12
27
19
22
24
62
75
Saccharide reference P@10 (15-gene-workflow representative)
Not selected
Retrieval only
Classifier only
Both criteria
Selection evidence pooled over all 11/15-gene windows
B
242 Viruses
154 Duplodnaviria
154 Uroviricota
154 Caudoviricetes
129 Caudoviricetes family unresolved†
9 Pantevenvirales
8 Kyanoviridae
3 Other families of Caudoviricetes (group)
1 Crassvirales
1 Darmviridae
1 Schitoviridae predicted placement
12 Genera of Caudoviricetes family unassigned (group)
67 Varidnaviria
67 Nucleocytoviricota
67 Megaviricetes
33 Imitervirales
20 Mimiviridae
9 Imitervirales family unresolved†
4 Imitervirales other families (group)
20 Algavirales
19 Phycodnaviridae
1 Algavirales family unresolved†
12 Pimascovirales
8 Pithoviridae
4 Pimascovirales other families (group)
2 Megaviricetes family unresolved†
3 Adnaviria
3 Taleaviricota
3 Tokiviricetes
2 Primavirales
2 Tristromaviridae
1 Ligamenvirales
1 Lipothrixviridae
3 Realm unassigned
3 Naldaviricetes
3 Lefavirales
3 Baculoviridae
15 Unplaced (display only)
Viruses
Realm
Phylum
Class
Order
Family / deepest assignment
H
Reference glycan chemotypes
Viral GT-associated communities
UMAP 1
UMAP 2
Reference BGC chemotype
Capsule
EPS/matrix polysaccharide
LPS/LOS/O-antigen
S-layer/other surface glycan
Aminoglycoside/aminocyclitol
Glycosylated specialized metabolite
Other/unresolved
Other reference BGCs (2,307)
Viral Leiden community
VC01 (n = 68)
VC02 (n = 55)
VC03 (n = 74)
VC04 (n = 35)
VC05 (n = 55)
Saccharide reference landmarks
Viral locus context, left map

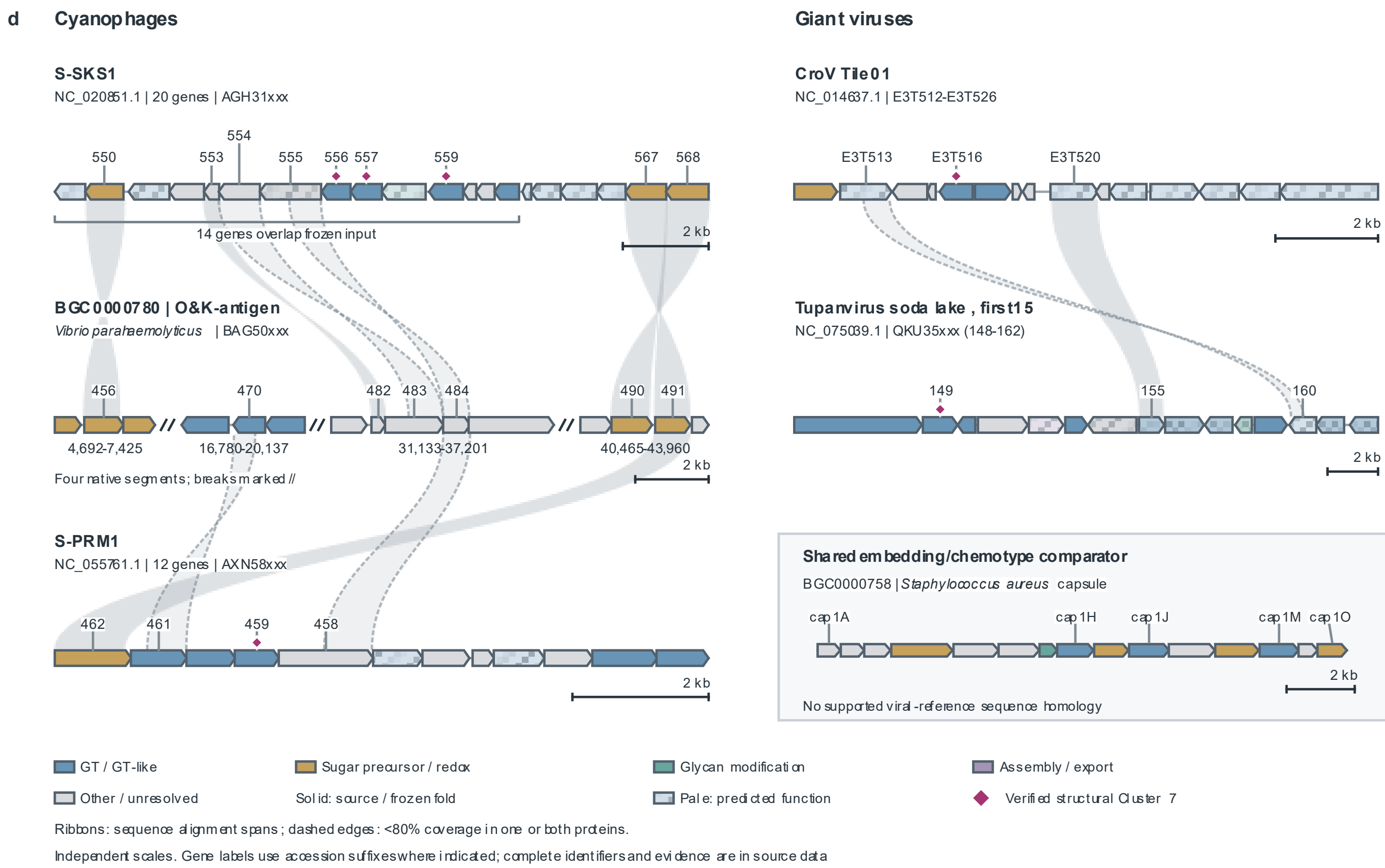

**Figure 7 .**

A. Identification of putative glycan-associated viral GT loci. GT-centred 11- and 15-gene neighbourhoods (±5 and ±7 coding sequences, respectively) were analysed using BGC-Prophet to obtain 320-dimensional embeddings and biosynthetic class predictions. Candidate loci were identified by Saccharide classification and/or reference-based retrieval (P@10 ≥ 0.5) against 2,502 MIBiG BGCs. The histogram shows the Saccharide P@10 distribution of representative 15-gene windows across 556 evaluable viral loci, with colours indicating selection evidence pooled across both window sizes. A total of 287 loci met at least one selection criterion.

B. Taxonomic distribution of 242 viral genomes containing putative glycan-associated GT neighbourhoods, identified by BGC-Prophet Saccharide classification and/or embedding-based retrieval (≥5 Saccharide-annotated BGCs among the ten nearest MIBiG references; P@10 ≥ 0.5)

C. Distribution of reference glycan chemotypes and viral GT-associated communities in a shared biosynthetic gene cluster embedding space. Uniform manifold approximation and projection (UMAP) of BGC-Prophet embeddings, showing MIBiG reference BGCs and selected viral GT-associated loci. Left, saccharide-associated reference BGCs (n = 195) are coloured according to their annotated chemotypes: capsule, extracellular polysaccharide (EPS)/matrix polysaccharide, lipopolysaccharide/lipooligosaccharide/O-antigen (LPS/LOS), S-layer/other surface glycan, aminoglycoside/aminocyclitol, glycosylated specialized metabolite, and other/unresolved. The remaining reference BGCs (n = 2,307) are shown in light grey, with viral loci indicated by pale triangles. Right, 287 selected viral GT-associated loci are coloured according to five communities identified by exploratory Leiden clustering: VC01 (n = 68), VC02 (n = 55), VC03 (n = 74), VC04 (n = 35) and VC05 (n = 55). Saccharide-associated reference BGCs are shown as gold circles, and other reference BGCs as a grey background. Both panels use the same UMAP coordinates.

D.Comparative genomic contexts of structurally defined Cluster 7 viral glycosyltransferases.

Genomic organization of Cluster 7-containing regions in cyanophages (left) and giant viruses (right). The Synechococcus phages S-SKS1 and S-PRM1 are compared with selected genomic segments of BGC0000780, an O/K-antigen-associated biosynthetic locus from Vibrio parahaemolyticus. The giant-virus comparison includes Cafeteria roenbergensis virus (CroV) and Tupanvirus soda lake, with BGC0000758 from Staphylococcus aureus shown separately as a shared capsular-polysaccharide reference identified through embedding similarity. Arrows represent coding sequences and indicate gene orientation and relative length. Genes are coloured by predicted functional category: glycosyltransferase (blue), sugar-precursor biosynthesis and redox enzymes (ochre), glycan modification (teal), assembly/export (purple), and other or unresolved functions (grey). Pale, hatched arrows indicate putative functional assignments, whereas magenta diamonds mark structurally verified Cluster 7 proteins. Translucent ribbons connect proteins with supported sequence alignments; dashed ribbon boundaries indicate less than 80% coverage of at least one aligned protein. Double slashes indicate omitted intervening genomic regions, and scale bars are specific to each track. The bracket beneath S-SKS1 indicates the portion overlapping the original embedding input. BGC0000758 is included as an embedding-based reference without supported sequence homology to the displayed giant-virus proteins.

## Contribution

Lan Na: Conceptualization—lead; Methodology—lead; Software—lead; Formal analysis—lead; Data curation—lead; Visualization—lead; Investigation—lead; Writing—original draft—lead; Writing—review and editing—lead.

Rahil Tajule: Data curation.

Ruili Fang: Software.

## Acknowledgements

This work was supported by funding from NSF BioFoundry (2400220).

## Data Availability

The protein sequences, structural models, and genomic data analyzed in this study were obtained from publicly available databases, including BFVD, UniProt, the Protein Data Bank, the AlphaFold Protein Structure Database, CAZy, NCBI, UniRef90, and MIBiG, as described in the Methods. Curated datasets and figure source data supporting the initial structure-guided discovery and clustering of viral glycosyltransferases are deposited in Zenodo (DOI: 10.5281/zenodo.23626546). Additional computational results and supporting analyses are described in the manuscript and accompanying Supplementary Information.

## Code Availability

Codes are available for peer review at https://github.com/LanNa1229/viral-gt-research-record

## Competing Interests

The authors declare no competing interests.

## Method

### Structure-guided retrieval and coarse clustering

Viral protein structures in BFVD were searched using Foldseek with a 447-residue MGAT2 structure and a truncated GT-A core query. The 1,727 structures collected from the MGAT2 search were clustered using Foldseek (--min-seq-id 0.0 -c 0.05 --cov-mode 0; 64 threads). Cluster assignments were reconstructed from the complete representative–member membership table. Direct representative–member alignments were distinguished from membership-only connections using Foldseek alignment outputs. Protein lengths were derived from structure-associated sequences, and short (≤102 aa) and long (≥763 aa) proteins were examined across multi-member cluster-size classes.

### Positive-control benchmarking

Six viral GT proteins (A197, CvHAS, gp110, A111/114R, A064R and L230) were evaluated by repeating baseline clustering with seven GT structural entries, including two L230 models. Two nanobody chains associated with CvHAS (7SP6_B and 7SP6_C) were present in the spike-in dataset but excluded from the GT recovery denominator. For independent structural comparisons, the redundant L230 model was excluded, and six nonredundant controls were aligned against the original 1,727 structures using exhaustive Foldseek searches (--exhaustive-search 1 --alignment-type 2 --max-seqs 100000 -e 1000000 -c 0.0 --cov-mode 0). Structural support was evaluated against the dominant-cluster representative and the best-matching original giant-cluster member, prioritized using min(qTM, tTM).

**Structural evaluation of non-dominant clusters**

The 114 non-dominant multi-member cluster representatives were annotated using two Foldseek search batches against structural reference databases. High-confidence hits required probability ≥0.70, E-value $\leq 1 \times 10^{-3}$, query coverage ≥0.50 and alignment length ≥80 residues; non-GT assignments additionally required target coverage ≥0.30 where available. GT-like assignments required supporting matches from two distinct targets or databases, or a strong GT-annotated alignment (probability ≥0.95; query coverage ≥0.70). Representatives with consistent non-GT evidence and no comparable GT support were classified as non-GT; conflicting or insufficient evidence remained unresolved. Two representatives without matched batch records were examined manually.

Separately, exhaustive structural comparisons were examined for 383 unique protein pairs from 63 clusters lacking recorded direct representative–member alignments. Pairs were considered structurally supported when both min(qTM, tTM) and min(qcov, tcov) were ≥0.50. Supported-pair density was calculated as the proportion of within-cluster pairs satisfying both criteria.

**Singleton length and prediction-confidence analysis**

The 646 singleton proteins were characterized using structure-derived sequence length and median residue-level pLDDT. Singleton distributions were compared with those of other non-dominant multi-member proteins using two-sided Mann–Whitney U tests with Benjamini–Hochberg correction. Effect sizes were estimated using Cliff's delta with bootstrap confidence intervals. A length threshold of 122 amino acids, corresponding to the tenth percentile of the multi-member population, and median pLDDT ≥70 were used to identify a higher-confidence singleton subset.

**Curation of GT-A structures for structure-similarity network analysis**

GT-A protein entries were initially compiled from UniProt, and their corresponding structures were obtained from the AlphaFold Protein Structure Database and, where available, the Protein Data Bank (Varadi et al., 2022; Berman et al., 2000). To construct a nonredundant reference dataset while preserving broad sequence and taxonomic diversity, family-specific protein sequences associated with the initial structural collection were clustered with MMseqs2 across a range (Steinegger and Söding, 2017) of sequence-identity thresholds. Cluster-size distributions and taxonomic composition were examined to select family-appropriate clustering levels and representative sequences while retaining major sequence-diversity and taxonomic groups. To expand coverage of GT-A diversity, additional sequences belonging to families that were absent or underrepresented in the initial collection were identified using CAZy classifications (Drula et al., 2022) and retrieved using the corresponding NCBI or UniProt accessions. These sequences were clustered with MMseqs2, and representative proteins lacking suitable existing structures were subjected to structure prediction using Alphafold3 (Abramson et al., 2024) on the Sapelo2 high-performance computing cluster at the University of Georgia. The resulting nonredundant and coverage-expanded structural reference panel was used for subsequent structure-similarity network analysis. Dataset curation was automated using Snakemake (Köster and Rahmann, 2012).

**Cross domain structural similarity network construction**

The analysis comprised 4,854 GT-A structures assembled from viral candidates and reference glycosyltransferases. All ordered structure pairs were supplied to Foldseek structurealign (van Kempen et al., 2024; version [version/commit]), bypassing heuristic prefiltering. Alignments were calculated using TM-align (Zhang and Skolnick, 2005; --alignment-type 1), with an alignment-normalised TM-score threshold of 0.65 (--tmscore-threshold-mode 0) and minimum query and target coverage of 0.50 (--cov-mode 0). No additional E-value filter was applied (-e inf). Self-alignments were removed, and directional records were collapsed into unordered protein pairs. For each directional alignment, global structural similarity was defined as the smaller of the TM-scores normalised by the full query and target lengths. The undirected edge weight was the arithmetic mean of this value across the available orientations; pairs recovered in only one orientation were retained. The resulting network contained 224,825 edges. All input proteins were retained as nodes, including 580 isolates with no non-self edge.

### Weighted community detection

Community detection was performed using python-igraph [version] and leidenalg [version] with the Reichardt–Bornholdt configuration-model objective (RBConfigurationVertexPartition) and the Leiden algorithm (Traag et al., 2019). The primary partition used the global structural-similarity weights defined above, a resolution parameter of 1.0 and random seed 1. Optimisation continued until no further improvement was obtained (n_iterations = -1). This weighting accounts for similarity relative to the full lengths of both proteins. Unweighted and mean alignment-normalised TM-score-weighted partitions were also calculated for comparison. Community identifiers were assigned in decreasing order of community size. CAZy annotations, taxonomic origin and viral status were used for subsequent interpretation and were not included in community detection. CAZy annotations were not used to optimise the partition.

### CAZy family recapitulation

CAZy-family organisation was evaluated in the 80 multi-protein communities comprising 4,274 proteins. Isolates were excluded from this analysis. Family counts included only unambiguous single-family GT annotations; blank, unassigned and multi-family labels were excluded. Unannotated proteins remained included in community sizes. For community (c) and family (f), let $(n_{cf})$ denote the number of eligible annotated proteins. Community purity and family concentration were defined, respectively, as

[ $P_c = \frac{\max_f n_{cf}}{\sum_f n_{cf}}, Q_f = \frac{\max_c n_{cf}}{\sum_c n_{cf}}.$ ]

Purity measures the fraction of annotated community members belonging to its most frequent family, whereas concentration measures the fraction of a family's members occurring in the community containing its largest representation. Communities with at least three annotated members were classified as family-coherent at purity ≥0.80 and mixed below this threshold; communities with fewer annotations were designated annotation-limited. All 39 eligible families were included and classified as concentrated at concentration ≥0.80 or dispersed otherwise.

Protein-weighted community summaries included all members of each classified community. Protein-weighted family concentration was the sum of the largest community-specific count for each family divided by the total eligible family membership. Sensitivity analyses required at least five or ten annotations per community, or at least three, five or ten members per family, while retaining the 0.80 cutoffs. These summaries were descriptive; no null-hypothesis tests were applied.

### Node topology

All topology metrics were calculated from the complete weighted network. Node degree was the number of neighbours, and strength was the sum of incident edge weights. Within-community strength was the weight of edges to members of the same Leiden community; its fraction was calculated relative to total node strength. The weighted participation coefficient was

[ $P_i = 1 - \sum_c \left(\frac{s_{ic}}{s_i}\right)^2,$ ]

where $(s_{ic})$ is the summed weight of edges from node (i) to community (c) and $(s_i)$ is total node strength. Isolates were assigned zero strength and participation. These metrics were used to distinguish nodes with predominantly internal relationships from those connected across communities.

**Classification of viral structural placement**

Viral placement used the primary partition and CAZy-family metadata from CAZy_family, with GTFam used as a fallback . Communities without annotated members were designated annotation-depleted, and those with one to four annotated members sparsely annotated. An established-family context required at least five annotated members and dominant-family purity ≥0.80. Communities with at least five annotated members and lower purity were classified as supported mixed-family contexts when the second-most-abundant family contained at least two proteins and represented at least 20% of annotated members.

For placement classification, viral enrichment was evaluated by a one-sided hypergeometric test against the complete 4,854-node network, containing [verified viral count] viral proteins. P values were adjusted across all 660 communities using the Benjamini–Hochberg procedure. Annotation-depleted communities with adjusted ($q<0.05$) were designated virus-enriched dark communities.

Within established-family communities, proteins with within-community strength below the community-specific 25th percentile were classified as family-peripheral. Intercommunity connectors were non-singleton nodes with degree ≥3, at least one internal and one external neighbour, at least 20% of total strength directed outside their community and a participation coefficient ≥0.30. Placement categories were applied in the following precedence order: singleton or isolate, connector, family-peripheral context, established-family context, supported mixed-family context, sparsely annotated community, virus-enriched dark community and other annotation-depleted community.

**CAZy family evidence for viral proteins**

For CAZy-assigned viral proteins, the assigned family was compared with the dominant family of their Leiden community to identify concordant or discordant contexts. This comparison describes agreement with community composition and is not an independent validation of the existing annotation.

For CAZy-unassigned viral proteins, edge weights to directly connected, CAZy-annotated neighbours within the same community were summed separately for each family. The family with the largest cumulative weight was considered best supported. Direct support required at least two neighbours from that family, at least 67% of the total annotated-neighbour strength and a normalised margin ≥0.20, defined as

$$[M_i = \frac{S_{i1} - S_{i2}}{\sum_f S_{if}},]$$

where $(S_{i1})$ and $(S_{i2})$ are the largest and second-largest family-specific strengths and the denominator sums these strengths over all represented families. A protein was classified as directly supported when these criteria were met and the best-supported family matched the community-dominant family. Other outcomes with annotated neighbours were grouped as limited, ambiguous or alternative support, encompassing insufficient neighbour counts, failure of the strength-fraction or margin criteria, or support for a different family. Proteins without an annotated neighbour in their community were classified as lacking direct support.

Existing annotation, structural placement and family evidence were displayed in an alluvial diagram with ribbon widths proportional to protein counts. Direct support for an unassigned protein was interpreted as a candidate structural association with a CAZy family.

**Topology guided identification of viral structural modules**

Candidate modules were screened among the 4,274 non-isolated network proteins. Previously characterised groups designated clusters 7 and 8 were mapped from the earlier network by protein accession and used as reference modules. Candidate seeds were viral, CAZy-unassigned proteins with within-community strength fraction ≥0.85 and participation coefficient ≤0.26 that were not classified as connectors. These thresholds were derived empirically from the reference groups. Connected groups of at least four qualifying nodes within a community were evaluated as candidate modules .

Module edge density was the fraction of possible internal edges that were present. For each member, the internal-strength fraction was the weight of edges to other module members divided by total node strength; module summaries used the median of these fractions. Weighted conductance was defined as

$$[\frac{W_{\text{out}}}{2W_{\text{in}} + W_{\text{out}}},]$$

where $(W_{\text{in}})$ and $(W_{\text{out}})$ are the summed internal and boundary edge weights, respectively. Modules were compared with the reference groups using these metrics, median structural-similarity weights, mutual query-target coverage and the proportion of edges recovered in both orientations. Source-virus diversity was examined to assess redundancy within individual genomes. Directly connected viral neighbours supported over a shared structural domain were retained as peripheral affiliates **.**

A separate screen prioritised multi-protein communities without CAZy-assigned members that were enriched in viral proteins at Benjamini–Hochberg-adjusted ($q<0.05$). The hypergeometric background comprised [background node count] proteins, including [background viral count] viral proteins, with correction across [number of tested communities] communities . Source-virus labels were checked against organism records. Selected communities were assessed using the module metrics above. Catalytic motifs, active-site architecture, sequence-profile coherence and genomic context were examined after topology-based selection.

**Network visualisation**

For visualisation, each connected node's five highest-weight edges were selected from the complete network, with ties resolved by the alphanumeric identifier of the neighbouring protein. The union of these selections formed a 15,576-edge backbone retaining all 4,274 connected proteins. Full-network Leiden assignments were retained. The layout was calculated once in Cytoscape [version] (Shannon et al., 2003), using [layout algorithm and parameters], and the coordinates were reused for views coloured by CAZy family or taxonomic origin. Isolates were excluded from the layout. Summary tables and figures were generated using pandas and Matplotlib.

**Profile hidden Markov model construction for the C027 viral protein family.** To search for remote sequence homologues of the structurally defined C027 community, the amino-acid sequences of the seven C027 proteins were aligned using Clustal Omega (Sievers et al., 2011) (version X.X.X) with default parameters. The resulting multiple-sequence alignment contained 546 alignment columns. A profile hidden Markov model (profile HMM) was constructed from the alignment using hmmbuild in HMMER (version X.X; Eddy, 2011), with the consensus-column threshold set to --symfrac 0.5. Thus, an alignment position was represented as a consensus match state when at least 50% of the weighted seed sequences contained a residue at that position. This procedure generated a 416-position C027 profile (C027_core_v1.hmm). Model integrity was assessed by searching the profile against the original seven-sequence seed set using hmmsearch; all seven seed proteins were recovered. The seed sequences and profile were subsequently held fixed during searches of external sequence databases to avoid iterative model drift.

**C027 homologue identification and phylogenetic analysis.** Seven structurally coherent viral proteins assigned to the isolated C027 community were aligned to construct a profile hidden Markov model using HMMER. The resulting C027 profile was validated by self-search against the seven seed proteins and searched against UniRef90, comprising 121,389,642 representative protein sequences. HMMER searches were performed with report thresholds of $E \le 10^{-2}$ for both full-sequence and domain matches and inclusion thresholds of $E \le 10^{-5}$. Candidate homologues were evaluated using independent domain E-value, domain bit score, profile coverage, protein length and annotation. Twenty-seven full- or near-full-core homologues were retained, whereas four strongly scoring but truncated viral proteins, one weak cellular boundary candidate and one non-significant prenyltransferase match were excluded from the primary phylogeny. The 27 UniRef90 representatives were combined with four nonredundant structural seeds, yielding 31 unique proteins. Sequences were aligned to the C027 profile using hmmalign --trim, producing a 467-site core alignment. Maximum-likelihood phylogeny was inferred with IQ-TREE v3.0.1 (Wong et al., 2026) using ModelFinder Plus, which selected Q.PFAM+F+I+G4 as the best-fitting model. Branch support was assessed with 1,000 SH-aLRT replicates and 1,000 ultrafast bootstrap replicates with BNNI optimization. Because no defensible homologous outgroup was identified, the tree was interpreted as unrooted. Viral taxonomy was subsequently mapped to the tree using UniProt and NCBI taxonomy metadata.

**Gene-neighborhood heatmap construction.** Cluster 3-containing viral neighborhoods were compared using a confidence-weighted gene-class matrix derived from Foldseek annotation outputs. For each candidate locus, neighboring ORFs were structurally searched against PDB100, AFDB-SwissProt, AFDB50, and CATH50, and the resulting named Foldseek hit tables were parsed using a custom rule-based classifier. ORFs were assigned to functional gene classes associated with hydroxymethyl-base precursor formation, pyrimidine/nucleotide metabolism, DNA metabolism, glycan biosynthesis, virion structure, or conserved unknown proteins. Annotation confidence was encoded as a discrete score from 0 to 3 based on Foldseek probability, e-value, query coverage, alignment length, database source, and hit rank. High-confidence hits required probability ≥ 0.90, e-value ≤ $1 \times 10^{-5}$, query coverage ≥ 0.50, and alignment length ≥ 100 amino acids; medium-confidence hits required probability ≥ 0.70, e-value ≤ $1 \times 10^{-3}$, query coverage ≥ 0.35, and alignment length ≥ 80 amino acids; weak hits required e-value ≤ $1 \times 10^{-2}$ and alignment length ≥ 50 amino acids. CATH-only fold-level hits were capped at medium confidence, and low-ranked hits were penalized to reduce functional overinterpretation.

For each neighborhood, the heatmap value for a given gene class represents the maximum evidence score among all neighboring ORFs assigned to that class, rather than a raw gene count. Generic transferases were not counted as glycosyltransferases unless the annotation specifically supported glycosyltransferase-like function. TET-associated GT-like annotations were also not considered direct evidence of DNA glycosylation; DNA-modification context was inferred from independent neighboring genes such as C5-cytosine methyltransferases, 5mYOX/TET-like dioxygenases, dCMP hydroxymethylases, ThyX/ThyS-like enzymes, or 5hmC/5hmdU-related enzymes. Rows were sorted by inferred neighborhood class and then by DNA-context, glycan-pathway, and virion-structural scores. Heatmaps were plotted in Python using pandas and matplotlib.

**Taxonomic reconciliation and visualization of viral GT-associated genomes**

We used 466 GT-anchor-containing viral genome units from the paired 11- and 15-gene neighbourhood screen, collapsing identical nucleotide records with the same source taxid. A genome was designated a candidate if any associated window was Saccharide-positive under the frozen BGC-Prophet classifier or had P@10 ≥ 0.5, defined as the fraction of Saccharide-labelled entries among its ten nearest MIBiG reference embeddings. Accession-specific taxonomic assignments were reconciled against NCBI Taxonomy and ICTV MSL41/VMR records. Unresolved genomes were further examined using vConTACT3 (v.3.1.6, reference v230) and geNomad (v.1.13.0, database v1.9). Full-genome and selected-GT-anchor-excluded evidence were compared; unsupported family assignments were not imposed, and marker-only higher-rank placements remained provisional. We excluded three single-CDS Poxviridae records from the visualization only. The remaining 242 candidate genome units were mapped to compact and full taxonomic Sankey diagrams, with 18 and 31 mutually exclusive terminal categories, respectively, using custom Python scripts and Matplotlib (v.3.11.2). Unresolved terminals were treated as display categories, not taxa.

**Embedding-based analysis of viral GT-associated biosynthetic loci**

Previously generated 320-dimensional BGC-Prophet embeddings were used to compare viral glycosyltransferase (GT)-associated gene neighbourhoods with 2,502 MIBiG reference biosynthetic gene clusters (BGCs). The reference set included 195 Saccharide-annotated BGCs, including hybrids. From 556 evaluable viral loci, 287 were retained based on a prior screen in which at least one GT-centred 11- or 15-gene window was Saccharide-classifier positive (score ≥ 0.5 and ≥ the Other-class score) or had a Saccharide reference-neighbour fraction of at least 0.5 among the ten nearest reference BGCs (P@10). Archived representative embeddings were retained without model refitting or reselection. Viral embeddings were compared using cosine similarity, and an undirected weighted graph was formed from the union of 20-nearest-neighbour relationships, excluding self-edges and non-positive weights. Communities were detected with weighted Leiden optimization of the RB configuration-model objective (resolution 1.0; random seed 42). Clustering robustness was evaluated across parameter settings and through comparison with HDBSCAN. Previously fitted joint UMAP coordinates were used only to visualize reference chemotypes and viral communities. Neither chemotype annotations nor UMAP coordinates were used for clustering.

### GT31 structural conservation and mapping to F2Y1U4

Predicted structures of F2Y1U4 and 17 GT31 references from the archived C012 dataset were analysed using Foldseek and FoldMason. References comprised four annotation-defined groups, with taxonomic labels retained from accompanying metadata. Each group was represented by the protein with the highest median within-group structural similarity. Core positions were evaluated in a 17-reference subset of the FoldMason alignment, excluding F2Y1U4. Residue-level conservation was quantified using Cα-based local lDDT-like scores within groups and between representatives. Eligible columns required ≥70% occupancy in every group and pLDDT ≥70 for all four representative residues. Strict cores required all four representatives to pass within-group (≥0.75) and between-group (≥0.65) tests; supported cores required at least three passes per test. Additional criteria are detailed in Supplementary Methods. Supported-only positions comprised supported positions outside the strict core. Core classes were mapped onto F2Y1U4 and visualized in PyMOL. Separately, FoldMason sequence identities and Foldseek query coverage and alignment TM-scores were summarized as the unweighted mean of the four group medians.

### R699 Analysis

Protein structures from the C005 (13 proteins) and C009 (15 proteins) network reference sets were analysed alongside R699 (Protein Data Bank accession 9DYT, chain A). Reference structures were compared all against all using Foldseek, and each community medoid was selected as the structure with the highest median symmetric TM-score against other members. Separate reference-only structural alignments were generated using FoldMason. Reference cores comprised medoid residues passing the within-community quality filter, with a within-community local structural similarity score ≥0.75 and residue presence ≥0.70. Cores were defined without R699 and held fixed during projection. Each reference set was then realigned with R699, and core positions were transferred through the medoid's ungapped residue indices to corresponding R699 alignment positions. Individual R699 residues were classified as mapping to both cores, C005 only, C009 only or neither. Residue classes were mapped to chain-A coordinate numbering and visualized in PyMOL. Here, support denotes alignment correspondence to a reference core.

## Supplementary Figures & Methods

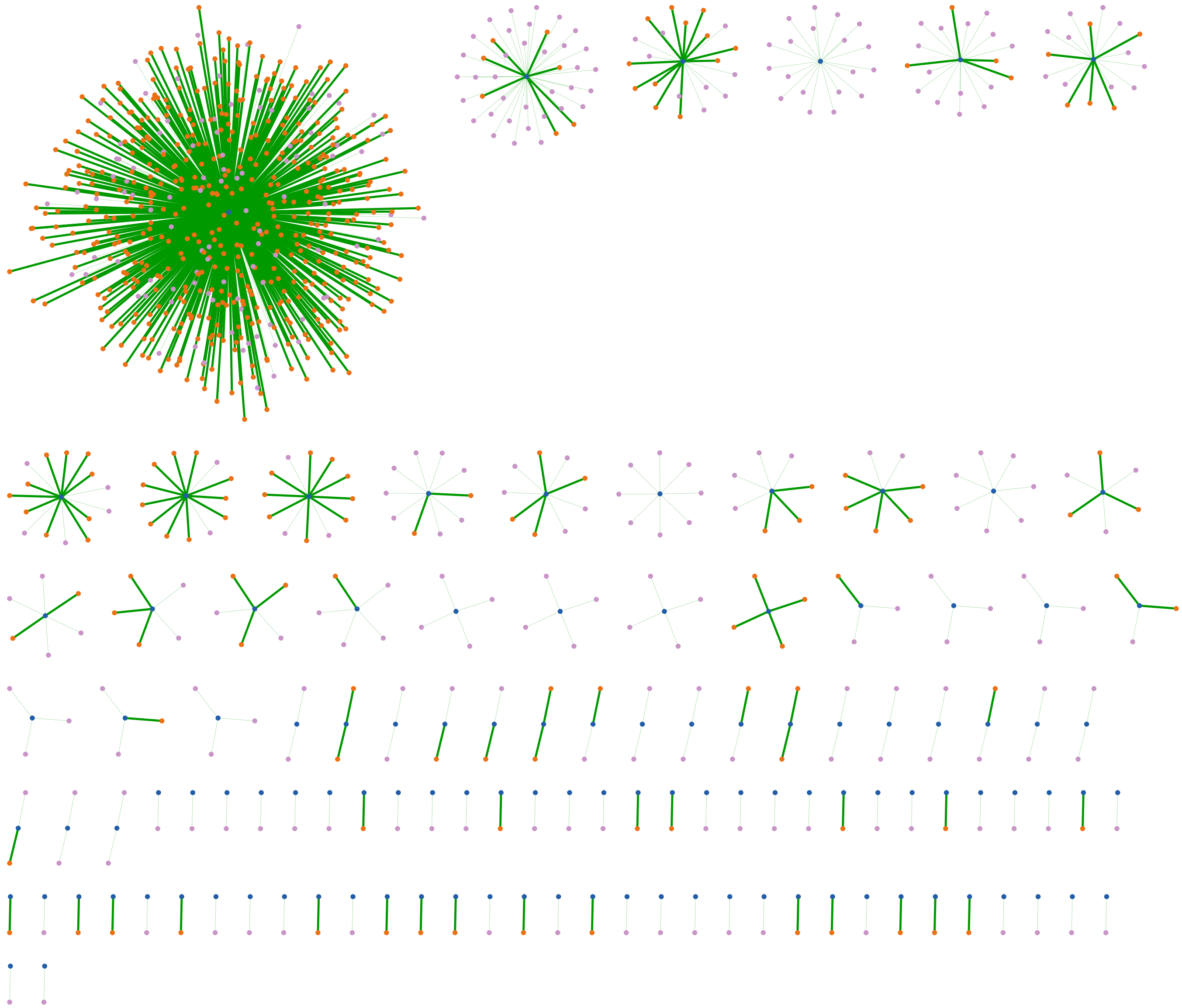

**Supplementary Figure 1. Structural clustering of viral GT candidates.**

The 1,081 viral proteins assigned to 115 multi-member clusters are shown as nodes, with each connected component representing one final Foldseek cluster. Blue nodes denote the 115 final cluster representatives, orange nodes denote 620 direct members with an explicit non-self representative–member alignment in the Foldseek m8 output, and purple nodes denote 346 indirect members assigned to the same final cluster but lacking an explicit alignment to the final representative in the current alignment output. Solid edges represent explicit representative–member alignments, whereas dashed edges were for indirect members and do not represent additional pairwise structural alignments. The dominant cluster contained 594 proteins. The 646 singleton proteins are not shown. Node positions and edge lengths reflect the Cytoscape layout.

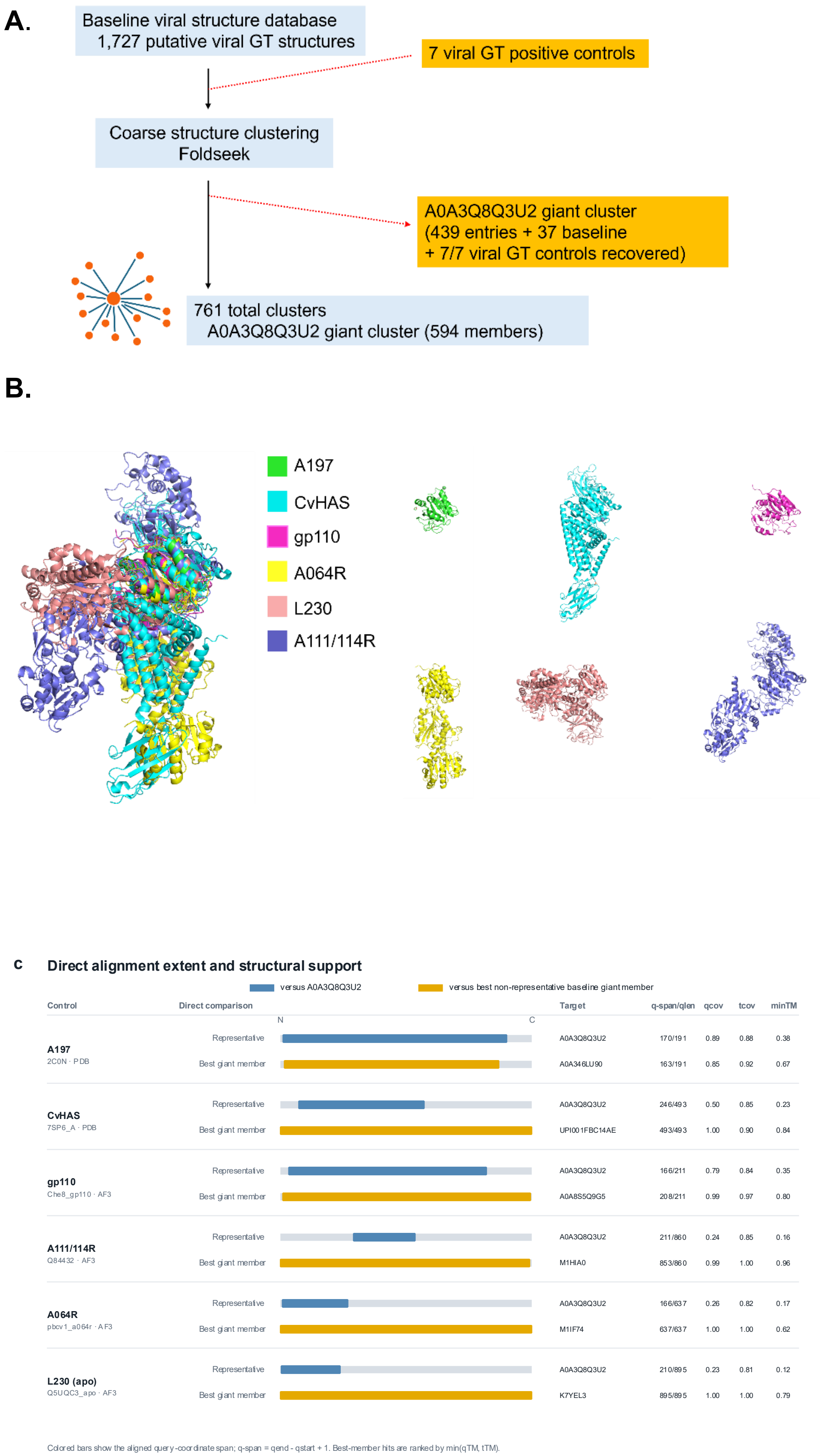

**Supplementary Figure 2. Positive-control spike-in benchmark of permissive Foldseek coarse clustering.** **A.** Foldseek clustering of 1,727 viral GT candidates produced 761 clusters, including a dominant cluster of 594 proteins represented by A0A3Q8Q3U2. After adding seven structural entries representing six viral GT controls,

clustering with unchanged parameters recovered all seven entries in the cluster represented by A0A3Q8Q3U2. The resulting cluster contained 483 entries, comprising 439 original dominant-cluster members, 37 proteins from other clusters and seven controls. Two nanobody chains co-deposited with CvHAS (PDB 7SP6) were excluded.

**B.** Structural superposition (left) and indidividual structures (right) of the six viral GT controls: A197 from *Sulfolobus turreted icosahedral virus 1* (green; PDB 2C0N); Chlorella virus hyaluronan synthase (CvHAS) from *Paramecium bursaria Chlorella virus CZ-2* (cyan; PDB 7SP6, chain A); the virion glycosylation-associated glycosyltransferase gp110 from *Mycobacterium phage Che8* (magenta; AlphaFold 3 model); the multidomain glycosyltransferase A111/114R from *Paramecium bursaria Chlorella virus 1* (yellow; UniProt Q84432; full-length AlphaFold 3 model); the multidomain rhamnosyltransferase/methyltransferase A064R from *Paramecium bursaria Chlorella virus 1* (salmon; full-length AlphaFold 3 model); and the bifunctional lysyl hydroxylase/glycosyltransferase L230 from *Acanthamoeba polyphaga mimivirus* (purple; UniProt Q5UQC3; full-length AlphaFold 3 models).

**C.** Direct structural alignments of each control with A0A3Q8Q3U2 (blue) and its best-matching non-representative member of the original dominant cluster (orange). Grey tracks represent full-length control proteins from N to C. Values indicate aligned query span, query coverage (qcov), target coverage (tcov) and the conservative structural similarity score min(qTM, tTM).

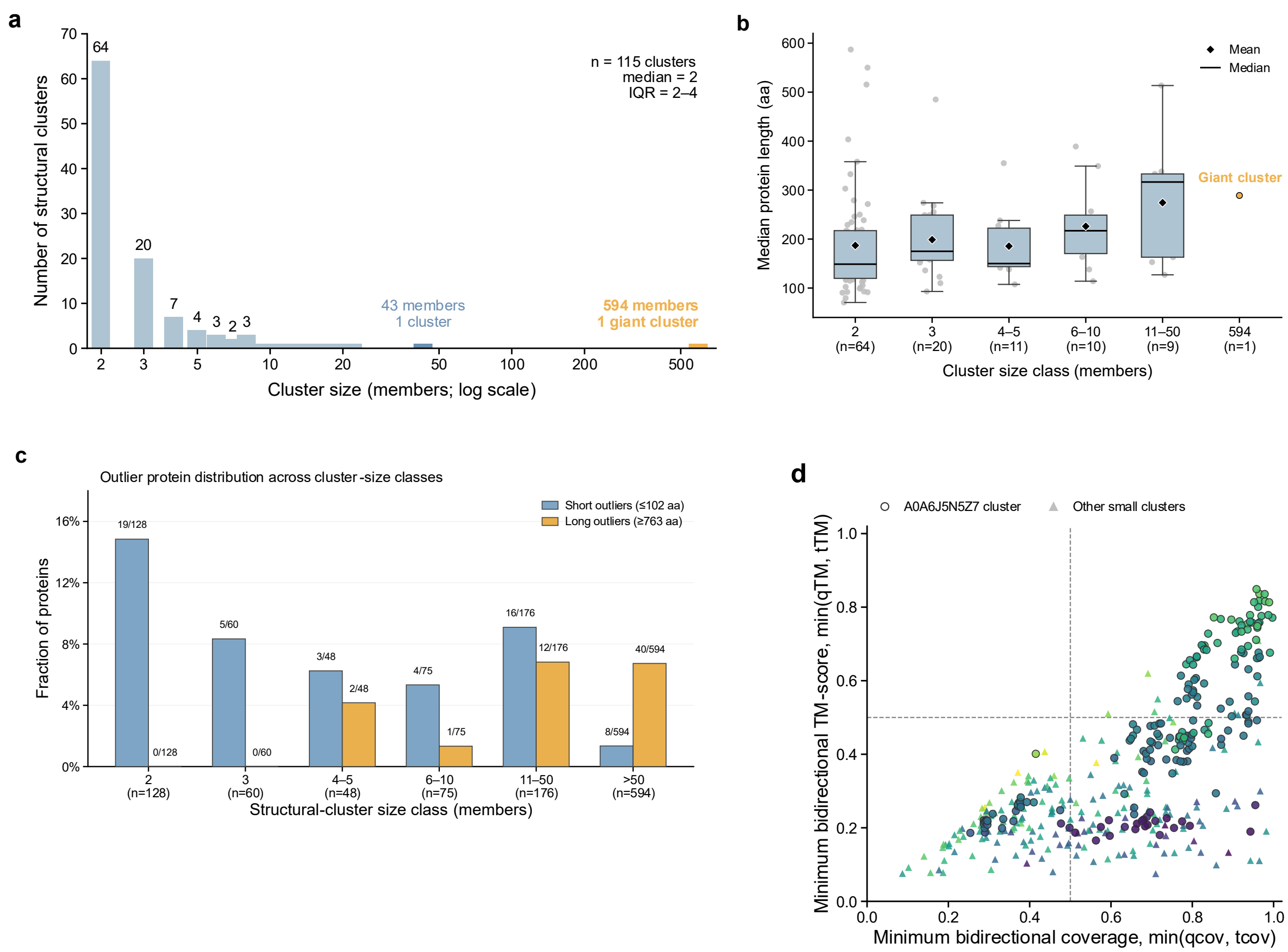

e

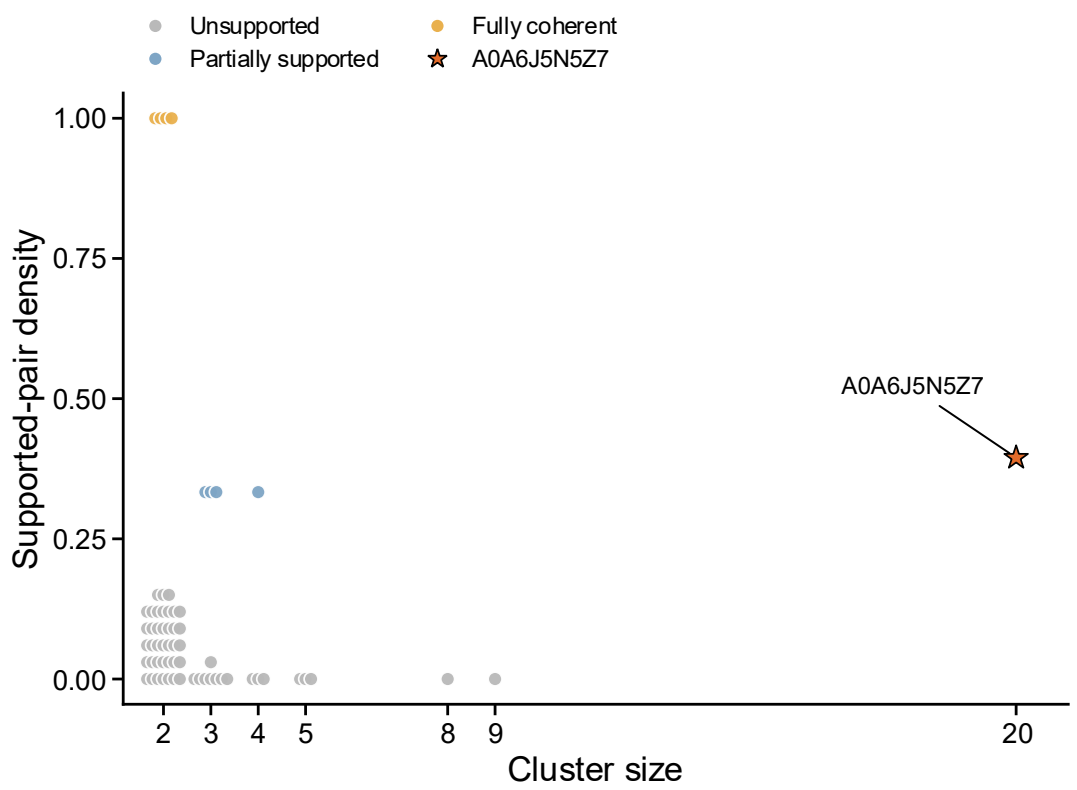

f

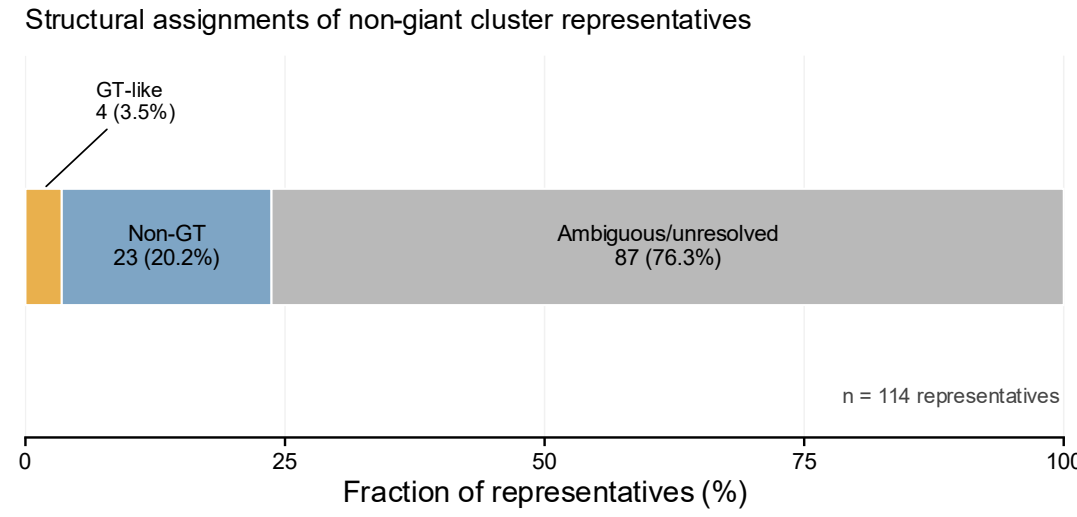

**Supplementary Figure 3. Distribution and structural assessment of viral GT candidate clusters.**

**a.** Exact size distribution of the 115 multi-member clusters containing 1,081 proteins. The 43-member and dominant 594-member clusters are highlighted. The median cluster size was 2 proteins (IQR, 2–4).
**b.** Distribution of cluster-level median protein lengths across cluster-size classes. Each point represents one cluster; boxes show the IQR, centre lines the median, whiskers 1.5 × IQR and diamonds the arithmetic mean. The 594-member cluster is shown as a single point.
**c.** Fractions of short (≤102 amino acids) and long (≥763 amino acids) proteins across cluster-size classes. Values indicate outlier counts relative to the total number of proteins in each class.
**d.** Pairwise structural comparisons of 383 protein pairs from 63 smaller clusters lacking direct representative–member alignments in the initial Foldseek output. Axes show the minimum query/target coverage, min(qcov, tcov), and minimum query/target-normalized TM-score, min(qTM, tTM); colour indicates LDDT. Comparisons from the 20-member A0A6J5N5Z7 cluster are shown as circles and all others as triangles. Dashed lines indicate thresholds of 0.50.

**e.** Supported-pair density for the same 63 clusters, defined as the fraction of all possible within-cluster pairs satisfying both min(qTM, tTM) ≥ 0.50 and min(qcov, tcov) ≥ 0.50. Coincident observations are shown as stacked dots; A0A6J5N5Z7 is highlighted.
**f.** Representative-level Foldseek assignments for all 114 non-giant clusters. Four representatives were classified as GT-like, 23 as non-GT and 87 as ambiguous or unresolved.

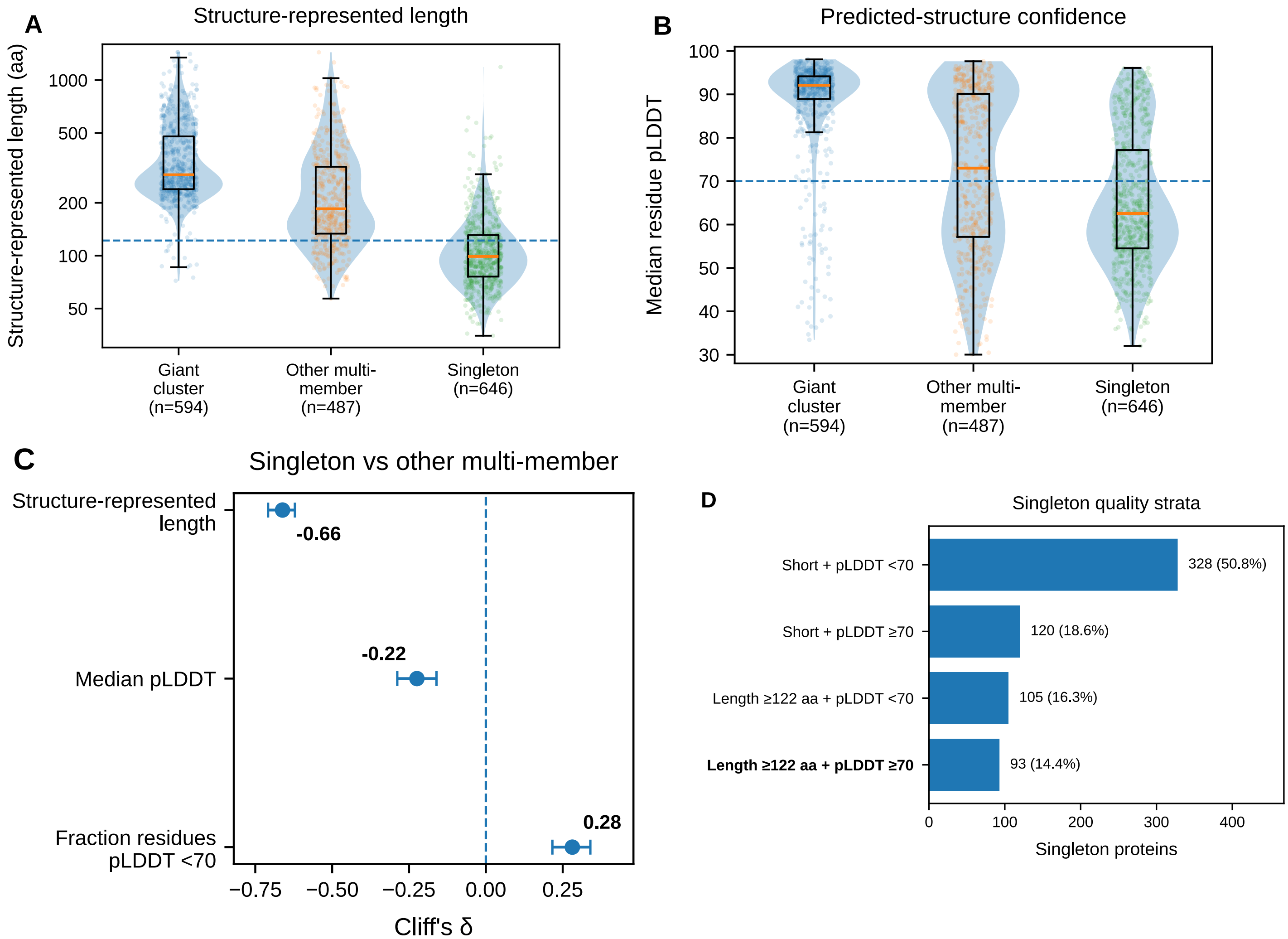

**Supplementary Figure 4. Protein length and prediction confidence of viral GT candidate singletons.**

**A.** Distribution of structure-represented protein lengths among the dominant cluster, other multi-member clusters and singletons. Numbers above the box plots indicate group medians. The dashed line marks the 10th percentile of the multi-member protein-length distribution (122 aa).

**b.** Distribution of median residue-level pLDDT for the same groups. The dashed line indicates pLDDT = 70. In a and b, boxes indicate the interquartile range (IQR), centre lines the median, whiskers 1.5 × IQR and individual points represent proteins.

**c.** Cliff's δ comparing singleton proteins with proteins from other non-giant multi-member clusters. Points show Cliff's δ and horizontal bars indicate 95% bootstrap confidence intervals. Negative values indicate lower values among singletons. Effect sizes were −0.661 for protein length (95% CI, −0.708 to −0.621), −0.224 for median pLDDT (95% CI, −0.288 to −0.161) and 0.281 for the fraction of residues with pLDDT below 70 (95% CI, 0.216 to 0.340). Statistical comparisons used two-sided Mann–Whitney U tests with Benjamini–Hochberg correction (adjusted $P = 1.24 \times 10^{-80}$, $1.17 \times 10^{-10}$ and $5.80 \times 10^{-16}$, respectively).

**D.** Grouping of 646 singleton proteins by structure-represented length (122 aa threshold) and median pLDDT (70 threshold). Of these, 328 (50.8%) were shorter than 122 aa with median pLDDT below 70, whereas 93 (14.4%) had lengths of at least 122 aa and median pLDDT of at least 70.

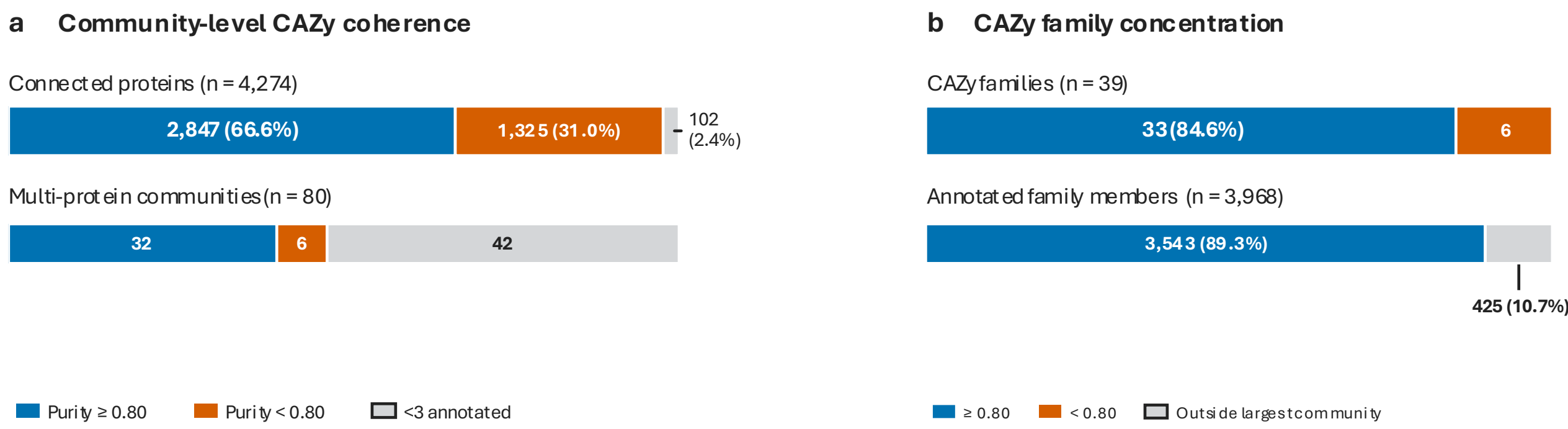

**Supplementary Figure 5. Multi-protein structural communities substantially recapitulate CAZy family-level organization.**

**A. Community-level CAZy purity.** The 80 multi-protein structural communities (4,274 proteins) were classified according to the proportion of CAZy-annotated members belonging to the most abundant family within each community. Communities containing at least three unambiguously annotated proteins were classified as family-coherent (purity ≥0.80; 32 communities) or mixed (purity <0.80; 6 communities). The remaining 42 communities were designated annotation-limited. The upper bar shows the number of communities in each category; the lower bar shows the corresponding protein counts, including proteins without CAZy assignments.

**B. CAZy family concentration.** For each of 39 CAZy families, concentration was defined as the fraction of its annotated members found in the structural community containing the largest number of members of that family. Thirty-three families (84.6%) had a concentration of at least 0.80 (upper bar). Across 3,968 unambiguously CAZy-annotated proteins, 3,543 (89.3%) occurred in their respective family's largest community, whereas 425 (10.7%) were distributed elsewhere (lower bar). The 580 singleton proteins were excluded from both analyses. CAZy annotations were applied after structural community detection and did not influence community assignment.

**A.**

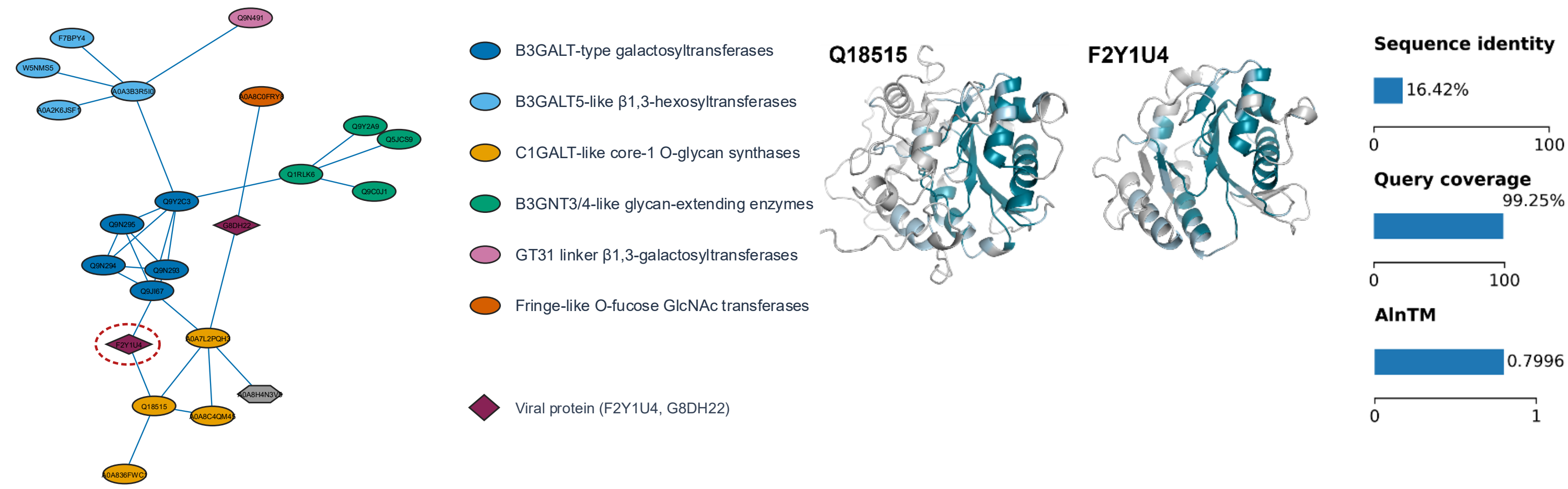

**B.**

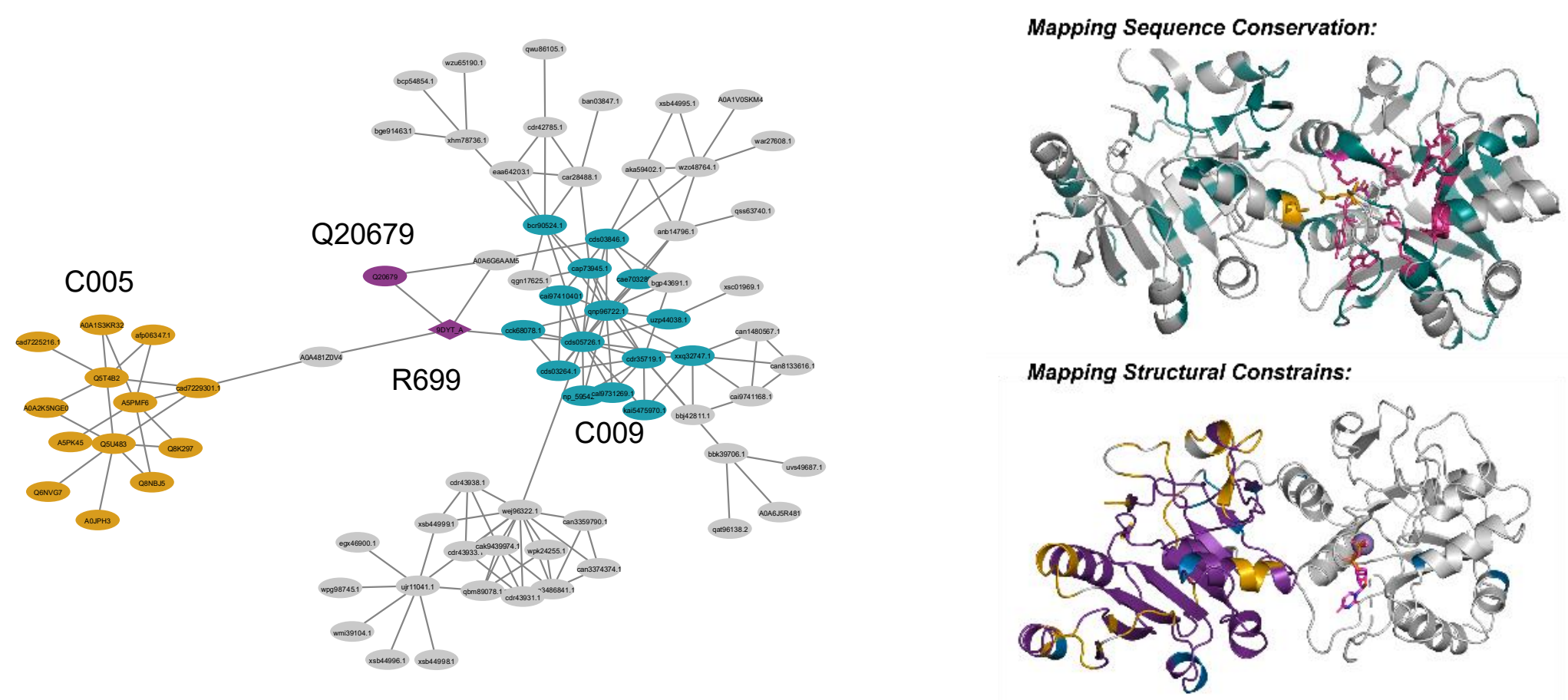

**Supplementary Figure 6. Distinct structural relationships of viral glycosyltransferases within the GT-A landscape.**

**A. Structural placement of CAZy-unassigned viral GTs (F2YIU4 and G8DH22) within the GT31 family**. Local structure-similarity network showing GT31 reference proteins coloured by functional subgroup, as indicated in the key. CAZy-unassigned viral proteins F2Y1U4 and G8DH22 are shown as purple diamonds, with F2Y1U4 highlighted by a dashed red circle. Structural comparison of the GT31 reference Q18515 and F2Y1U4 shows the strict pan-GT31 core (dark teal), additional structurally supported positions (light blue) and non-core regions (grey). Bars summarize structural comparisons of F2Y1U4 across four GT31 reference subclusters, showing mean subcluster-median sequence identity (16.42%), query coverage (99.25%) and alignment TM-score (AlnTM, 0.7996).

**B. Domain-specific structural relationships of R699 across communities C005 and C009.** Local structure-similarity network highlighting C005 (orange), C009 (turquoise), R699 (purple diamond) and the PLOD-like glycosyltransferase Q20679 (purple oval); remaining proteins are shown in grey. Upper, sequence identities between R699 and Q20679 mapped onto the R699 structure following FoldMason alignment. Identical residues are shown in teal, identical active-site residues in magenta, and non-identical active-site residues in orange. Other residues are shown in grey. Lower, conserved structural cores independently defined from C005 and C009 reference proteins mapped onto R699. Both communities showed extensive structural correspondence within the non-UDP-binding domain, whereas the UDP-binding catalytic domain showed limited structural support.

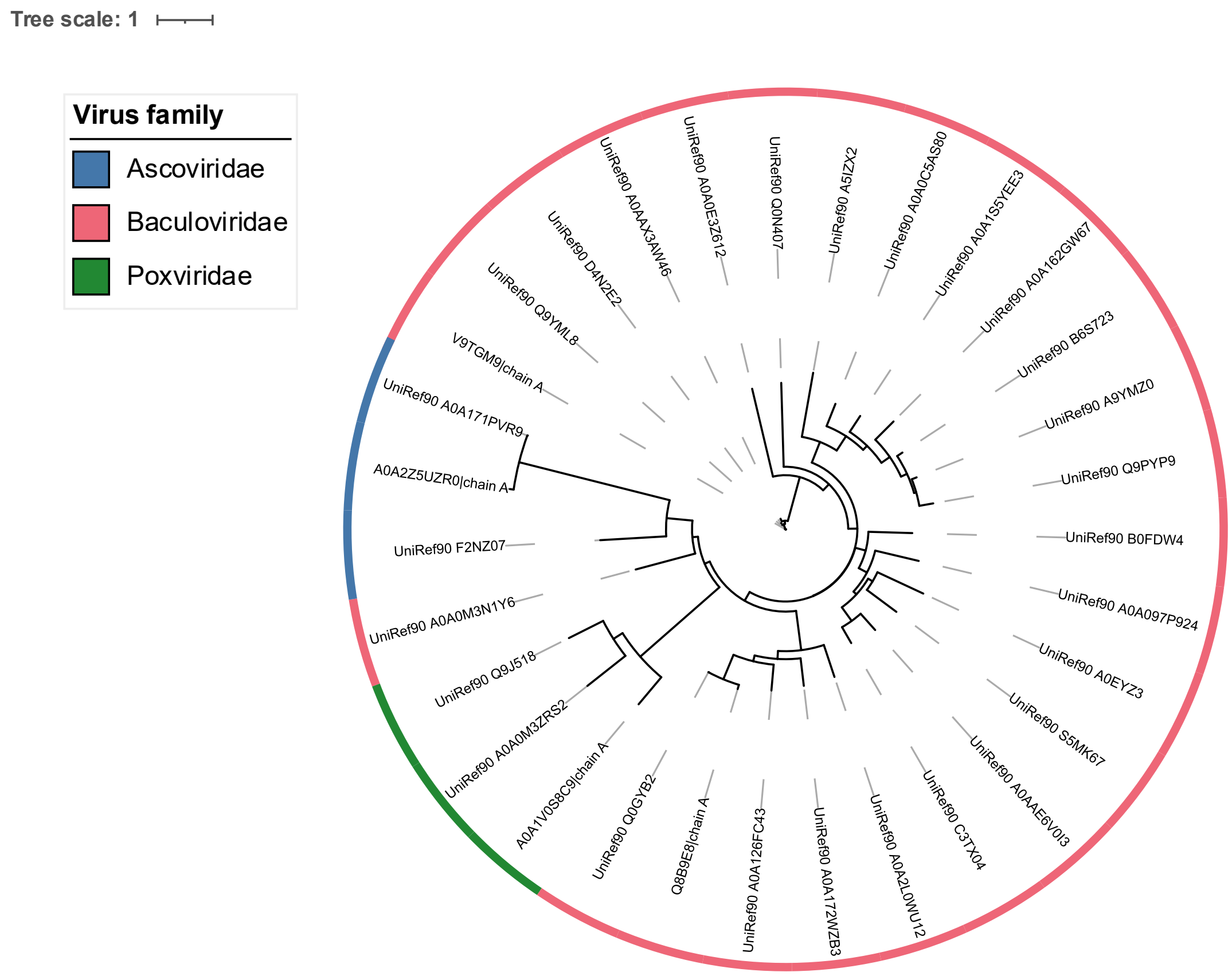

**Supplementary Figure 7. Phylogenetic distribution of C027 homologues across viral families.** Unrooted phylogenetic tree of 31 C027 homologues. The outer color strip denotes viral-family assignment: Ascoviridae (blue; *n* = 3), Baculoviridae (pink; *n* = 25), and Poxviridae (green; *n* = 3). Branch lengths indicate amino-acid substitutions per site;

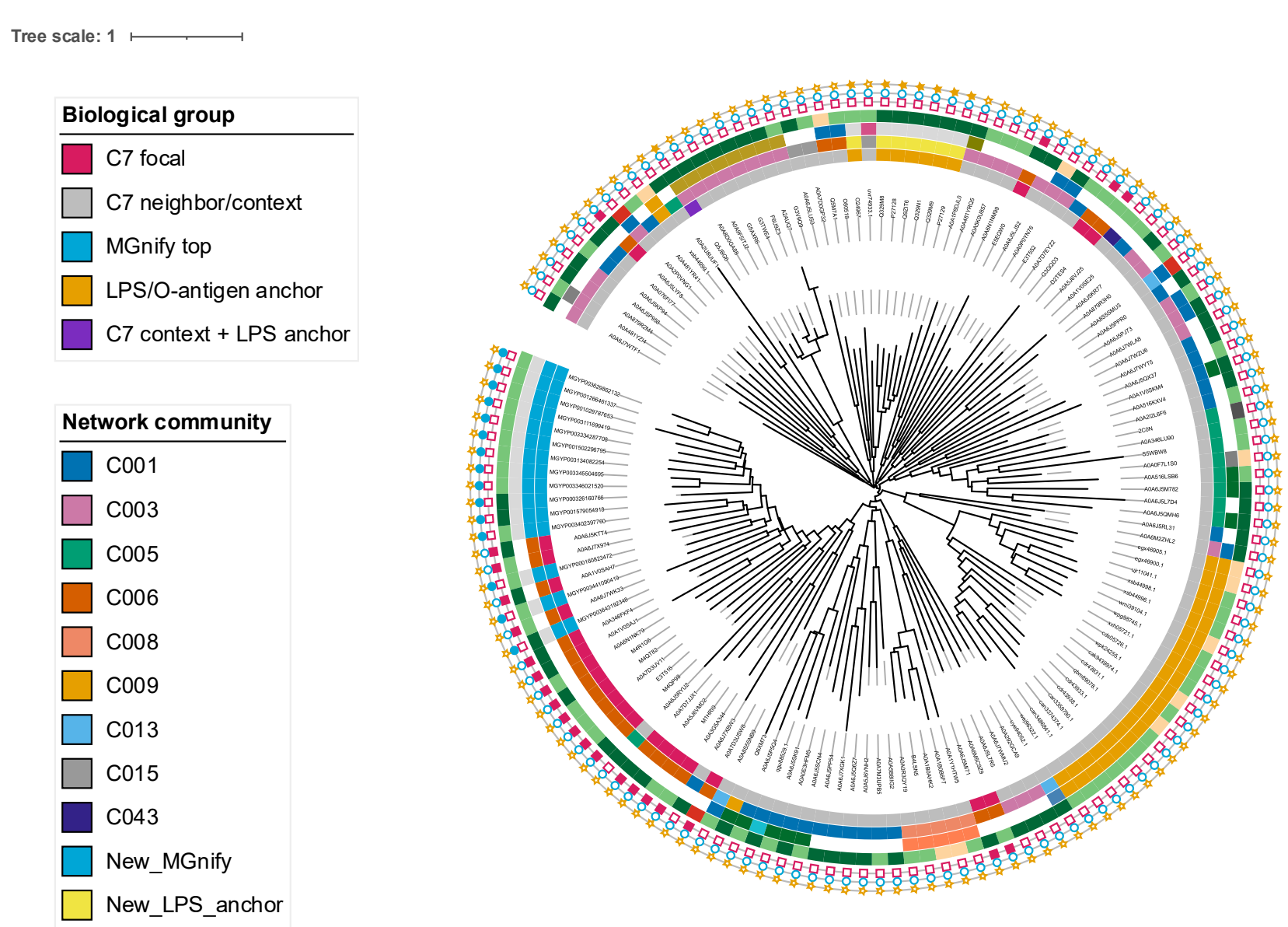

**Supplementary Fig 8. Structural phylogeny of viral Cluster 7 and its extended neighbourhood.** MAD-rooted FoldTree of 143 non-redundant structures spanning Cluster 7 focal proteins, network-neighbour proteins, MGnify structural hits and LPS/O-antigen-associated glycosyltransferase anchors. Branch lengths represent FoldTree-corrected structural distance. Concentric tracks indicate biological group, structure-similarity-network community, CAZy-family assignment and structure-confidence class; outer symbols mark Cluster 7 focal proteins, MGnify hits and LPS/O-antigen anchors.

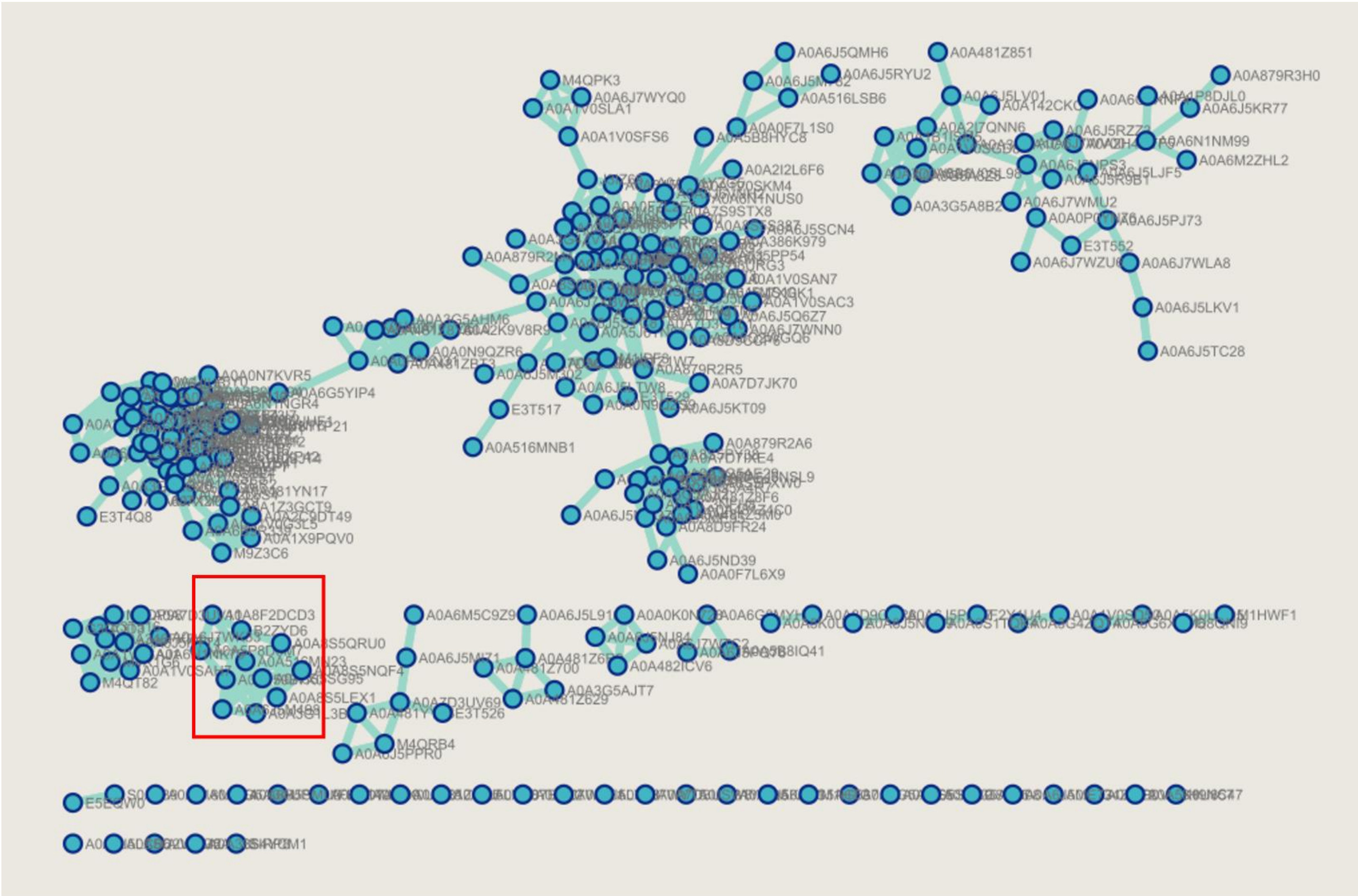

**Supplementary Figure 9. Iterative structural clustering identifies cluster 8 as a dark viral GT subnetwork.**

Structure similarity network of viral glycosyltransferase candidates generated from all-vs-all structural comparison. Nodes represent predicted viral GT-like proteins and edges represent pairwise structural similarity above the selected threshold. Iterative refinement of the initial clustering result revealed a small, distinct subnetwork, designated cluster 8 (red box). Cluster 8 was separated from the major viral GT communities and was prioritized for further analysis because its members lacked confident conventional sequence-based annotation but retained GT-like structural features and were subsequently found in DNA-modification-associated gene neighborhoods.

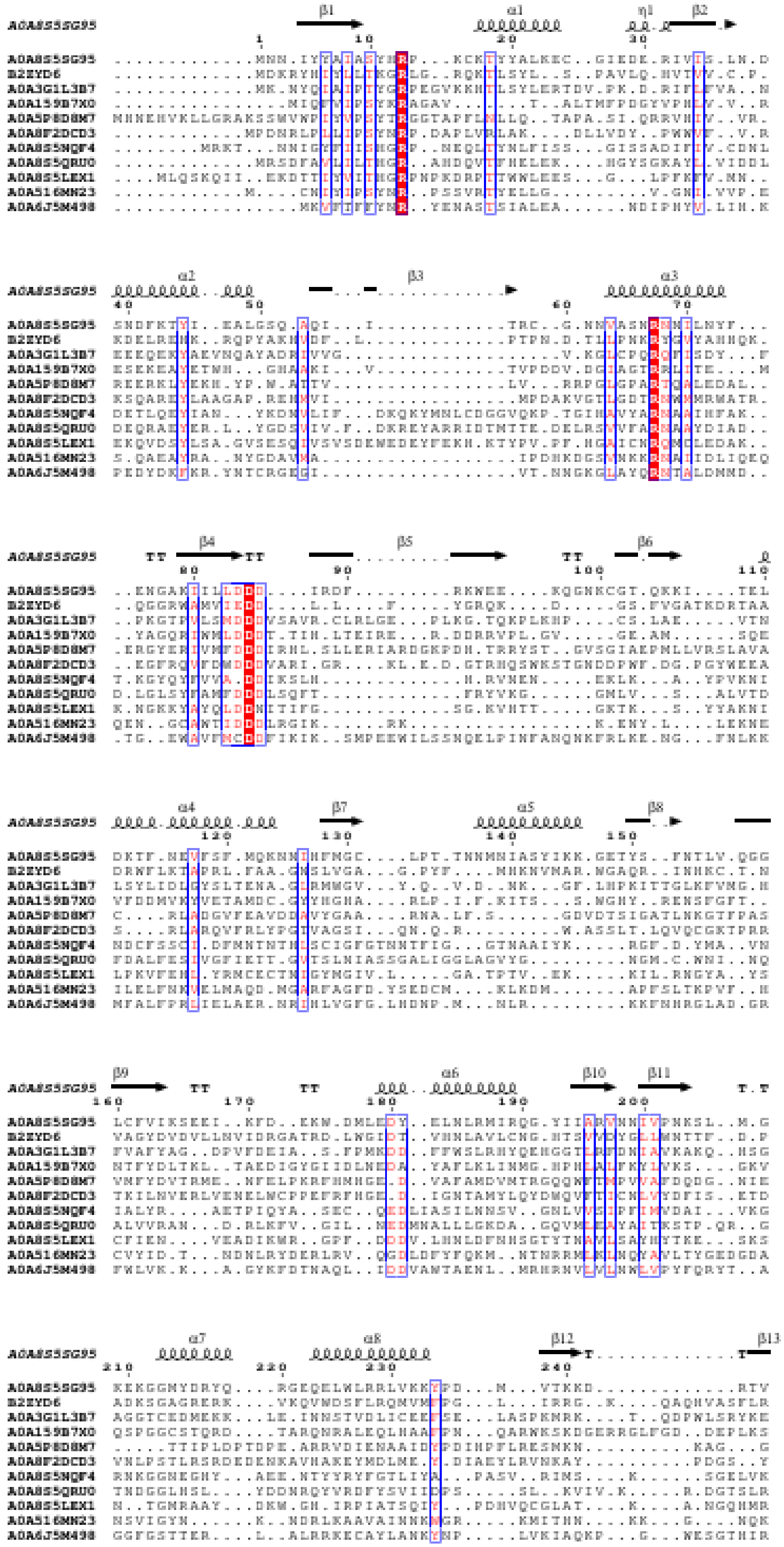

A0A8S5SG95
B2ZYD6
A0A3G1L3B7
A0A159B7X0
A0A5P8D8M7
A0A8F2DCD3
A0A8S5NQF4
A0A8S5QRU0
A0A8S5LEX1
A0A516MN23
A0A6J5M498

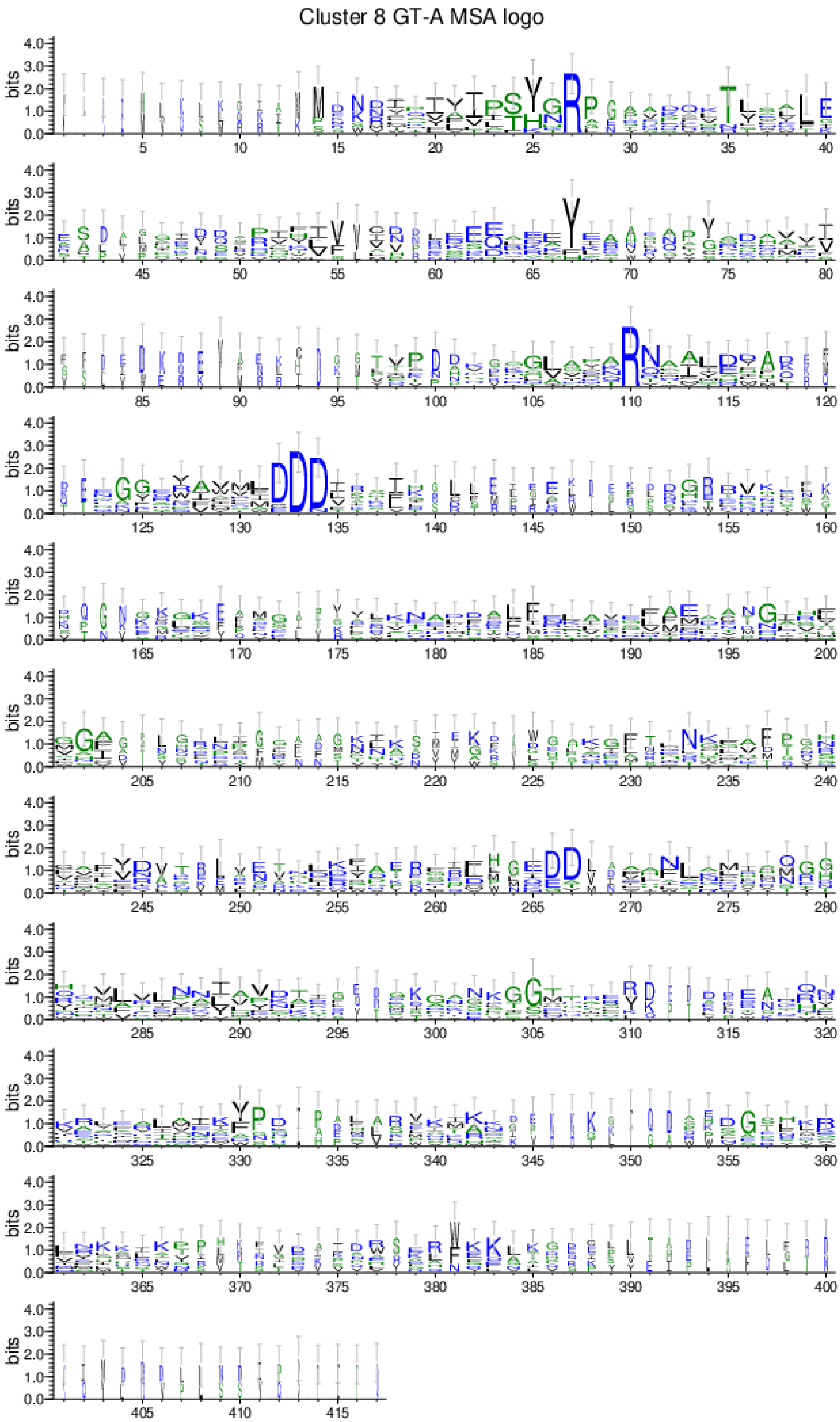

**Supplementary Figure 10.** Multiple sequence alignment of cluster 8

Multiple sequence alignment between cluster 8 members were generated by Foldmason. Visualization were performed by Esprit2 (ref). Sequence conservation were generated by Weblogo.

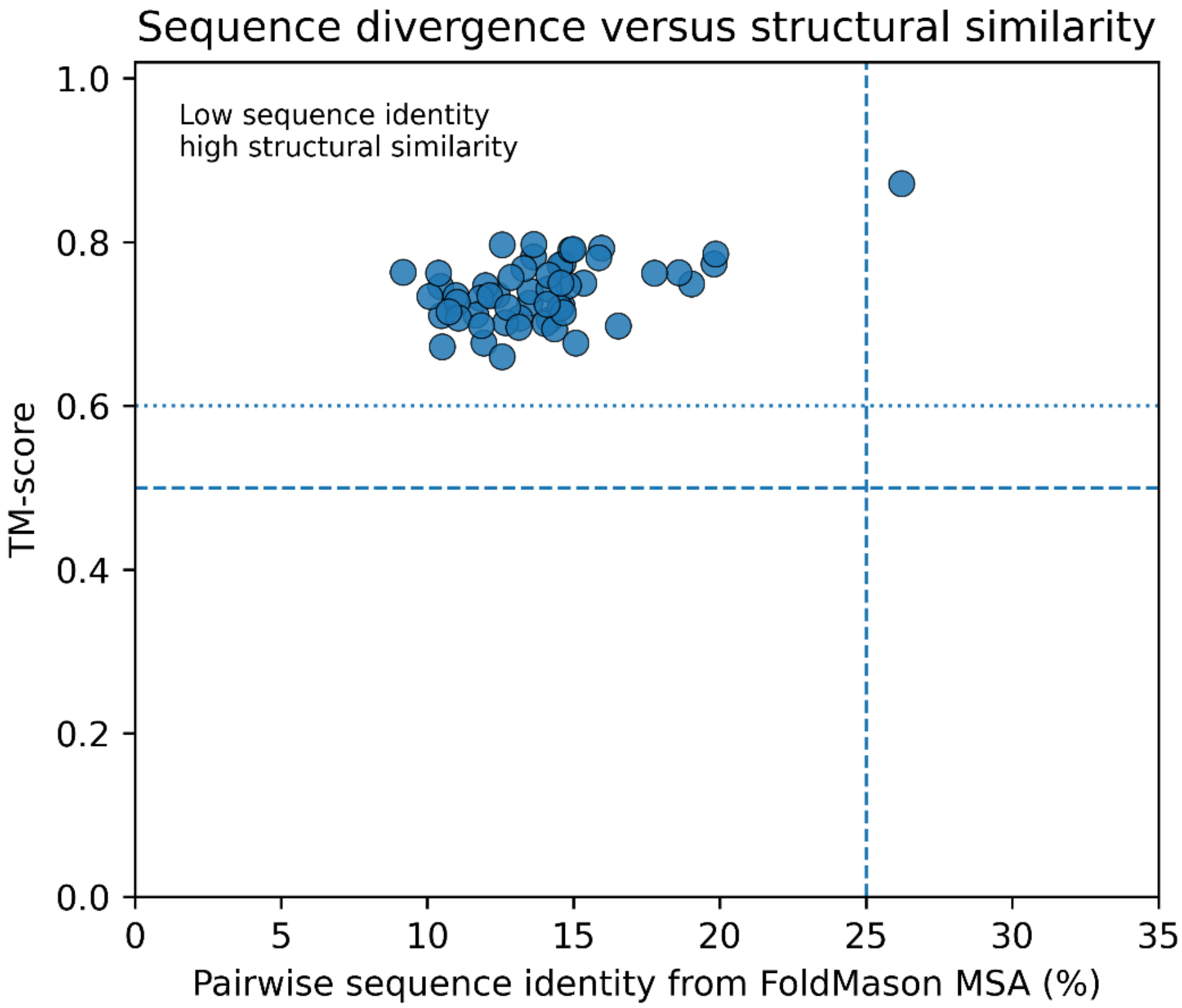

**Supplementary Figure 11. Multiple sequence alignment of cluster 8**

Each point represents one pairwise comparison between Cluster 8 proteins. The axis shows pairwise sequence identity derived from the FoldMason structure-guided amino-acid MSA, calculated over columns where both proteins contain residues. The y-axis shows the average TM-score from pairwise TM-align structural superposition.

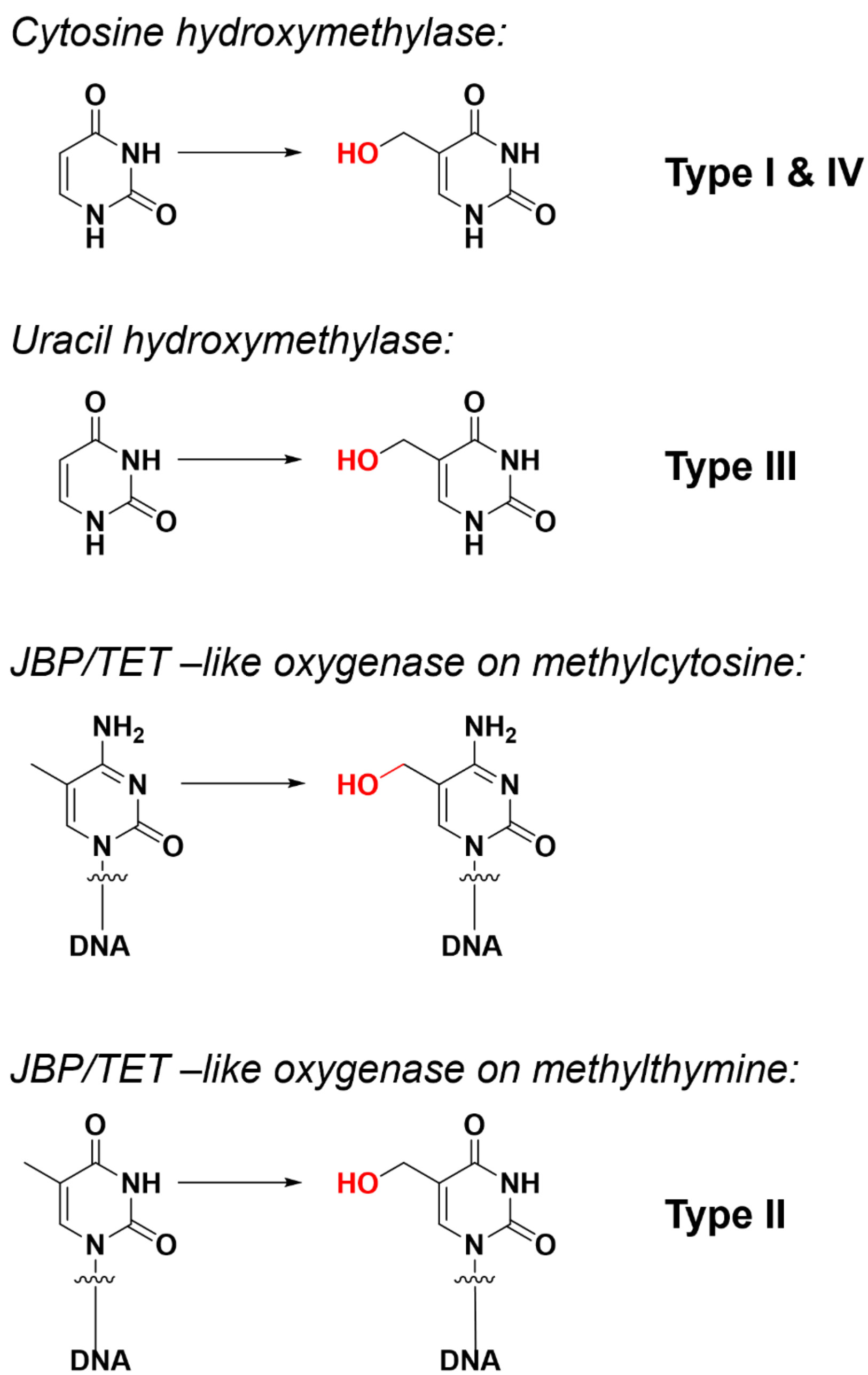

**Supplementary Fig. 12. Hydroxymethyl-pyrimidine precursor chemistries associated with DNA-linked cluster 8 architectures.** Schematic overview of the predicted precursor-generating reactions associated with the DNA-modification-linked cluster 8 locus types. Type I and Type IV architectures are predicted to use cytosine-derived nucleotide precursor routes, in which dCMP hydroxymethylase-like enzymes generate 5-hydroxymethyl-dCMP/5hmC-type precursors for downstream DNA glycosylation. Type II architectures are predicted to use a post-replicative TET/JBP-like 2-oxoglutarate/Fe(II)-dependent oxygenase route, in which thymine/5-methyluracil already embedded in DNA is oxidized to 5-hydroxymethyluracil. Type III architectures are predicted to use a uracil-derived nucleotide precursor route, in which a dUMP hydroxymethylase-like enzyme generates a 5-hydroxymethyl-dUMP/5hmdU-type precursor. Red highlights indicate the hydroxymethyl handle introduced or generated by each precursor-forming enzyme. These hydroxymethyl-pyrimidine handles provide the proposed nucleophilic acceptors for downstream cluster 3-associated DNA glycosylation.

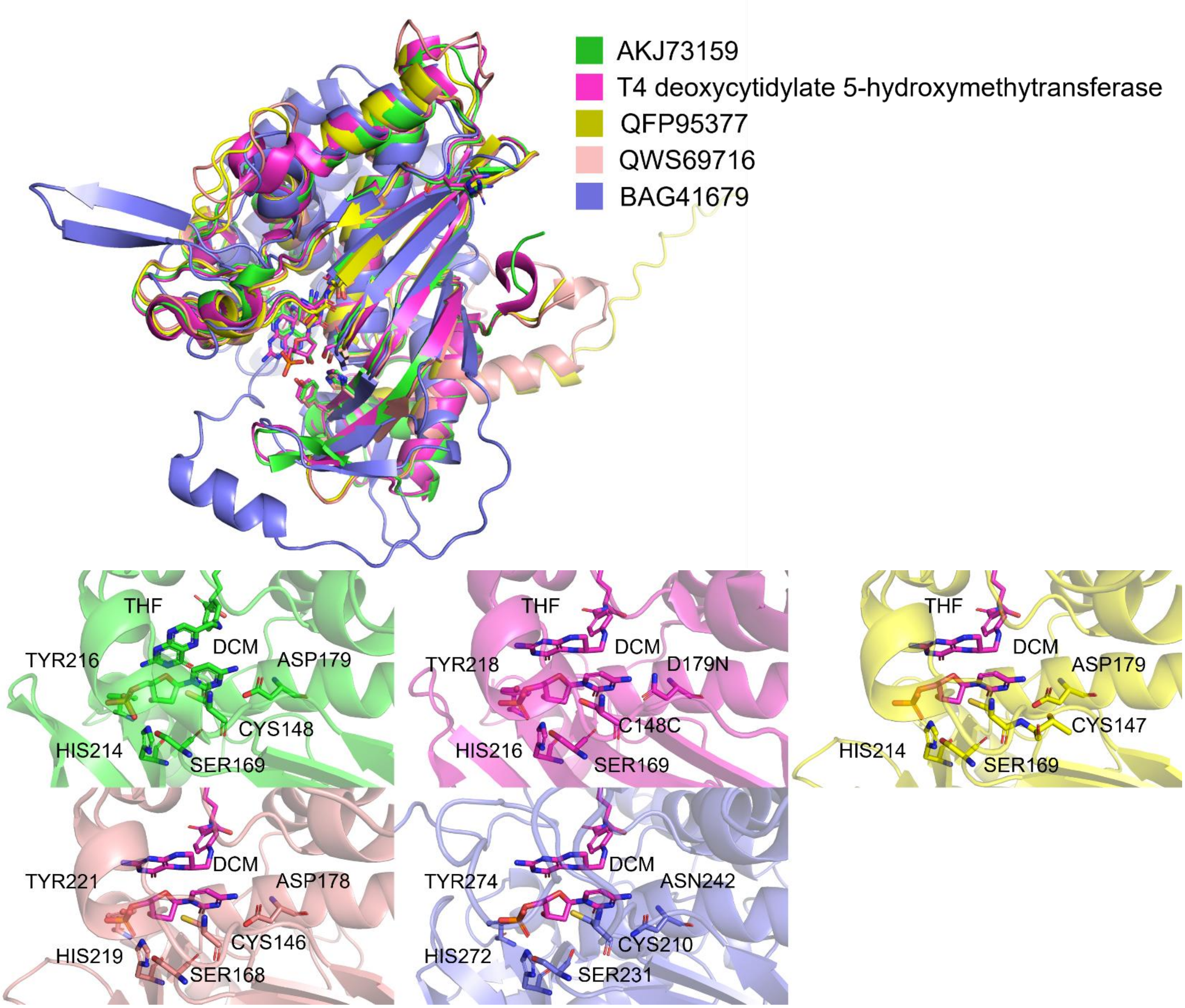

**Supplementary Fig. 13. Structure-guided annotation of hydroxymethyl-pyrimidine precursor enzymes in DNA-associated cluster 8 loci.**
Structural and active-site comparison of predicted folate-dependent pyrimidine C5-hydroxymethyltransferases encoded near DNA-associated cluster 8 loci. The top panel shows structural superposition of representative cluster 3-neighborhood enzymes with the crystallographic reference T4 deoxycytidylate 5-hydroxymethyltransferase. The reference structure contains bound tetrahydrofolate (THF) and deoxycytidylate/dCMP-site ligand, providing an active-site template for substrate-recognition analysis. Predicted enzymes from Type I and Type IV loci, including AKJ73159, QFP95377, and QWS69716, preserve the cytosine-directed active-site architecture of T4-like dCMP hydroxymethyltransferases, including the catalytic cysteine, conserved histidine/serine residues, aromatic base-positioning residue, and an Asp residue positioned for cytosine recognition. These features support annotation of these proteins as dCMP C5-hydroxymethyltransferase-like enzymes that can generate 5hmC-type precursors for downstream DNA glycosylation. In contrast, the Type III enzyme BAG41679 preserves the same global fold and catalytic cysteine but contains Asn242 at the position equivalent to the cytosine-specific Asp residue in dCMP-directed enzymes. This Asp-to-Asn active-site substitution is consistent with altered pyrimidine specificity and supports annotation of BAG41679 as a dUMP hydroxymethyltransferase-like enzyme predicted to generate a 5-hydroxymethyluracil-type precursor. Together, these comparisons justify the assignment of Type I/IV loci as 5hmC-linked pathways and the Type III locus as a distinct 5hmdU-linked hydroxymethyl-pyrimidine pathway.

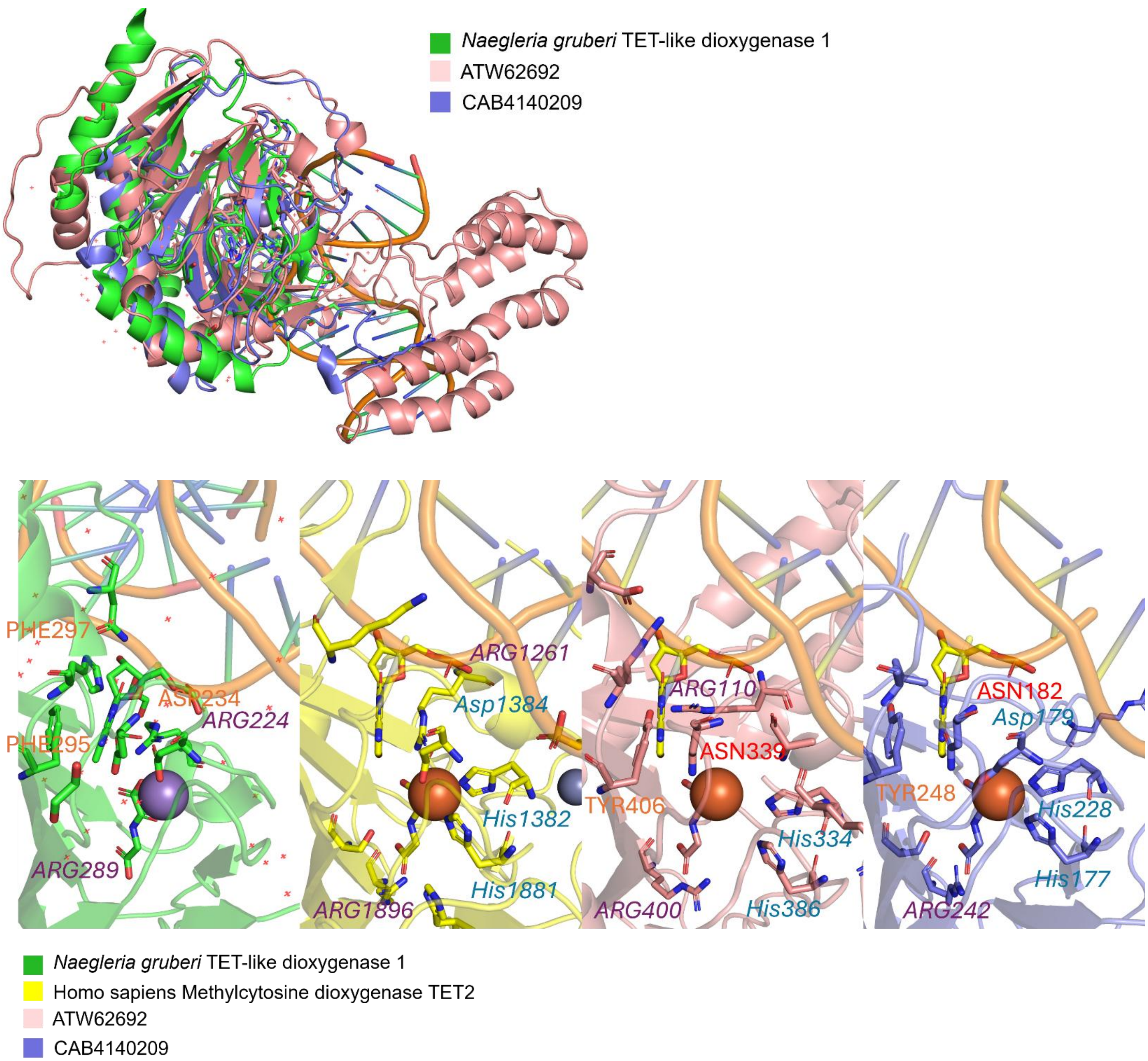

**Supplementary Fig. 14. Structure-guided annotation of Type II TET/JBP-like oxygenases supports a thymine/5-methyluracil hydroxylation module.** Predicted Type II oxygenases encoded in DNA-associated cluster 8 loci were compared with structurally characterized TET/JBP-like 2-oxoglutarate/Fe(II)-dependent dioxygenases. The left panel shows structural superposition of Type II oxygenases with reference Naegleria gruberi TET-like dioxygenase 1 and human TET-family oxygenase structures. The right panels show active-site comparisons centered on the catalytic metal, 2-oxoglutarate-binding region, and flipped-base

recognition pocket. Residues involved in Fe(II) coordination are highlighted in blue, whereas residues involved in 2-oxoglutarate binding are highlighted in magenta. These catalytic features are spatially conserved across the Type II oxygenases and reference enzymes, supporting annotation of the Type II proteins as TET/JBP-like 2OG/Fe(II) dioxygenases. The flipped-base recognition pocket is also conserved, including spatially aligned aromatic residues positioned to stack against or stabilize the extruded pyrimidine base. Notably, both Type II oxygenases contain an Asp-to-Asn substitution at the base-recognition position corresponding to the cytosine/5mC-specific Asp in TET-like enzymes. This substitution is consistent with a shift toward thymine/5-methyluracil recognition and supports a model in which the Type II oxygenases generate 5-hydroxymethyluracil-like DNA bases for downstream glycosylation by neighboring cluster 8 GTs.

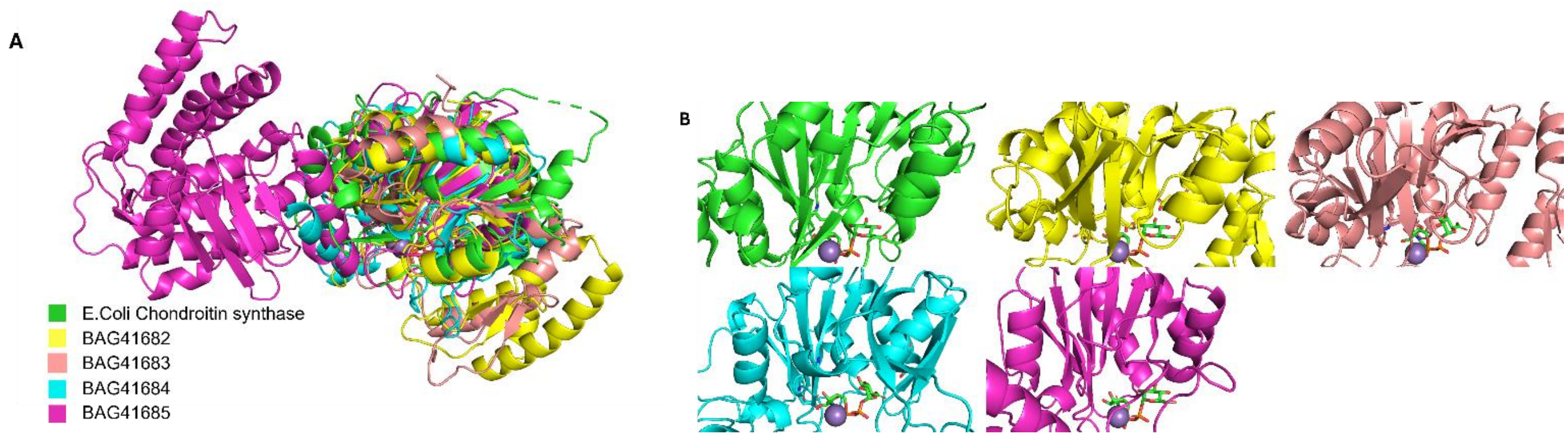

**Supplementary Fig. 15. Structural annotation of additional GT-like enzymes in the B2ZYD6 Type III locus.**
Additional proteins including BAG41682, 41683, 41684, 41685 encoded in the B2ZYD6 Type III locus were compared structurally with *E. coli* chondroitin synthase, a GT2-fold glycosyltransferase reference. Global structural superposition and active-site inspection show that several B2ZYD6 proteins adopt GT-like folds with conserved nucleotide-sugar donor-binding architecture, supporting their annotation as putative glycosyltransferases which contribute to downstream elaboration of a DNA-linked glycan.

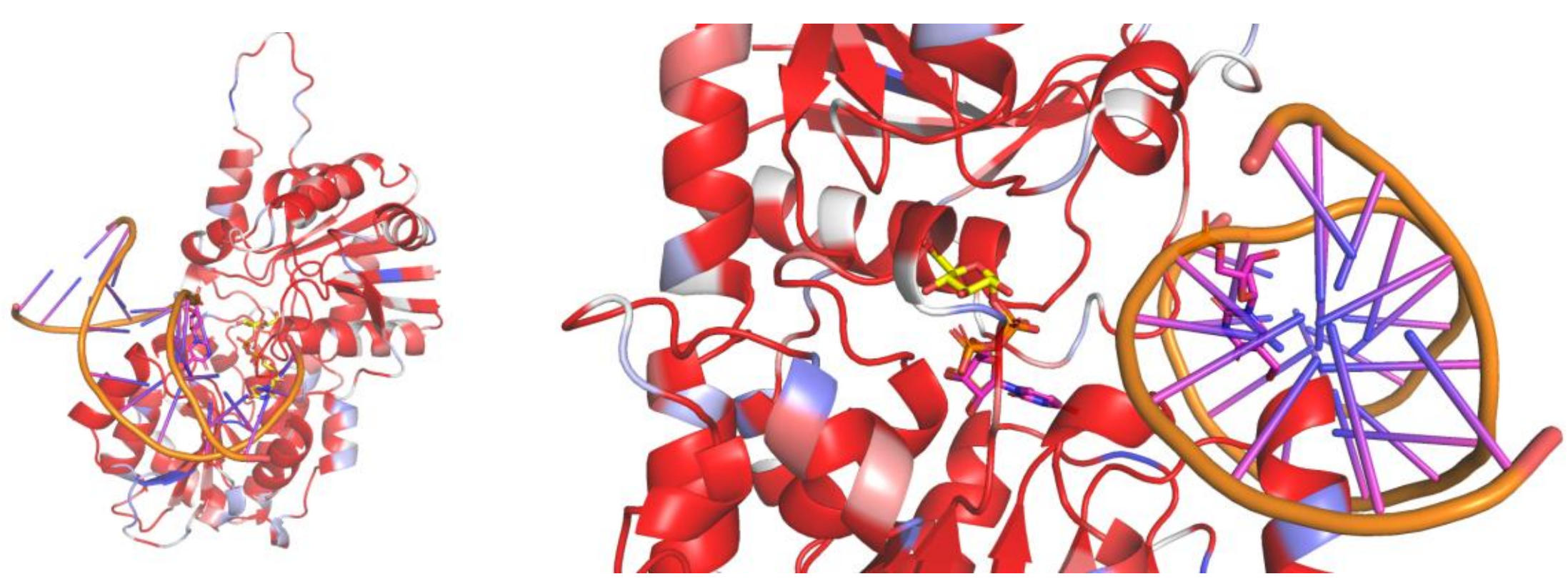

**Supplementary Fig. 16 .AKJ73146 preserves conserved features of T4-like α-glucosyltransferases.**
Sequence alignment result between Enterobacteria phage T4 α-glucosyltransferases and AKJ73146 was mapped onto the predicted AKJ73146 structure. Conserved residues are shown in red and variable residues in blue. The global view (left) shows conservation across the catalytic cleft and DNA-binding surface, while the active-site zoom (right) highlights strong conservation around the predicted catalytic pocket. This pattern supports AKJ73146 as a T4-like α-glucosyltransferase that likely performs the first glucosylation step on 5hmC-DNA.

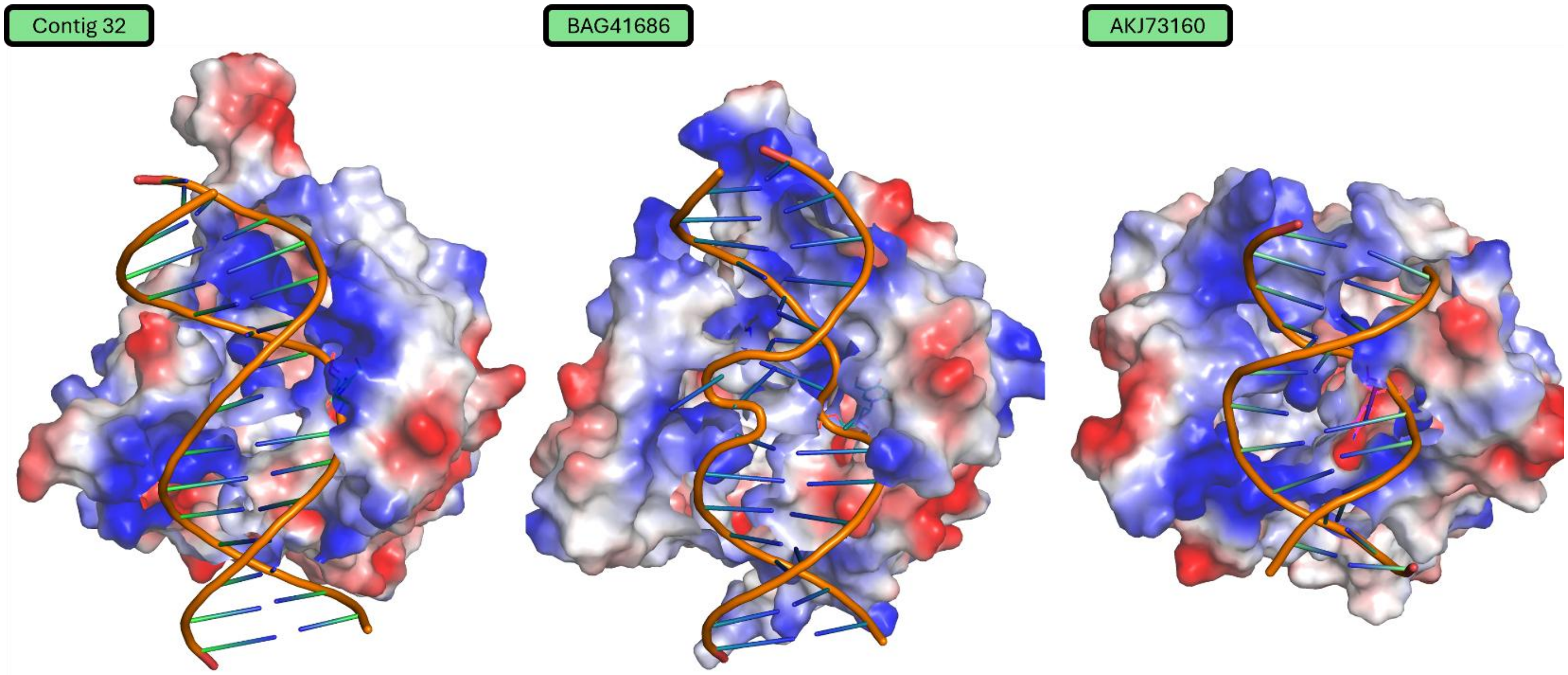

**Supplementary Fig. 17. BAG41686 and AKJ73160 retain electropositive DNA-binding grooves.** Electrostatic surface maps of contig 32, BAG41686, and AKJ73160 modeled with modified DNA acceptors. Blue indicates electropositive surface potential and red indicates electronegative potential. The reference DNA GT contig 32 contains a positively charged groove that accommodates the DNA phosphate backbone. BAG41686 preserves a comparable groove, supporting direct binding of hydroxymethylated DNA. AKJ73160 also retains an electropositive DNA-binding surface, consistent with DNA-associated substrate recognition, although its active site is remodeled to recognize Glc-5hmC rather than naked 5hmC. These results support distinct acceptor-recognition modes within a shared DNA-binding framework.

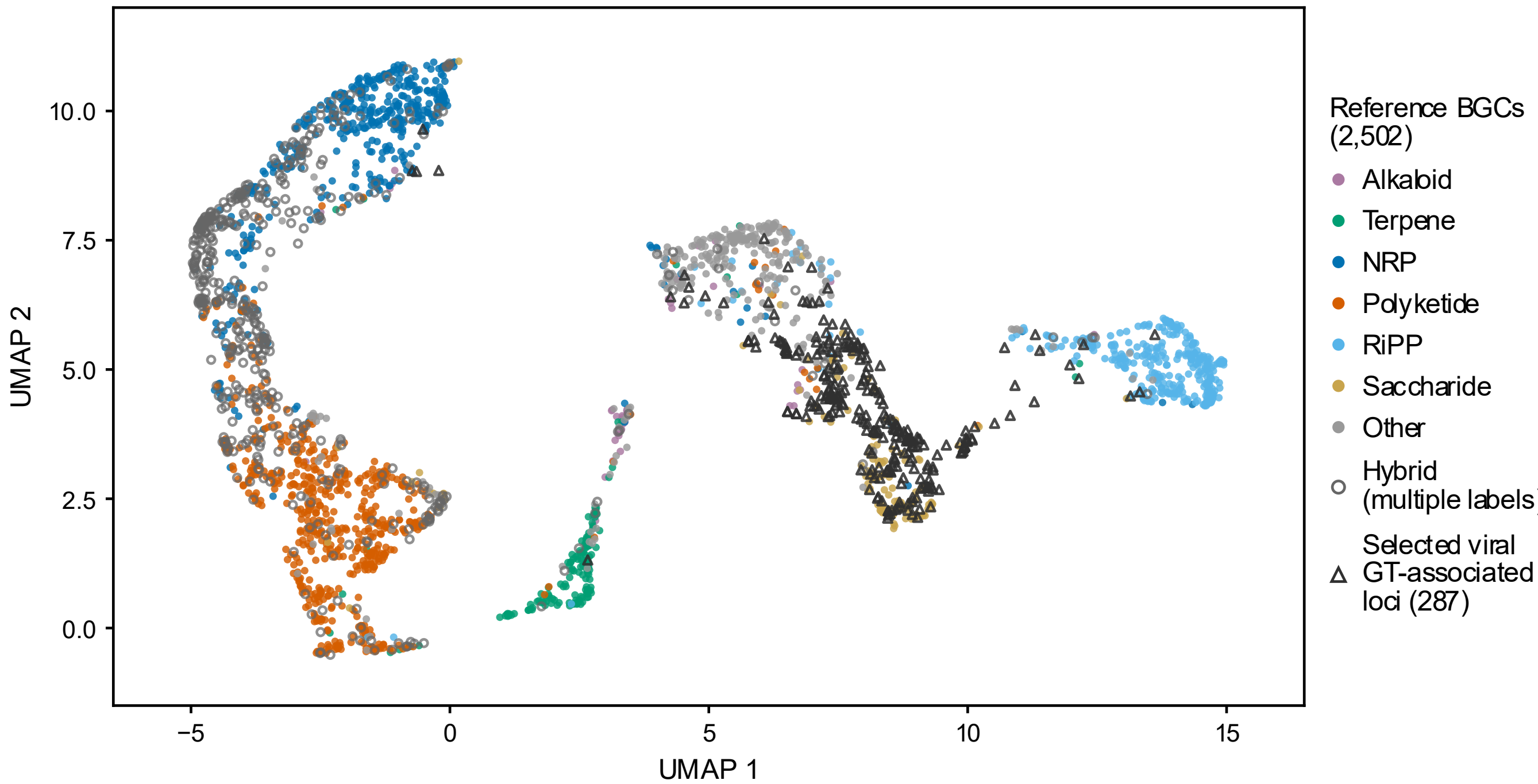

**Supplementary Figure 18.** Uniform manifold approximation and projection (UMAP) of 320-dimensional BGC-Prophet embeddings comprising 2,502 MIBiG reference biosynthetic gene clusters (BGCs) and 556 viral GT-associated loci. Reference BGCs are shown as circles coloured according to their annotated biosynthetic classes: alkaloid (n = 53, 2.1%), terpene (n = 150, 6.0%), non-ribosomal peptide (NRP; n = 468, 18.7%), polyketide (n = 606, 24.2%), ribosomally synthesized and post-translationally modified peptide (RiPP; n = 330, 13.2%), saccharide (n = 130, 5.2%) and other (n = 298, 11.9%). Open circles represent hybrid BGCs assigned to multiple biosynthetic classes (n = 467, 18.7%). Open triangles indicate 287 selected viral GT-associated loci, identified through Saccharide-class prediction and/or nearest-reference retrieval (P@10 ≥ 0.5). The remaining 269 viral loci were included in the global embedding but are not displayed. Percentages are calculated relative to the 2,502 reference BGCs.

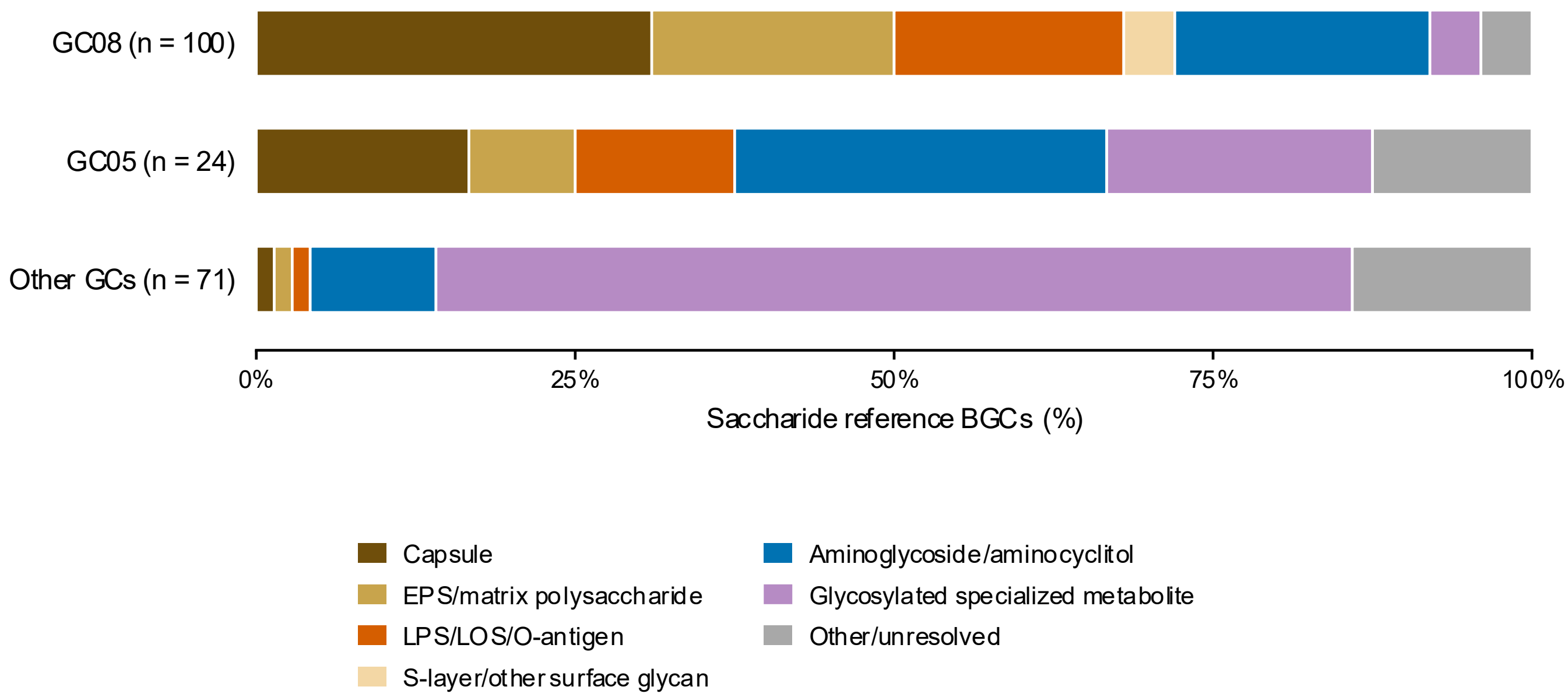

**Supplementary Figure 19. Chemotype composition of Saccharide-associated reference BGCs across global Leiden communities.** Stacked bars show the relative proportions of annotated saccharide-related chemotypes among MIBiG reference biosynthetic gene clusters (BGCs) assigned to GC08 (n = 100), GC05 (n = 24) and the remaining global communities combined (Other GCs; n = 71). Global communities were identified by Leiden clustering of the joint embedding space comprising reference BGCs and viral GT-associated loci. Reference BGCs were categorized as capsule, extracellular polysaccharide (EPS)/matrix polysaccharide, lipopolysaccharide/lipooligosaccharide/O-antigen (LPS/LOS), S-layer/other surface glycan, aminoglycoside/aminocyclitol, glycosylated specialized metabolite, or other/unresolved. Percentages were calculated within each group, and n denotes the number of Saccharide-associated reference BGCs, including those with hybrid class annotations.

## Supplementary Methods

### Definition and mapping of the pan-GT31 structural core

A common FoldMason alignment (Gilchrist et al., 2026) was generated for 17 GT31 reference structures assigned to four multi-member structural subclusters: SC01 ($n = 5$), SC02 ($n = 4$), SC03 ($n = 4$) and SC04 ($n = 4$). Candidate pan-GT31 core positions were evaluated at homologous alignment columns using subcluster occupancy, model confidence and medoid-anchored local structural conservation. A strict pan-GT31 core position was required to satisfy all prespecified criteria in each of the four subclusters: occupancy ≥0.70; medoid-residue pLDDT ≥70; within-subcluster local lDDT ≥0.75 with neighbourhood coverage ≥0.60; and cross-subcluster local lDDT ≥0.65 with neighbourhood coverage ≥0.40 and focal-position presence ≥0.50. These criteria identified 98 strict core columns. Supported core positions satisfied the same occupancy and confidence requirements but were permitted to retain structural support in three of the four subclusters while satisfying the median geometric-support criteria. The supported set contained 147 columns, comprising the 98 strict positions and 49 additional supported-only positions.

Reference-derived core columns were mapped to Q18515 through its coordinates in the reference-only FoldMason alignment. F2Y1U4 was excluded from core discovery and mapped subsequently through the query-inclusive alignment, using Q9N294 residue indices as the coordinate bridge. Q18515 contained all 98 strict and 49 supported-only positions; F2Y1U4 mapped to 93 strict and 41 supported-only positions. Group-level placement metrics were calculated separately for SC01–SC04. The displayed values represent the arithmetic mean of the four subcluster-level medians.

### Structure-guided mapping of sequence identities between R699 and Q20679

For structure-guided comparison of R699 with a metazoan PLOD-like collagen glycosyltransferase, the glycosyltransferase-relevant region of Q20679 was isolated by removing the lysyl-hydroxylase domain prior to alignment. R699 and the trimmed Q20679 construct were then aligned using FoldMason (Gilchrist et al., 2026), and the resulting pairwise amino-acid alignment was used as the coordinate framework for residue classification.

Alignment columns were parsed sequentially and converted to ungapped residue indices for each protein. These indices were then mapped onto the corresponding structure coordinates of R699 and trimmed Q20679 to recover the residue numbering used for visualization. Exact sequence identity was assigned when the aligned residues in the two proteins were identical. R699 active-site residues were defined from the experimentally characterized donor-binding, metal-coordinating and acceptor-binding network of R699 and included residues 48, 54, 62, 83, 85, 86, 114, 118, 138, 162, 163, 164, 216, 218, 222 and 438.

Each aligned residue position was assigned to one of four visualization classes: (i) active-site identity, when an exact sequence identity overlapped an R699 active-site residue; (ii) sequence identity, for exact identities outside the R699 active-site set; (iii) active-site non-identity, for R699 active-site residues aligned to non-identical residues in Q20679; and (iv) aligned non-identity, for all other aligned non-identical positions. Unaligned positions and background residues were rendered separately as non-highlighted regions. Residue classes were displayed on the R699 and Q20679 structures in PyMOL, using magenta for active-site identity, teal for non-active-site sequence identity, orange for active-site non-identity, and grey tones for other aligned or background regions. For structural context, UDP from the R699 structure was shown as grey sticks.